\documentclass[twocolumn]{aastex62}

\usepackage{amsmath, mathrsfs, amssymb}
\usepackage{graphicx}
\usepackage{rotating}
\usepackage{morefloats}
\usepackage{color}
\usepackage{enumitem}
\usepackage{verbatim}
\usepackage{hyperref}
\usepackage{booktabs}
\usepackage{xfrac}

\setlist[itemize]{leftmargin=*}

\DeclareMathOperator*{\argmin}{arg\,min}

\received{August 20, 2020}
\revised{October 08, 2020}
\accepted{October 15, 2020}
\submitjournal{\textit{Monthly Notices of the Royal Astronomical Society}}

\newcommand{\fr}{f_{r}}

\definecolor{gimppurple}{HTML}{AD26FB}
\definecolor{lightgrey}{HTML}{E0E0E0}
\definecolor{deepblue}{rgb}{0,0,0.5}
\definecolor{deepred}{rgb}{0.6,0,0}
\definecolor{deepgreen}{rgb}{0,0.5,0}

\begin{document}
\label{firstpage}

\shorttitle{Giving Uncertainties an $f$}
\shortauthors{Salvesen \& Miller (2020)}

\title{\LARGE{Black Hole Spin in X-ray Binaries: Giving Uncertainties an $f$}}

\correspondingauthor{Greg Salvesen}
\email{gregsalvesen@gmail.com}

\author[0000-0002-9535-4914]{Greg Salvesen}
\affil{CCS-2, Los Alamos National Laboratory, P.O. Box 1663, Los Alamos, NM 87545, USA.}
\affil{Center for Theoretical Astrophysics, Los Alamos National Laboratory, Los Alamos, NM 87545, USA.}

\author[0000-0001-6432-7860]{Jonah M. Miller}
\affil{CCS-2, Los Alamos National Laboratory, P.O. Box 1663, Los Alamos, NM 87545, USA.}
\affil{Center for Theoretical Astrophysics, Los Alamos National Laboratory, Los Alamos, NM 87545, USA.}

\begin{abstract}
The two established techniques for measuring black hole spin in X-ray binaries often yield conflicting results, which must be resolved before either method may be deemed robust. In practice, black hole spin measurements based on fitting the accretion disc continuum effectively do not marginalize over the colour correction factor $f_{\mathrm{col}}$. This factor parametrizes spectral hardening of the disc continuum by the disc atmosphere, whose true properties are poorly constrained. We incorporate reasonable systematic uncertainties in $f_{\mathrm{col}}$ into the eight (non-maximal) black hole spin measurements vetted by the disc continuum fitting community. In most cases, an $f_{\rm col}$ uncertainty of $\pm$0.2--0.3 dominates the black hole spin error budget. We go on to demonstrate that plausible departures in $f_{\rm col}$ values from those adopted by the disc continuum fitting practitioners can bring the discrepant black hole spins into agreement with those from iron line modeling. Systematic uncertainties in $f_{\rm col}$, such as the effects of strong magnetization, should be better understood before dismissing their potentially dominant impact on the black hole spin error budget.
\end{abstract}


\section{Introduction}
\label{sec:intro}
Black hole spin is a driver and diagnostic probe of numerous astrophysical phenomena: Relativistic jets that inject energy into their surroundings may be powered by tapping into the black hole spin energy reservoir \citep{Penrose1969, BlandfordZnajek1977}. The cosmological evolution of the coupled growth between supermassive black holes and their host galaxies is imprinted on their spin distribution \citep{Volonteri2005}. Stellar mass black hole spins provide insights into core-collapse supernovae, as the natal spin is not expected to change appreciably over the lifetime of an X-ray binary system \citep{KingKolb1999}. Accretion discs are also subject to interesting dynamics in the presence of a spinning black hole. When the black hole spin axis is misaligned to the accretion disc rotational axis, the disc experiences a differential precession \citep{LenseThirring1918}, which may manifest as quasi-periodic variability features \citep[e.g.,][]{StellaVietri1998, Fragile2007, Ingram2009}. If strong enough, this precession can cause the disc to break into distinct interacting rings of gas \citep{Nixon2012b}, which is one candidate mechanism for black hole state transitions \citep{NixonSalvesen2014}. Trustworthy measurements of black hole spin are necessary to realize these far-reaching astrophysical implications.

Two sophisticated techniques are at the forefront of the effort to measure black hole spin. One method is based on modeling the profile of the relativistically broadened iron emission line, which is a feature in the disc reflection spectrum \citep{Fabian1989, Laor1991, Miller2009a}. The other method appeals to modeling the accretion disc continuum \citep{Zhang1997, McClintock2006}.
In practice, the spins of supermassive black holes are only accessible with the iron line diagnostic, but both approaches can be used to measure the spins of stellar mass black holes in X-ray binaries. Cross comparing black hole spin measurements on the same set of X-ray binaries from these two independent techniques is crucial before deeming either to be robust.

\begin{figure}
    \begin{center}
        \includegraphics[width=0.495\textwidth]{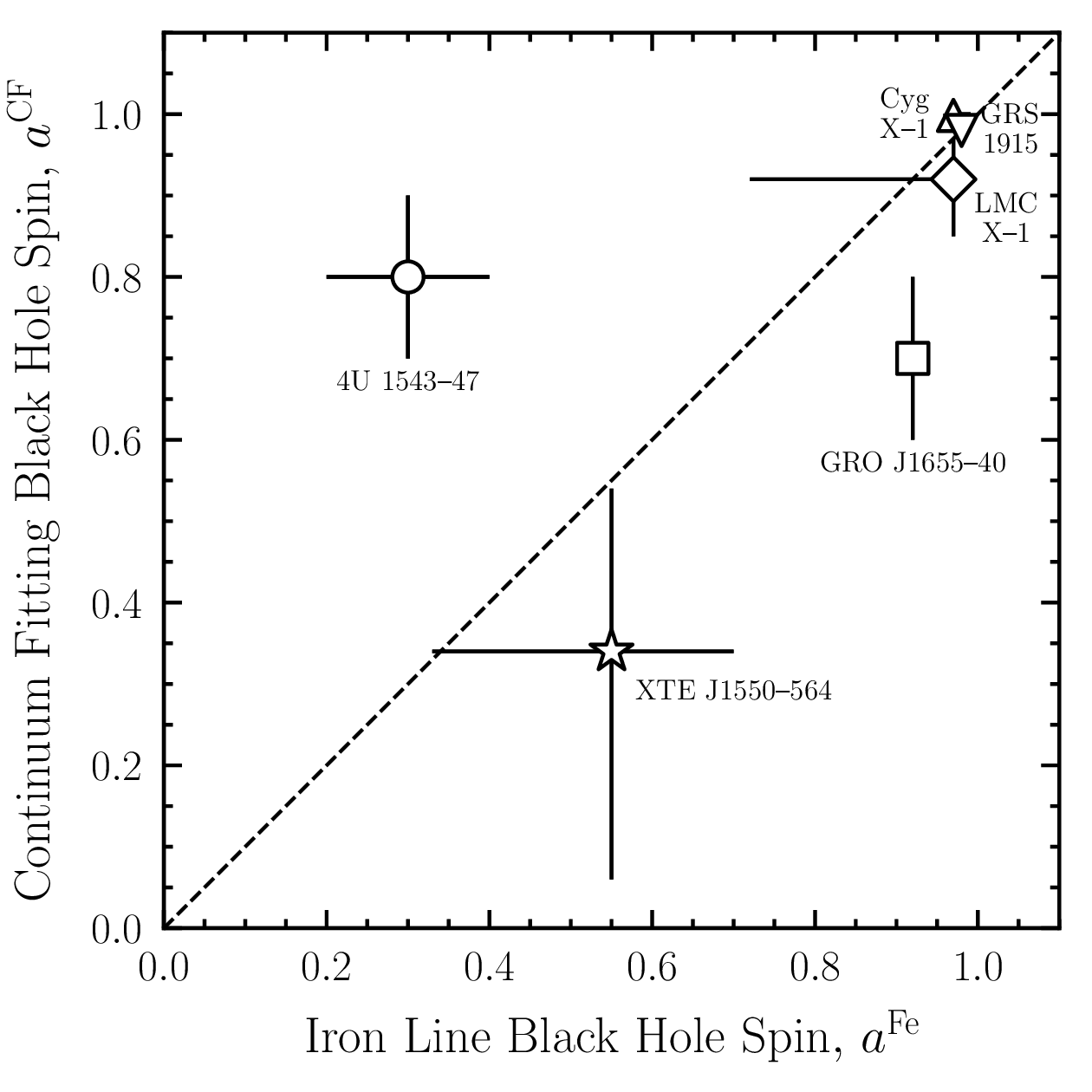}
    \end{center}
    \vspace{-6mm}
    \caption{Black hole spin measurements from the iron line ($a^{\mathrm{Fe}}$) and continuum fitting ($a^{\mathrm{CF}}$) techniques plotted against one another for the sources in Table \ref{tab:obs}. Error bars are at the $68\%$-level. Spin measurements for 4U 1543--47, GRO J1655--40, and XTE J1550--564 are in disagreement, as these points do not lie on the \textit{dashed line} that indicates agreement between the two techniques.}
    \label{fig:bhspins}
\end{figure}

Now mature, both methods report black hole spins for a modest population of X-ray binaries \citep{Reynolds2014, McClintock2014}. For the six systems with spins vetted by both the iron line and disc continuum fitting communities, Figure \ref{fig:bhspins} shows that the techniques agree when the spin is measured to be near-maximal, but disagree otherwise. \citet{Steiner2011} measured the spin of XTE J1550--564 with each technique independently, finding weak agreement as a consequence of large uncertainties. \citet{Miller2009a} employed both techniques simultaneously for a sample of X-ray binaries by linking the spin parameter between the disc reflection and disc continuum spectral components. While broad agreement was found with previous spin measurements from the iron line technique alone, the systems 4U 1543--47 and GRO J1655--40 showed strong discrepancies with previous disc continuum fitting results. Notably, \citet{Dong2020} modeled the reflection spectrum of 4U 1543--47 and report a spin measurement consistent with that from continuum fitting. Significant discrepancy in the black hole spin obtained with two different methods for the same source is troublesome. This warrants an investigation into the validity of the assumptions behind both approaches, but in this paper we focus on disc continuum fitting.

The disc continuum fitting method applies to black hole X-ray binaries in high-luminosity ($L_{X} \sim 0.1~L_{\rm Edd}$), soft-spectral states that produce predominantly thermal emission, which is remarkably well-described by a colour-corrected, multi-temperature, blackbody model for a geometrically thin, optically thick accretion disc \citep[e.g.,][]{ShakuraSunyaev1973, GierlinskiDone2004}. This disc continuum model approximates the local (i.e., at radius $R$) specific flux as a colour-corrected blackbody
\begin{equation}
F_{\nu} = \frac{\pi}{f_{\rm col}^{4}} B_{\nu}\left( f_{\rm col} T_{\rm eff} \right), \label{eqn:Fccbbody}
\end{equation}
where $B_{\nu}$ is the Planck function, $T_{\rm eff}$ is the effective temperature, and $f_{\rm col}$ is the colour correction. Local disc atmosphere models in the density and temperature regimes applicable to the high/soft state of X-ray binaries generally support this colour-corrected blackbody approximation, and its extension to a disc-integrated (i.e., continuum) model with a single $f_{\rm col}$ value, at least for photon energies in the range $\sim 1$--$10~{\rm keV}$ \citep[e.g.,][]{ShimuraTakahara1995b, Davis2005}.

The colour-corrected, multi-temperature, disc blackbody model can be cast in terms of only two parameters \citep[e.g.,][]{Mitsuda1984}, such that spectral fits directly yield a characteristic colour temperature $T_{\rm col}^{\ast}$ and the disc flux normalization $K_{\rm flux}$. Worked into $K_{\rm flux}$ are the static system parameters: black hole mass $M$, distance $D$, inner disc inclination\footnote{The inner disc geometry can change on observational timescales in disc warping \citep{Pringle1996} or disc tearing \citep{Nixon2012b} scenarios, so $i_{\rm disc}$ may not be a static parameter.} $i_{\rm disc}$; and the dynamic parameters: inner disc radius $r_{\rm in}$ and colour correction factor $f_{\rm col}$, which parametrizes spectral hardening of the disc continuum by atmospheric physics (see Appendix \ref{app:physics}). Combining independent constraints on the static parameters with knowledge of $f_{\rm col}$, one can convert $K_{\rm flux}$ into a measurement of $r_{\rm in}$, from which the black hole spin parameter $a$ follows by assuming $r_{\rm in}$ coincides with the location of the innermost stable circular orbit \citep[ISCO; e.g.,][]{Zhang1997}. Over time, several refinements led to the modern disc continuum fitting technique discussed in \S\ref{sec:meth}, which decomposes $K_{\rm flux}$ into its constituent parameters \citep[e.g.,][]{McClintock2006}.

At the crux of measuring black hole spin from disc continuum fitting is a good estimate of $f_{\rm col}$ for the specific observation under consideration. Physical processes operating in the disc atmosphere (e.g., electron scattering, emissive/absorptive opacities) collectively conspire to cause the observed thermal disc continuum, with characteristic colour temperature $T_{\rm col}^{\ast}$, to be harder than that expected from a disc blackbody model with characteristic effective temperature $T_{\rm eff}^{\ast}$. In Appendix \ref{app:physics}, we describe the physics behind disc spectral hardening. The degree of spectral hardening is phenomenologically parametrized by the constant colour correction factor $f_{\rm col} = T_{\rm col} / T_{\rm eff}$ \citep{ShimuraTakahara1995b}.\footnote{Sometimes the colour correction factor is called the spectral hardening factor and/or the symbol $h_{\rm d}$ or $f$ is used instead of $f_{\rm col}$.}

Obtaining observational constraints on $f_{\rm col}$ is not easy because a featureless disc continuum model can only constrain two free parameters --- a flux amplitude and an energy shift. Notwithstanding, attempts to measure $f_{\rm col}$ suggest it is inappropriate to draw any firm conclusions from spectral fits that blindly adopt the canonical and static value of $f_{\rm col} = 1.7$ \citep{ShimuraTakahara1995b}. Jointly fitting 26 \textit{Swift}/XRT observations of the particularly clean disc component of 4U 1957+11 in the high/soft state with a relativistic thin disc model, \citet{Maitra2014} obtained absolute measurements of $f_{\rm col}$ having best-fit values of 1.9--2.1 with large 90\% confidence ($\pm$) errors, typically $+1.9$ and $-0.3$. Even for a relatively unabsorbed source like 4U 1957+11, the $f_{\rm col}$ measurements are sensitive to a strong degeneracy with the black hole spin, the presence of a high-energy Comptonization component, and the instrument bandpass \citep{Nowak2008, Nowak2012}.\footnote{\citet{Pahari2018} report $f_{\rm col} = 1.54^{+0.14}_{-0.06}$ in 4U 1630--47 based on fixing $M$, $D$, $i_{\rm disc}$ to their poorly constrained values with no consideration of their uncertainties.}

Evidence for dynamic evolution of $f_{\rm col}$ comes from systematic analyses of X-ray binaries over many observations and across different accretion states \citep[e.g.,][]{Nowak2008}.\footnote{\citet{Belloni2010} gives a comprehensive description of the spectral/timing state classifications of X-ray binaries.} Surveying all stellar-mass black holes observed with \textit{Swift} through June 2010, \citet{ReynoldsMiller2013} studied the evolution of the disc component down to low luminosities ($L_{X} \gtrsim10^{-3}~L_{\rm Edd}$) and found that $r_{\rm in}$ \textit{increases} as the source softens, if $f_{\rm col}$ is constant. This is at odds with the disc truncation school of thought, which predicts that the inner disc radius decreases as the source transitions from the low/hard state to the high/soft state \citep[e.g.,][]{Esin1997}. Supposing instead that the inner disc, which is clearly detected in the \textit{Swift} data, is not severely truncated in the low/hard state, this requires $f_{\rm col}$ to decrease as the source softens \citep{ReynoldsMiller2013}. Analyzing multiple state transitions of GX 339--4 observed with \textit{RXTE}, \citet{Salvesen2013} found that $f_{\rm col}$ variations by factors of $\sim$2.0--3.5 could explain the spectral evolution of the soft thermal component without invoking disc truncation.

The inability of disc continuum fitting to constrain actual values for $f_{\rm col}$ from the data necessitates an appeal to theoretical models. As described in \S\ref{sec:fcol}, the disc continuum fitting technique fits the data with a colour-corrected blackbody model \citep{Li2005}, with each fit adopting an informed choice for $f_{\rm col}$ based on the disc spectra predicted by a sophisticated disc atmosphere model \citep{DavisHubeny2006}. However, the reported black hole spin error budget effectively does not include systematic uncertainties associated with this choice for $f_{\rm col}$, which are potentially large enough to render the spin measurements moot.

We organize our paper by first introducing and re-parametrizing the disc continuum model used to measure black hole spin (\S\ref{sec:meth}; Appendix \ref{app:MCD}). Next, we reverse-engineer the eight non-maximal spin measurements in Table \ref{tab:obs} and revise them to include reasonable levels of systematic uncertainty in $f_{\rm col}$ (\S\ref{sec:fcolErr}; Appendix \ref{app:meth_PDFs}). Our main result is that incorporating $f_{\mathrm{col}}$ uncertainties at the level of $\pm0.2$--0.3 dominates the black hole spin error budget (Figure \ref{fig:spinPDFs}). We go on to demonstrate that the choice for $f_{\rm col}$ strongly influences the measured black hole spin (\S\ref{sec:fcolVal}). In \S\ref{sec:disc}, we discuss the various uncertainties associated with $f_{\rm col}$ to (i) convince the reader that the $f_{\rm col}$ uncertainties we considered are plausible and (ii) motivate the path forward towards a better understanding of $f_{\rm col}$ and disc atmospheric physics, with an emphasis on moderate-to-strong disc magnetization. Finally, we summarize and conclude our work in \S\ref{sec:sumconc}.

\section{Disc Continuum Fitting Method}
\label{sec:meth}
\begin{figure}
    \begin{center}
        \includegraphics[width=0.495\textwidth]{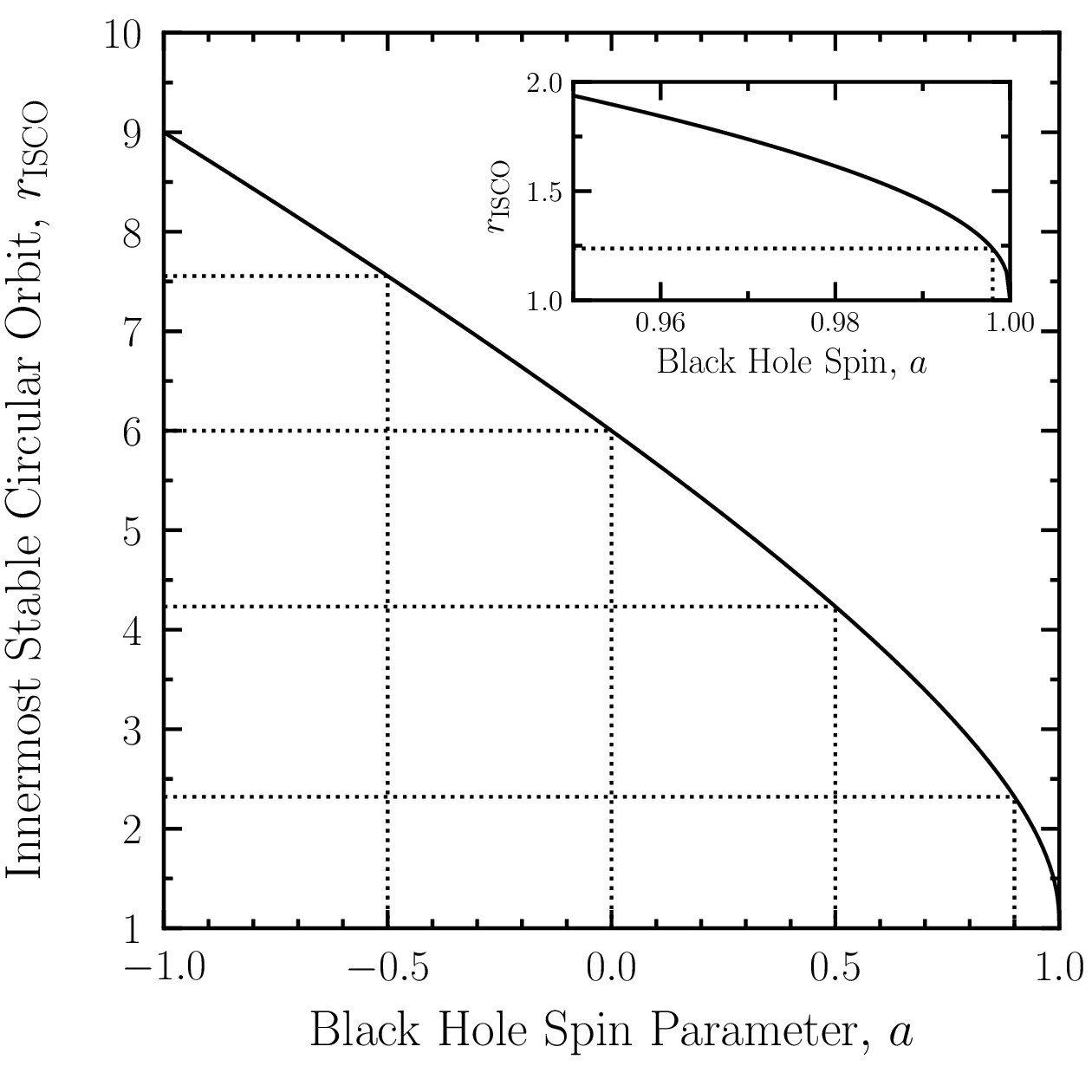}
    \end{center}
    \vspace{-6mm}
    \caption{Radial location of the innermost stable circular orbit $r_{\mathrm{ISCO}} \equiv R_{\mathrm{ISCO}} / R_{\mathrm{g}}$, as a function of the dimensionless black hole spin parameter $a$ \citep{Bardeen1972}. The \textit{inset} zooms in on the region of near-maximal black hole spin, where the monotonic behaviour of $r_{\rm ISCO}$ with $a$ is very non-linear. {\it Dotted lines} mark the location of the ISCO for black hole spins: $a = -0.5, 0, 0.5, 0.9$, and $0.998$ (\textit{inset}).}
    \label{fig:rISCO_spin}
\end{figure}

In practice, a featureless disc continuum model can only constrain two parameters when fitting an observed X-ray spectrum. The modern continuum fitting practitioners adopt the \texttt{kerrbb} model \citep{Li2005} and choose these two free parameters to be the mass accretion rate $\dot{M}$ and the black hole spin $a$, which is equivalent to the inner disc radius $r_{\rm in} \equiv R_{\rm in} / R_{\rm g}$ through its assumed association with the innermost stable circular orbit (ISCO).\footnote{The iron line technique also assumes $r_{\rm in} = r_{\rm ISCO}$.} Figure \ref{fig:rISCO_spin} shows that the location of the ISCO is a monotonically decreasing function of the black hole spin, so measuring $r_{\rm in} = r_{\rm ISCO}$ is tantamount to measuring $a$. The remaining \texttt{kerrbb} parameters are the colour correction factor $f_{\rm col}$, black hole mass $M$, distance $D$, and inner disc inclination $i_{\rm disc}$, all of which are held fixed during spectral fitting.\footnote{Often, the iron line technique adopts a simpler disc model (e.g., \texttt{diskbb}) that can adequately fit the data without needing to specify $f_{\rm col}$ explicitly, and so does not suffer from its uncertainty.} A marginal density for $a$ is then built up by repeating this fitting procedure for many fixed sets of $\{ f_{\rm col}, M, D, i_{\rm disc} \}$, where $M$, $D$, $i_{\rm disc}$ are sampled from their appropriate joint density (\S\ref{sec:MDi}) and an informed choice is made for $f_{\rm col}$ (\S\ref{sec:fcol}).

As detailed in Appendix \ref{app:MCD}, our approach is to recast \texttt{kerrbb} as a two-parameter disc model in terms of a characteristic colour temperature $T_{\rm col}^{\ast}$ and a disc flux normalization
\begin{align}
K_{\rm flux} &\equiv \frac{r_{\rm in}^{2}}{f_{\rm col}^{4}} \left( \frac{G M / c^{2}}{D} \right)^{2} \cos\left( i_{\rm disc} \right) \nonumber \\
&\times \Upsilon\left( i_{\rm disc} \right) g_{\rm GR}\left( r_{\rm in}, i_{\rm disc} \right) g_{\rm NT}\left( r_{\rm in} \right). \label{eqn:Kflux_body}
\end{align}
The cornerstone of our analysis is $K_{\rm flux}$, while $T_{\rm col}^{\ast}$ is inconsequential.\footnote{That is, deriving $r_{\rm in}$ from $T_{\rm col}^{\ast}$ would require knowledge of $\dot{M}$ (see equation \ref{eqn:Tcol_x}), which is a dynamic parameter that cannot be independently determined, unlike the parameters $\{ f_{\rm col}, M, D, i_{\rm disc} \}$ needed to derive $r_{\rm in}$ from $K_{\rm flux}$.} In addition to explicit dependencies on $\{ r_{\rm in}, f_{\rm col}, M, D, i_{\rm disc} \}$, $K_{\rm flux}$ incorporates a limb-darkening law $\Upsilon$ on the emitted intensity and two disc flux correction factors: $g_{\rm GR}$ and $g_{\rm NT}$, which account for relativistic effects on photon propagation and on the radial disc structure, respectively. Following the early disc continuum fitting modelers \citep[e.g.,][]{Sobczak1999}, we treat $K_{\rm flux}$ as the observable and $r_{\rm in}$ as the derived variable. Importantly, we define $K_{\rm flux}$ in a way that is consistent with the \texttt{kerrbb} model used by the modern disc continuum fitting technique.

To be clear, in our subsequent analysis we do not perform any spectral fits to the disc continuum. Instead, we manipulate probability densities to reverse-engineer a published black hole spin marginal density $f_{a}^{\rm CF}( a )$ to obtain the disc flux normalization marginal density $f_{K}( K_{\rm flux} )$, which is the observable in our re-parametrized continuum fitting method. With $f_{K}( K_{\rm flux} )$ in hand, we can then explore how the choice for $f_{\rm col}$ and its uncertainty affects the measured black hole spin. The actual disc continuum fitting method appeals to spectral fitting, which possesses constraining power to favor or disfavor different combinations of model parameters. Our approach allows us to explore how choosing different model parameters affects the measured black hole spin, but we cannot speak to whether these choices are in tension with the data.

Just as we took care to re-parametrize the disc continuum model in a way that is consistent with \texttt{kerrbb}, we must also handle the fixed \texttt{kerrbb} parameters $\{ f_{\rm col}, M, D, i_{\rm disc} \}$ in a way that is consistent with the modern continuum fitting technique, as described next.

\subsection{Black Hole Mass, Distance, Inner Disc Inclination}
\label{sec:MDi}
For the black hole X-ray binaries in Table \ref{tab:obs}, good independent observational constraints exist for the black hole mass $M$ and distance $D$, but the continuum fitting practitioners equate the inner disc inclination $i_{\rm disc}$ to the measured inclination of either the binary orbital axis $i_{\rm orb}$ or the jet axis $i_{\rm jet}$, which is a controversial assumption \citep[see Table 1 of][]{SalvesenPokawanvit2020}. Most continuum fitting spin measurements assume $M$, $D$, $i_{\rm disc}$ to be normally and independently distributed, the latter meaning that their joint density can be decomposed into the product of their marginals,
\begin{equation}
f_{M, D, i}\left( M, D, i_{\rm disc} \right) = f_{M}\left( M \right) f_{D}\left( D \right) f_{i}\left( i_{\rm disc} \right).
\end{equation}
Of the systems in Table \ref{tab:obs}, only GRS 1915+105 \citep{McClintock2006} and Cygnus X--1 \citep{Gou2011} did not assume $M$, $D$, $i_{\rm disc}$ independence. Calculating the appropriate joint density in this case is straightforward, but not necessary because we do not include GRS 1915+105 or Cygnus X--1 in our analysis for reasons explained in \S\ref{sec:analres}.

In a few cases, the marginal densities used to construct $f_{M, D, i}( M, D, i_{\rm disc} )$ were not taken to be normally distributed. For XTE J1550--564, \citet{Steiner2011} adopted an asymmetric Gaussian for the distance marginal density $f_{D}( D )$. For H1743--322, \citet{Steiner2012a} modeled the jet to derive marginal densities for the distance $f_{D}( D )$ and the jet / inner disc inclination $f_{i}( i_{\rm disc} )$, and they adopted a mass marginal density $f_{M}( M )$ based on Galactic stellar mass black hole populations.

When constructing $f_{M, D, i}( M, D, i_{\rm disc} )$ for each system in Table \ref{tab:obs}, we always use the same marginal densities and assumptions about parameter correlations / independence as the disc continuum fitting practitioners who performed the analysis for that system.

\subsection{Colour Correction Factor}
\label{sec:fcol}
Good observational constraints do not exist for $f_{\rm col}$, so fitting a disc continuum with a colour-corrected disc model like \texttt{kerrbb} necessarily relies on selecting a theoretically-motivated $f_{\rm col}$ value. To this end, the modern continuum fitting technique appeals to the \texttt{bhspec} model, which is based on calculations of the local emergent spectrum from a one-dimensional slab with non-local thermodynamic equilibrium radiative transfer, including the effects of thermal Compton scattering and free-free/bound-free opacities of abundant ion species \citep{Davis2005, DavisHubeny2006}. Construction of \texttt{bhspec} involved partitioning the same relativistic global ``$\alpha$-disc'' model used by \texttt{kerrbb} into many axisymmetric radial zones, self-consistently computing the local vertical structure and radiative transfer in each radial zone, and then integrating over all annuli to calculate the observed disc spectrum --- accounting for relativistic photon transfer without returning radiation.

For fixed $M$, $D$, $i_{\rm disc}$ and an assumed effective viscosity parameter $\alpha$, \texttt{bhspec} has only two model parameters: the black hole spin $a$ and the Eddington-scaled disc luminosity $l \equiv L_{\rm disc} / L_{\rm Edd}$, which is interchangeable with mass accretion rate $\dot{M}$,\footnote{By defining $\dot{m} \equiv \dot{M} / \dot{M}_{\rm Edd}$, where $\dot{M}_{\rm Edd} = L_{\rm Edd} / (\eta c^{2})$, $L_{\rm Edd} = 4 \pi G M m_{\rm p} c / \sigma_{\rm T}$, and $L_{\rm disc} = \eta \dot{M} c^{2}$, the equivalence $l = \dot{m}$ is apparent (see also \S \ref{sec:uncertain}).} whereas \texttt{kerrbb} requires $f_{\rm col}$ as a third model parameter. Despite this extra parameter, the continuum fitting practitioners elect to fit the data using \texttt{kerrbb} because \texttt{bhspec} does not include returning radiation \citep[e.g.,][]{McClintock2014}. Plus, \texttt{bhspec} sometimes gives worse fits than \texttt{kerrbb} despite being a more sophisticated model that dispenses with the phenomenological $f_{\rm col}$ \citep[e.g.,][]{Kubota2010}.

To take advantage of \texttt{bhspec}, the modern disc continuum fitting practitioners constructed the hybrid model \texttt{kerrbb2} to obtain an informed $f_{\rm col}$ for each spectral fit, as follows \citep[e.g.,][]{McClintock2006, McClintock2010}. Holding the sampled values $\{ M, D, i_{\rm disc} \}$ from the joint density $f_{M, D, i}( M, D, i_{\rm disc} )$ fixed, one tabulates $f_{\rm col}$ on an $(l, a)$-grid by fitting the colour-corrected \texttt{kerrbb} model to the \texttt{bhspec} disc atmosphere model. When tabulating $f_{\rm col}$, care is taken to use the same detector response matrix and energy range that will be used to fit the actual data with \texttt{kerrbb}.\footnote{Also, \texttt{kerrbb}'s \texttt{rflag} is turned off when fitting \texttt{bhspec} models, which do not include returning radiation. It is unclear if interstellar absorption is accounted for when tabulating $f_{\rm col}$.} This is important because the $f_{\rm col}$ value obtained from \texttt{kerrbb} fits to disc spectra calculated with \texttt{bhspec} is sensitive to the bandpass being considered \citep[e.g.,][]{DoneDavis2008}.

For the disc luminosities relevant to the high/soft state ($l \sim 0.01$--$0.1$), the \texttt{bhspec} models generally find that $f_{\rm col}$ lies in the range 1.4--1.8 and for a given set of input parameters, $f_{\rm col}$ tends to increase only slightly with $\dot{M}$ \citep{Davis2005, Shafee2006, DavisElAbd2019}. Consequently, when modeling a set of observations over which $\dot{M}$ is approximately constant, \texttt{kerrbb2} effectively confines $f_{\rm col}$ to a very narrow range. Unfortunately, the range of adopted $f_{\rm col}$ values usually goes unreported in the disc continuum fitting literature, except for LMC X--1 and M33 X--7, where $f_{\rm col} = 1.562 \pm 0.008$ and $f_{\rm col} = 1.61 \pm 0.01$, respectively \citep{Gou2009, Liu2008}. \textit{This means that the black hole spin error budget incorporates uncertainties in $f_{\rm col}$ at the $\lesssim 1\%$-level.} In our approach, we will treat this tiny $f_{\rm col}$ uncertainty range implicitly adopted by the modern disc continuum fitting technique as being synonymous with a Dirac delta function for the marginal density $f_{f}( f_{\rm col} )$.

While the success of the \texttt{kerrbb} model in describing thermal spectra from X-ray binaries lends credence to the notion of a colour-corrected, multi-temperature, disc blackbody model, the appropriate uncertainty on $f_{\rm col}$ is undeniably much greater than the $1\%$-level. In addition to neglecting the various uncertainties associated with $f_{\rm col}$ discussed in \S\ref{sec:uncertain}, \texttt{kerrbb2} simply selects the best-fit $f_{\rm col}$ value and does not estimate any uncertainties associated with fitting \texttt{kerrbb} to \texttt{bhspec}. Assigning a concrete uncertainty to $f_{\rm col}$ is difficult, so instead we explore the sensitivity of continuum fitting black hole spin measurements to different levels of $f_{\rm col}$ uncertainty.

\setlength{\tabcolsep}{1.5pt}
\begin{table*}
\centering
\begin{tabular}{c c c c c c c}
\hline
\hline
Source & $a^{\rm Fe}$ & $a^{\rm CF}$ & $f_{\rm col}$ & $M$ & $D$ & $i_{\rm orb}$ \\
\smallskip
 & & & & $[M_{\odot}]$ & $[{\rm kpc}]$ & $[^{\circ}]$ \\
\hline
\smallskip
GRS 1915+105 & $0.98 \pm 0.01$ (1) & $> 0.95$$^{\rm a}$ (12) & $1.6$$^{\rm b}$ & $14.0 \pm 4.4$ (9) & $11.2 \pm 0.8$ (4) & $66 \pm 2$$^{\rm c}$ (4) \\
\smallskip
Cygnus X--1 & $0.97^{+0.014}_{-0.02}$$^{\rm d}$ (3) & $0.9985$$^{+0.0005}_{-0.0148}$ (7) & $1.610$ (7) & $14.8 \pm 1.0$ (18) & $1.86 \pm 0.156^{\rm e}$ (22) & $27.1 \pm 0.8$ (18) \\
\smallskip
LMC X--1 & $0.97^{+0.02}_{-0.25}$ (27) & $0.92^{+0.05}_{-0.07}$ (5) & $1.562 \pm 0.008$ (5) & $10.91 \pm 1.54$ (17) & $48.10 \pm 2.22$ (17) & $36.38 \pm 2.02$ (17) \\
\smallskip
4U 1543--47 & $0.3 \pm 0.1$$^{\rm f}$ (13) & $0.80 \pm 0.10$$^{\rm a}$ (24) & 1.5$^{\rm g}$ (24) & $9.4 \pm 1.0$ (14, 15) & $7.5 \pm 1.0$$^{\rm h}$ (24) & $20.7 \pm 1.5$ (14, 15) \\
\smallskip
GRO J1655--40 & $0.92 \pm 0.02$$^{\rm d}$ (23) & $0.70 \pm 0.10$$^{\rm a}$ (24) & 1.65$^{\rm g}$ (24) & $6.30 \pm 0.27$ (8) & $3.2 \pm 0.2$ (10) & $70.2 \pm 1.2$ (8) \\
\smallskip
XTE J1550--564 & $0.55^{+0.15}_{-0.22}$$^{\rm i}$ (25) & $0.34^{+0.20}_{-0.28}$ (25) & 1.6$^{\rm b}$ & $9.10 \pm 0.61$ (19) & $4.38^{+0.58}_{-0.41}$ (19) & $74.7 \pm 3.8$ (19) \\
\smallskip
M33 X--7 & $-$ & $0.84 \pm 0.05$$^{\rm j}$ (11) & $1.61 \pm 0.01$$^{\rm j}$ (11) & $15.65 \pm 1.45$ (16) & $840 \pm 20$ (16) & $74.6 \pm 1.0$ (16) \\
\smallskip
LMC X--3 & $-$ & $0.25^{+0.13}_{-0.16}$ (28) & 1.6$^{\rm b}$ & $7.0 \pm 0.6$ (20) & $48.1 \pm 2.2$ (17) & $69.2 \pm 0.7$ (20) \\
\smallskip
H1743--322 & $-$ & $0.20^{+0.34}_{-0.33}$ (26) & 1.6$^{\rm b}$ & $-$$^{\rm k}$ (21) & $8.5 \pm 0.8$ (26) & $75 \pm 3$$^{\rm c}$ (26) \\
\smallskip
A0620--00 & $-$ & $0.12 \pm 0.19$ (6) & 1.6$^{\rm b}$ & $6.61 \pm 0.25$ (2) & $1.06 \pm 0.12$ (2) & $51.0 \pm 0.9$ (2) \\
\hline
\end{tabular}
\caption{Black hole X-ray binaries with spin measurements vetted by the disc continuum fitting community \citep{McClintock2014}. Six of these sources also have spins vetted by the iron line community \citep{Reynolds2014}. From {\it left} to {\it right} the columns list the source name, iron line spin $a^{\rm Fe}$, continuum fitting spin $a^{\rm CF}$, and the parameters adopted for the $a^{\rm CF}$ measurement: colour correction factor $f_{\rm col}$, black hole mass $M$, distance $D$, and binary orbital inclination $i_{\rm orb}$ (assumed to be the same as the inner disc inclination $i_{\rm disc}$). Uncertainties correspond to the $68\%$ confidence level unless indicated otherwise. Instances where the confidence interval was unclear are assumed to be at the $68\%$-level. Numbers in parentheses map to these references: (1) \citet{Blum2009}; (2) \citet{Cantrell2010}; (3) \citet{Fabian2012}; (4) \citet{Fender1999a}; (5) \citet{Gou2009}; (6) \citet{Gou2010}; (7) \citet{Gou2011}; (8) \citet{Greene2001}; (9) \citet{HarlaftisGreiner2004}; (10) \citet{HjellmingRupen1995}; (11) \citet{Liu2008}; (12) \citet{McClintock2006}; (13) \citet{Miller2009a}; (14) \citet{Orosz1998}; (15) \citet{Orosz2003}; (16) \citet{Orosz2007}; (17) \citet{Orosz2009}; (18) \citet{Orosz2011a}; (19) \citet{Orosz2011b}; (20) \citet{Orosz2014}; (21) \citet{Ozel2010}; (22) \citet{Reid2011}; (23) \citet{Reis2009}; (24) \citet{Shafee2006}; (25) \citet{Steiner2011}; (26) \citet{Steiner2012a}; (27) \citet{Steiner2012b}; (28) \citet{Steiner2014}.
$^{\rm a}$The wider error range suggested in Table 1 of \citet{McClintock2014} is adopted (lower limits are at the 99.7\% confidence level).
$^{\rm b}$The fiducial $f_{\rm col} = 1.6$ is adopted because a value is not given explicitly in the cited paper.
$^{\rm c}$The inclination angle of the jet axis $i_{\rm jet}$ is used (assumed to trace $i_{\rm orb}$) when measuring $a^{\rm CF}$.
$^{\rm d}$Errors are at the 90\% confidence level.
$^{\rm e}$The distance uncertainty incorporates a 10\% uncertainty in the absolute flux calibration to the Crab.
$^{\rm f}$See also \citet{Dong2020}, who find $a^{\rm Fe} = 0.67^{+0.15}_{-0.08}$ (90\% confidence), which is consistent with $a^{\rm CF}$.
$^{\rm g}$$f_{\rm col}$ is estimated from the figures in \citet{Shafee2006}.
$^{\rm h}$This distance appears to be unpublished, with \citet{Orosz1998} quoting $D = 9.1 \pm 1.1~{\rm kpc}$.
$^{\rm i}$See also \citet{Miller2009a}, who find $a^{\rm Fe} = 0.76 \pm 0.01$.
$^{\rm j}$Following the erratum to \citet{Liu2008}, we adjust their published black hole spin marginal density such that $a = 0.84 \pm 0.05$ and reduce their reported $f_{\rm col}$ values by 8.7\%.
$^{\rm k}$In the absence of a dynamical mass measurement, \citet{Steiner2012a} used the mass distribution of \citet{Ozel2010}.}
\label{tab:obs}
\end{table*}

\section{Analysis and Results}
\label{sec:analres}
Table \ref{tab:obs} lists the 10 black hole spin measurements vetted by the disc continuum fitting community \citep{McClintock2014}, along with the black hole mass $M$, distance $D$, and inner disc inclination $i_{\rm disc}$ used for each measurement. Unfortunately, the colour correction factor $f_{\rm col}$ tends to go unreported (see \S\ref{sec:fcol}), in which case we list a representative value in Table \ref{tab:obs}.

Measuring $r_{\rm in}$ from equation \eqref{eqn:Kflux_body} is the essence of our re-parametrized disc continuum fitting technique. Consequently, the main challenge in measuring the black hole spin $a$ is the need for tight constraints on $K_{\rm flux}$, $M$, $D$, $i_{\rm disc}$, and $f_{\rm col}$. Uncertainties associated with these parameters will all contribute to the uncertainty in the measured black hole spin. However, $f_{\rm col}$ is essentially treated as a known quantity in the reported spin measurements, which neglect both systematic uncertainties from dependence on the \texttt{bhspec} model and statistical uncertainties from fitting \texttt{kerrbb} to the \texttt{bhspec} models when determining $f_{\rm col}$.

In \S \ref{sec:fcolErr}, we suppose the $f_{\rm col}$ values in Table \ref{tab:obs} are correct and then fold reasonable $f_{\rm col}$ uncertainties into the black hole spin measurements. We find that for most of these X-ray binaries, an $f_{\rm col}$ uncertainty of $\pm$0.2--0.3 dominates the black hole spin error budget. In \S \ref{sec:fcolVal}, we show how choosing different $f_{\rm col}$ values from those in Table \ref{tab:obs} affects the spin measurements, which can help alleviate discrepancies with the iron line results.

We omit the maximally spinning black holes GRS 1915+105 and Cygnus X--1 from our analysis, primarily because our methodology struggles with the exponential shape of their measured spin marginal densities. Regardless, near-maximal spin measurements are not strongly affected by including $f_{\rm col}$ uncertainties because of the very non-linear dependence of the ISCO on the spin parameter in this regime (see Figure \ref{fig:rISCO_spin}). For instance, a factor of two change in the ISCO location from $r_{\rm ISCO} = 1 \rightarrow 2$ only changes the spin from $a = 1 \rightarrow 0.94$, whereas a factor of two change from $r_{\rm ISCO} = 3 \rightarrow 6$ changes the spin dramatically from $a = 0.78 \rightarrow 0$.

\subsection{Giving Uncertainties an $f$}
\label{sec:fcolErr}
\begin{figure*}
    \begin{center}
        \includegraphics[width=0.495\textwidth]{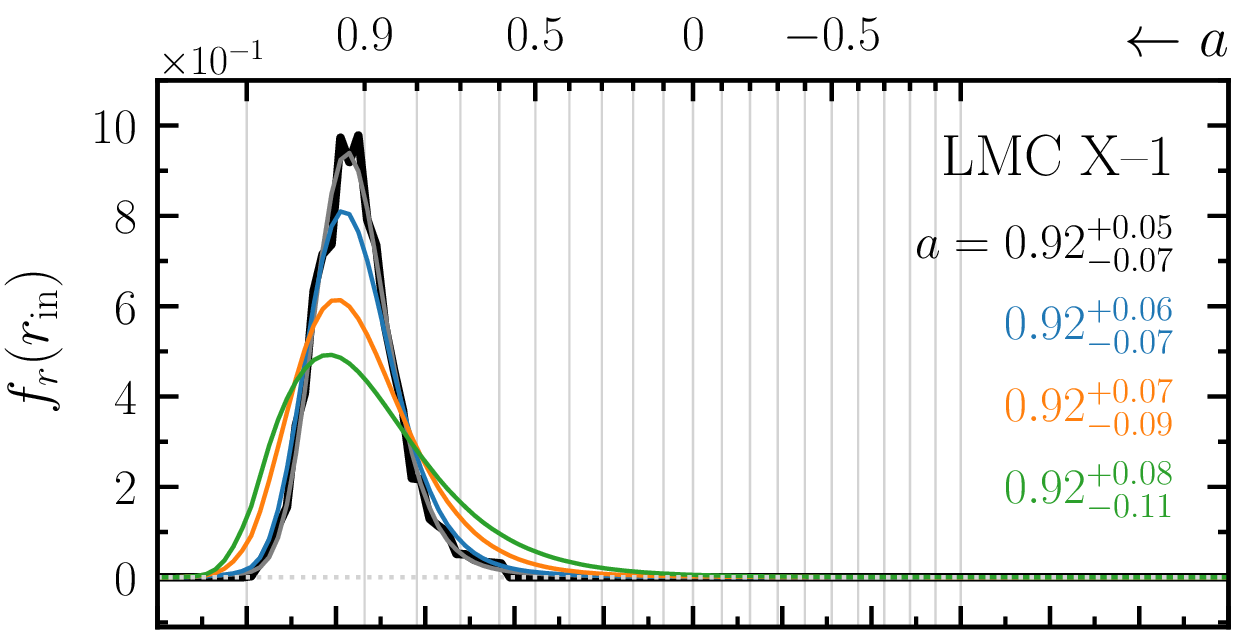}
        \includegraphics[width=0.495\textwidth]{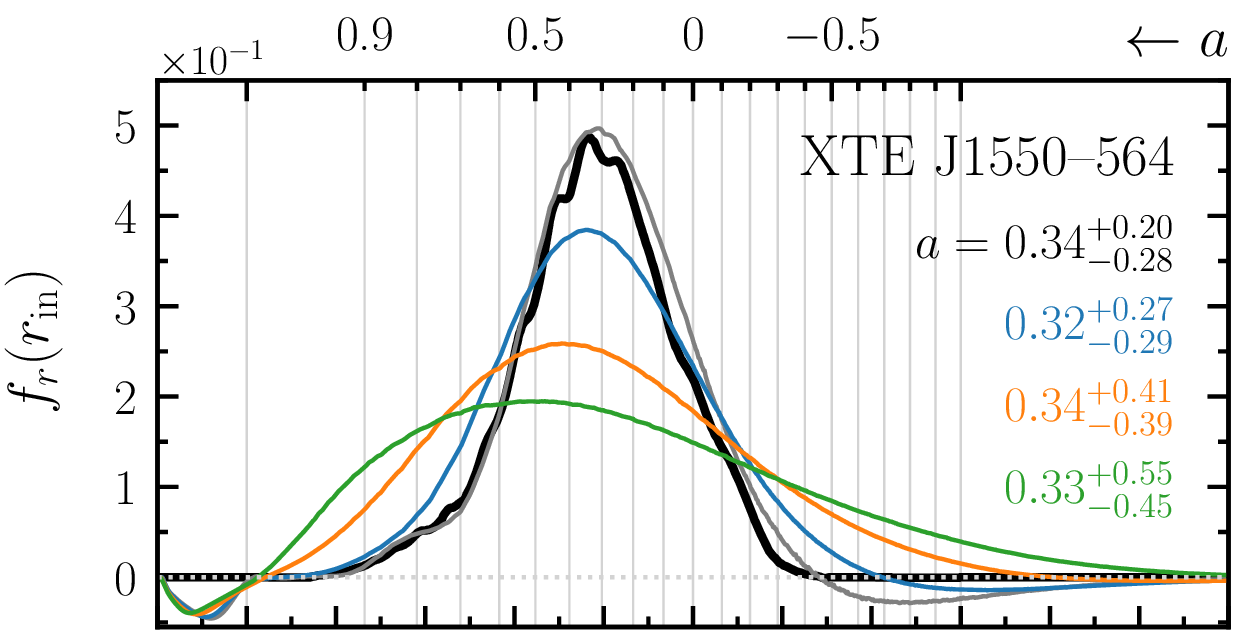}
        \includegraphics[width=0.495\textwidth]{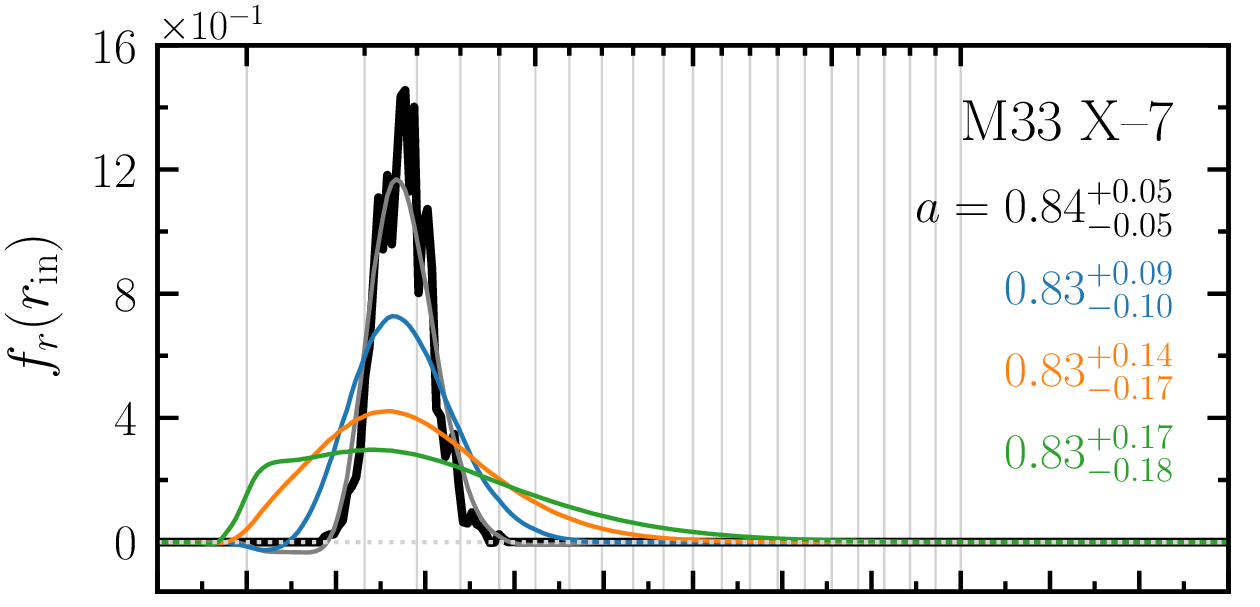}
        \includegraphics[width=0.495\textwidth]{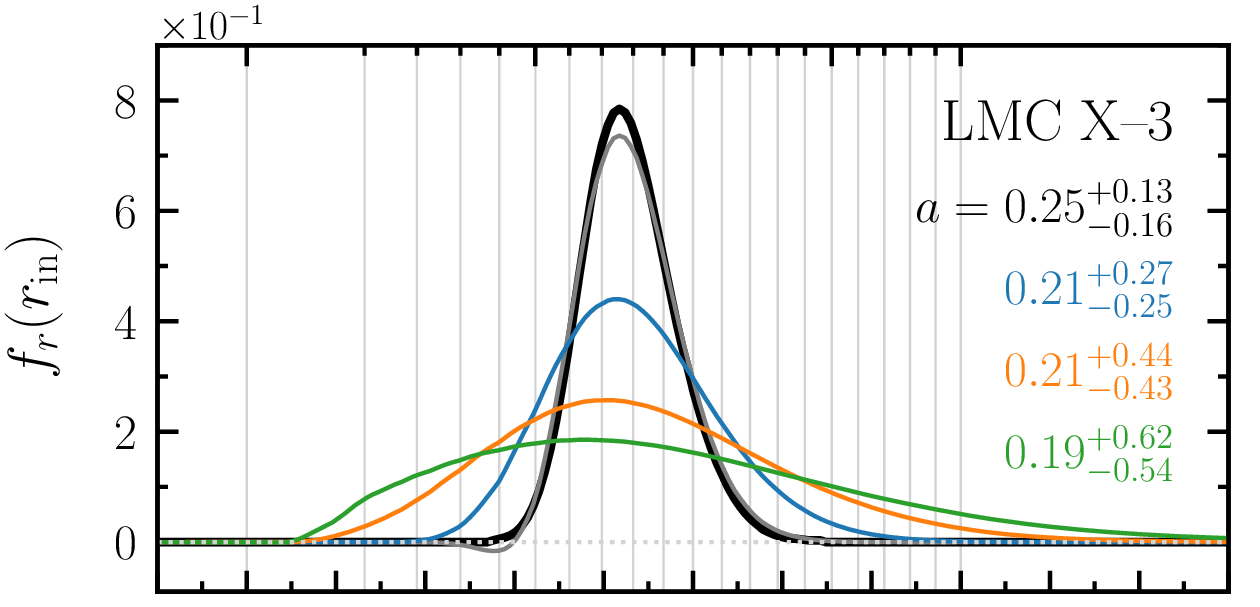}
        \includegraphics[width=0.495\textwidth]{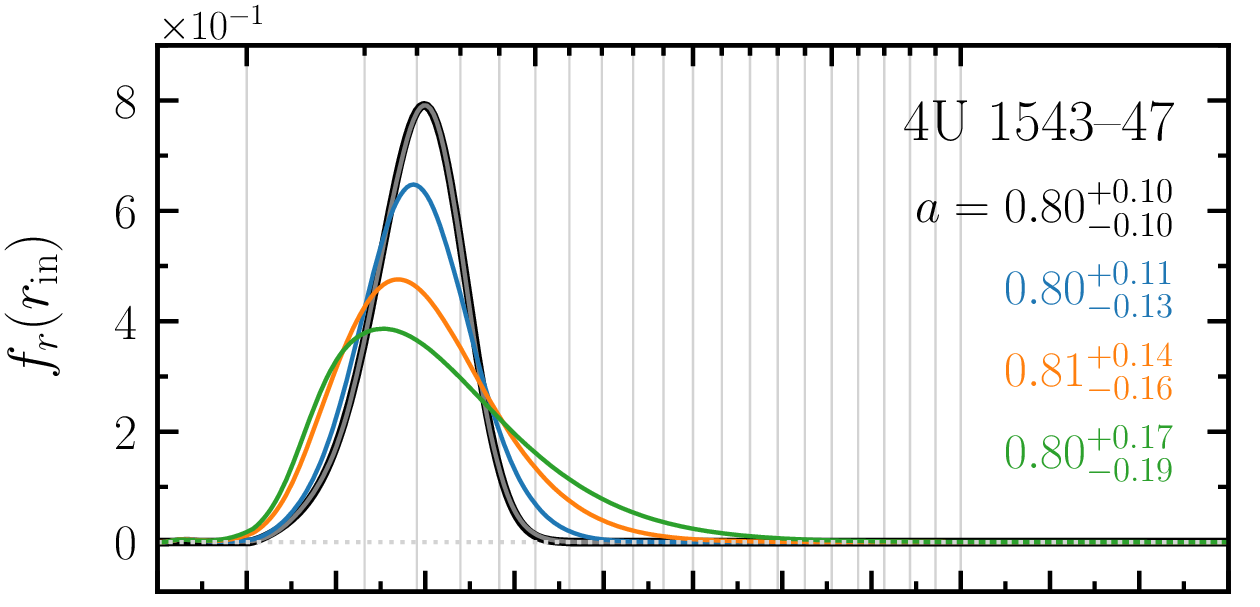}
        \includegraphics[width=0.495\textwidth]{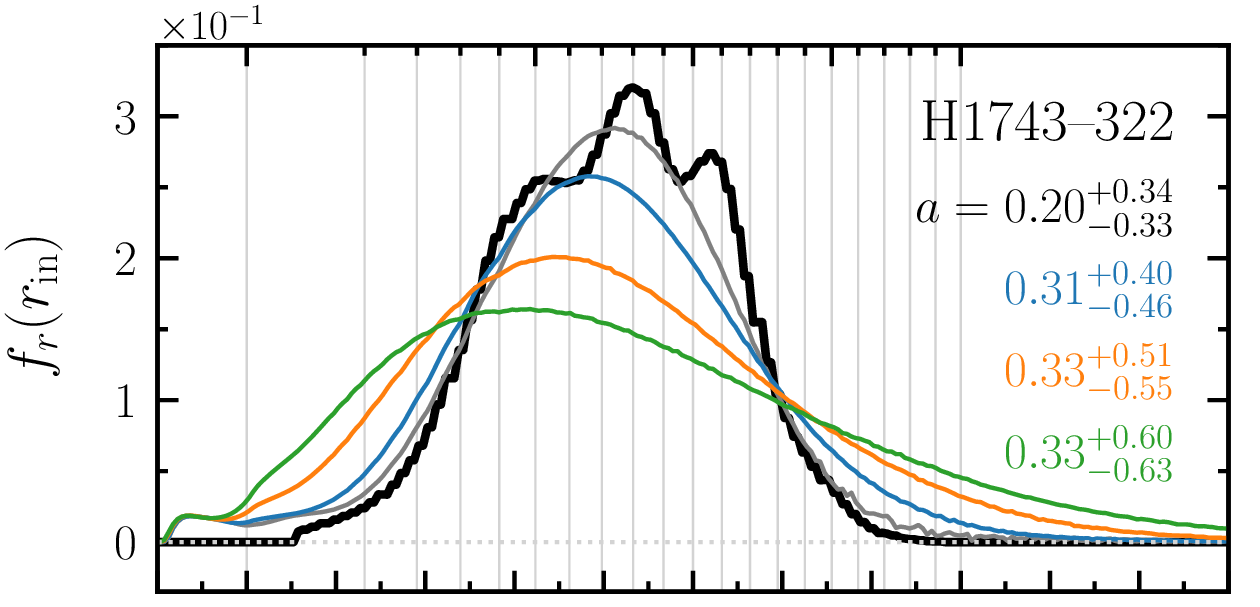}
        \includegraphics[width=0.495\textwidth]{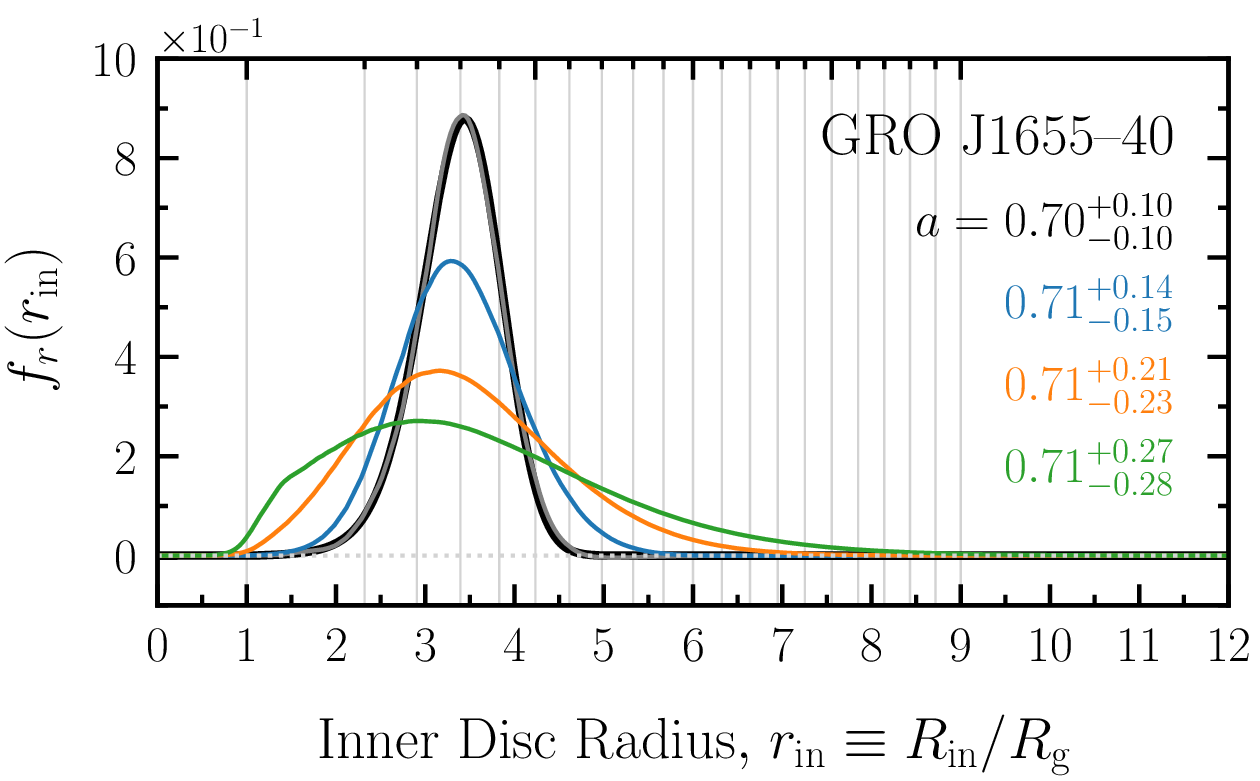}
        \includegraphics[width=0.495\textwidth]{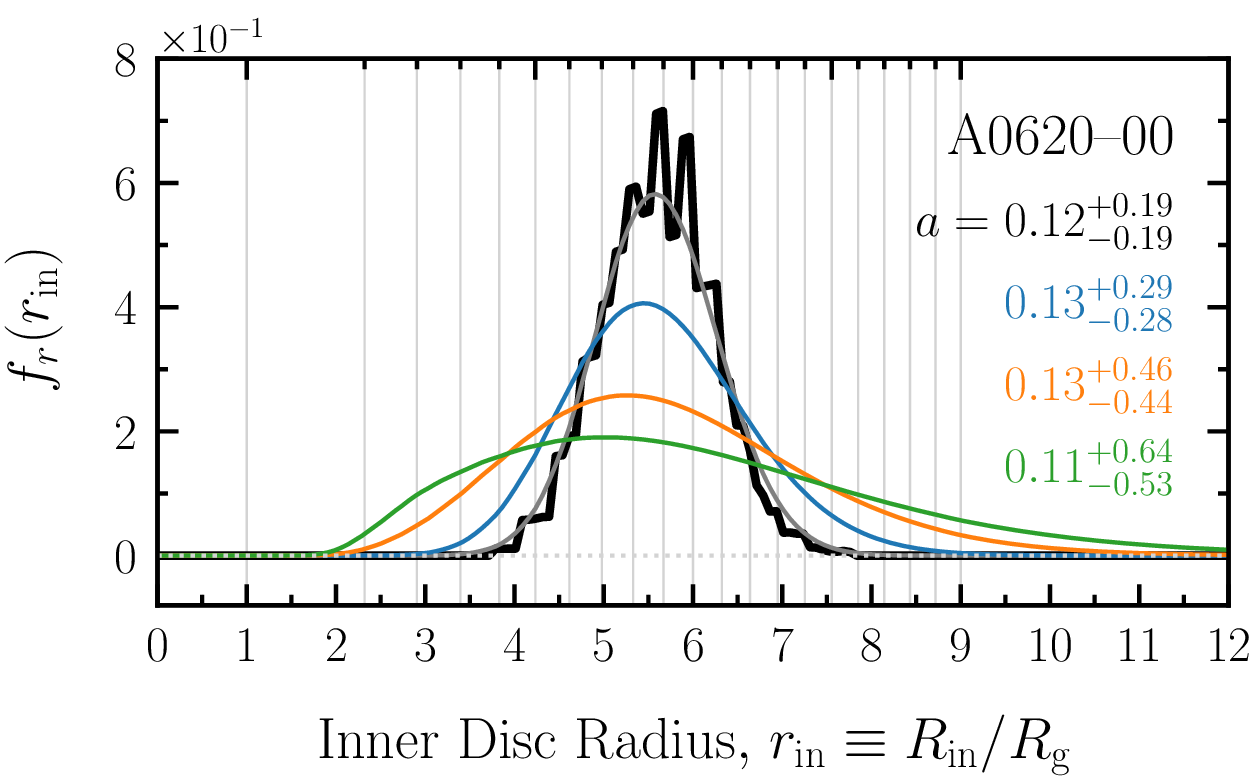}
    \end{center}
    \vspace{-4mm}
    \caption{Inner disc radius marginal density $f_{r}( r_{\rm in} )$ for each X-ray binary in Table \ref{tab:obs}, which maps to a measurement of the black hole spin parameter $a$ (\textit{top} $x$-axis) assuming that $r_{\rm in} = r_{\rm ISCO}$. The \textit{thick black line} shows the original disc continuum fitting measurement, which effectively adopts no uncertainty on the colour correction factor $f_{\rm col}$ listed in Table \ref{tab:obs}. The \textit{line colour} shows the revised black hole spin measurements after incorporating uncertainties in $f_{\rm col}$ of $\pm0.1$ (\textit{blue}), $\pm0.2$ (\textit{orange}), and $\pm0.3$ (\textit{green}), assuming a normal distribution for its marginal density $f_{f}( f_{\rm col} )$ centered on the value in Table \ref{tab:obs}. The \textit{grey lines} followed the same procedure as the \textit{coloured lines} (see \S\ref{sec:fcolErr}), but assumed no $f_{\rm col}$ uncertainty, and they successfully recover the original measurements. Each \textit{legend} gives the source name and quotes the spin measurement for each distribution, where the $a$ value corresponds to a cumulative density of \sfrac{1}{2} (i.e., the median) and the uncertainty in $a$ corresponds to the inter-68\% region of the marginal density. For most systems, including an $f_{\rm col}$ uncertainty of $\pm$0.2--0.3 more than doubles the original quoted black hole spin uncertainties; thus, dominating the error budget if such $f_{\rm col}$ uncertainty levels are reasonable (see \S\ref{sec:uncertain}).}
    \label{fig:spinPDFs}
\end{figure*}
For the eight X-ray binaries in Table \ref{tab:obs} with a non-maximal black hole spin measurement from disc continuum fitting, the \textit{thick black line} in Figure \ref{fig:spinPDFs} shows the inner disc radius marginal density $f_{r}^{\rm CF}( r_{\rm in} )$, which is synonymous with the black hole spin marginal density $f_{a}^{\rm CF}( a )$ under the usual $r_{\rm in} = r_{\rm ISCO}$ assumption. We extracted published black hole spin measurements --- that is, either $f_{a}^{\rm CF}( a )$ or $f_{r}^{\rm CF}( r_{\rm in} )$ --- using the \texttt{DataThief} program \citep{DataThief} for the X-ray binaries: LMC X--1, XTE J1550--564, M33 X--7, LMC X--3, H1743--322, A0620--00. A black hole spin marginal density is not published for either 4U 1543--47 or GRO J1655--40, so in these cases we construct normal distributions for $f_{a}^{\rm CF}( a )$ based on the reported $a^{\rm CF}$ values and uncertainties in Table \ref{tab:obs}.

Our aim is to incorporate uncertainties in the colour correction factor $f_{\rm col}$ into each black hole spin measurement derived from disc continuum fitting. We accomplish this by combining probability density transformations with an inverse integral transform method and constrained optimization methods, with the details relegated to Appendix \ref{app:meth_PDFs} and a summary provided below.

Starting with the black hole spin marginal density $f_{a}^{\rm CF}( a )$ reported from a disc continuum fitting measurement, we convert this to the inner disc radius marginal density $f_{r}^{\rm CF}( r_{\rm in} )$ by assuming that $r_{\rm in} = r_{\rm ISCO}$ and following \S\ref{app:a2risco}. Next, we reverse-engineer $f_{r}^{\rm CF}( r_{\rm in} )$ to obtain the marginal density of the disc flux normalization $f_{K}( K_{\rm flux} )$, which is the observable quantity in our re-parametrized disc continuum fitting approach described in \S\ref{sec:meth} and Appendix \ref{app:MCD}. This requires inverting the integral transform equation
\begin{align}
f_{r}^{\rm CF}\left( r_{\rm in} \right) &= \int\limits_{i_{\rm disc} = 0}^{\pi / 2} \int\limits_{D = 0}^{\infty} \int\limits_{M = 0}^{\infty} d M~d D~d i_{\rm disc} \nonumber \\
&\times f_{K}\left( h^{-1}\left( r_{\rm in}, f_{\rm col}^{\rm CF}, M, D, i_{\rm disc} \right) \right) \nonumber \\
&\times f_{M, D, i}\left( M, D, i_{\rm disc} \right) \nonumber \\
&\times \left| J\left( r_{\rm in}, f_{\rm col}^{\rm CF}, M, D, i_{\rm disc} \right) \right|, \label{eqn:frCF_fK}
\end{align}
which follows from performing a multi-variate change of variables and then marginalizing over $\{ f_{\rm col}, M, D, i_{\rm disc} \}$. Importantly, we treat the marginal density $f_{f}( f_{\rm col} )$ as a Dirac delta function centered on the Table \ref{tab:obs} value (denoted by $f_{\rm col}^{\rm CF}$) because that is effectively what the disc continuum fitting practitioners did when measuring $f_{r}^{\rm CF}( r_{\rm in} )$. In equation \eqref{eqn:frCF_fK},  $f_{M, D, i}( M, D, i_{\rm disc} )$ is the joint density of the black hole mass, distance, and inner disc inclination; $J( r_{\rm in}, f_{\rm col}^{\rm CF}, M, D, i_{\rm disc})$ is the Jacobian determinant for the change of variables from $r_{\rm in}$ to $K_{\rm flux}$; and $h^{-1}( r_{\rm in}, f_{\rm col}^{\rm CF}, M, D, i_{\rm disc} ) = K_{\rm flux}$ is the inverse transformation function, which is given by equation \eqref{eqn:Kflux_body}. Appendix \ref{app:rin2Kflux} derives equation \eqref{eqn:frCF_fK} and goes on to explain the inversion methods (\S\ref{app:basisfunc}; \S\ref{app:constopt}) used to obtain $f_{K}( K_{\rm flux} )$ in the integrand of equation \eqref{eqn:frCF_fK}.\footnote{We note that the marginalization over $\{ f_{\rm col}, M, D, i_{\rm disc} \}$ removes information. As such, the inversion of equation \eqref{eqn:frCF_fK} is under-determined and $f_{K}( K_{\rm flux} )$ is not unique.} Figure \ref{fig:fK} in the appendix shows the $f_{K}( K_{\rm flux} )$ distributions calculated from this process for each X-ray binary system.

At this stage, we have successfully constructed $f_{K}( K_{\rm flux} )$ in a way that is as consistent as possible with the reported disc continuum fitting black hole spin measurements. That is, our definition of $K_{\rm flux}$ came from a re-parametrized disc model that is equivalent to the \texttt{kerrbb} model (see \S\ref{sec:meth}; Appendix \ref{app:MCD}); we adopted the same joint density $f_{M, D, i}( M, D, i_{\rm disc} )$ used in making the reported spin measurements (see \S\ref{sec:MDi}); and we did not incorporate any $f_{\rm col}$ uncertainties (see \S\ref{sec:fcol}). With $f_{K}( K_{\rm flux} )$ in hand, we can now run the process described above in the opposite direction, but this time marginalizing over $f_{\rm col}$ to produce an updated inner disc radius marginal density $f_{r}( r_{\rm in} )$ that \textit{includes} $f_{\rm col}$ uncertainties,
\begin{align}
f_{r}\left( r_{\rm in} \right) &= \int\limits_{f_{\rm col} = 1}^{\infty} \int\limits_{i_{\rm disc} = 0}^{\pi / 2} \int\limits_{D = 0}^{\infty} \int\limits_{M = 0}^{\infty} d M~d D~d i_{\rm disc}~d f_{\rm col} \nonumber \\
&\times f_{K}\left( h^{-1}\left( r_{\rm in}, f_{\rm col}, M, D, i_{\rm disc} \right) \right) \nonumber \\
&\times f_{f}\left( f_{\rm col} \right) f_{M, D, i}\left( M, D, i_{\rm disc} \right) \nonumber \\
&\times \left| J\left( r_{\rm in}, f_{\rm col}, M, D, i_{\rm disc} \right) \right|. \label{eqn:fr_fK}
\end{align}
Notably, equation \eqref{eqn:fr_fK} assumes that $f_{\rm col}$ is independent of $\{ M, D, i_{\rm disc} \}$, which may not be strictly true but is consistent with the relatively weak dependence of $f_{\rm col}$ on these parameters in the \texttt{bhspec} model.\footnote{See footnote \ref{foot:fcol} in Appendix \ref{app:meth_PDFs}.} Appendix \ref{app:Kflux2rin} describes how to obtain this ``revised'' $f_{r}( r_{\rm in} )$ and its associated spin measurement $f_{a}( a )$ from equation \eqref{eqn:fr_fK}, which only differs from equation \eqref{eqn:frCF_fK} by including uncertainties in $f_{\rm col}$ through its marginal density $f_{f}( f_{\rm col} )$. The lower integration bound $f_{\rm col} = 1$ arises because values of $f_{\rm col} < 1$ are unphysical, so in practice we construct $f_{f}( f_{\rm col} )$ as a truncated distribution.

Evaluating the effects of $f_{\rm col}$ uncertainties on a black hole spin measurement requires specifying the $f_{f}( f_{\rm col} )$ distribution. Investigations into the effects of various atmospheric physics (e.g., magnetic support, density inhomogeneities, vertical dissipation of accretion power) on the emergent disc spectrum warrant considering a systematic uncertainty on $f_{\rm col}$ of $\pm$0.3 or perhaps even more (see \S\ref{sec:uncertain}). Lacking a good understanding of these uncertainties, we simply take $f_{\rm col}$ to be normally distributed and centered on its value in Table \ref{tab:obs} with a standard deviation of either 0.1, 0.2, or 0.3.\footnote{We adopt symmetric uncertainties for simplicity's sake and because different physical effects can conceivably drive $f_{\rm col}$ in either direction (see \S\ref{sec:uncertain}). Asymmetric uncertainties may turn out to be more realistic, which are straightforward to incorporate.}

Figure \ref{fig:spinPDFs} is the culmination of this exercise, approximating the black hole spin measurements that would have been obtained had the disc continuum fitting practitioners incorporated an uncertainty on $f_{\rm col}$ at the levels of $\pm0.1$ (\textit{blue lines}), $\pm0.2$ (\textit{orange lines}), or $\pm0.3$ (\textit{green lines}). As the uncertainty in $f_{\rm col}$ increases, the peak of $\fr$ shifts towards smaller $r_{\rm in}$, while the median of the distribution stays roughly the same. This has the effect of making $\fr$ more asymmetric, with a sharper rise and longer tail.

In most cases, an $f_{\rm col}$ uncertainty of $\pm$0.2--0.3 single-handedly dominates the black hole spin error budget,\footnote{That is, the revised uncertainty range on the black hole spin is more than twice the original reported uncertainty range.} which also includes uncertainties on $\{ M, D, i_{\rm disc} \}$ and $K_{\rm flux}$. Pairing the $f_{\rm col}$ uncertainty level required to at least double the spin error budget with each individual source, we have: $\pm0.1 \rightarrow$ None; $\pm 0.2 \rightarrow$ M33 X--7, GRO J1655--40, LMC X--3, A0620--00; $\pm0.3 \rightarrow$ XTE J1550--564; while for an $f_{\rm col}$ uncertainty of $\pm0.3$, the spin error budget increases by 83\% for H1743--322, 81\% for 4U 1543--47, and 61\% for LMC X--1. The \textit{grey lines}, which correspond to incorporating zero uncertainty on $f_{\rm col}$, validate our methodology by successfully recovering the original black hole spin measurements --- admittedly, XTE J1550--564 and H1743--322 are not perfect.

In \S\ref{sec:disc}, we make the case that a $\pm0.2$--0.3 level of systematic uncertainty in $f_{\rm col}$ is reasonable, given the current understanding of disc atmospheric physics.

\subsection{Black Hole Spin Measurement Sensitivity to $f_{\rm col}$}
\label{sec:fcolVal}
The previous section supposed that the $f_{\rm col}$ values listed in Table \ref{tab:obs} are correct and demonstrated that uncertainties in an adopted $f_{\rm col}$ value can significantly broaden the measured black hole spin marginal density. Here, we consider the possibility that an adopted $f_{\rm col}$ value is incorrect due to systematic uncertainties associated with the \texttt{bhspec} model used to inform the choice for $f_{\rm col}$. We will use the same approach outlined in \S\ref{sec:fcolErr} to determine the sensitivity of the black hole spin measurements to different $f_{\rm col}$ choices. Again, we stress that this approach cannot speak to whether an adopted $f_{\rm col}$ value is in tension with the actual data; however, degeneracies among the \texttt{kerrbb} parameters can conceivably compensate for different $f_{\rm col}$ choices to yield a statistically good fit to the data \citep[e.g.,][]{Nowak2008}.

The \textit{grey lines} in Figure \ref{fig:spinPDFs} necessarily recover the original black hole spin measurements because they adopt the same $f_{\rm col}$ values as in Table \ref{tab:obs} and neglect $f_{\rm col}$ uncertainty, as is effectively done by the disc continuum fitting practitioners. Maintaining this disregard for $f_{\rm col}$ uncertainty throughout this section, the inner disc radius marginal density $f_{r}( r_{\rm in} )$ from equation \eqref{eqn:fr_fK} becomes
\begin{align}
f_{r}\left( r_{\rm in} \right) &= \int\limits_{i_{\rm disc} = 0}^{\pi / 2} \int\limits_{D = 0}^{\infty} \int\limits_{M = 0}^{\infty} d M~d D~d i_{\rm disc} \nonumber \\
&\times f_{K}\left( h^{-1}\left( r_{\rm in}, f_{\rm col}, M, D, i_{\rm disc} \right) \right) \nonumber \\
&\times f_{M, D, i}\left( M, D, i_{\rm disc} \right) \nonumber \\
&\times \left| J\left( r_{\rm in}, f_{\rm col}, M, D, i_{\rm disc} \right) \right|, \label{eqn:fr_fK_fcol}
\end{align}
where we took $f_{f}( f_{\rm col} )$ to be a Dirac delta function centered on the new adopted value for $f_{\rm col}$.

\begin{figure}
    \begin{center}
        \includegraphics[width=0.495\textwidth]{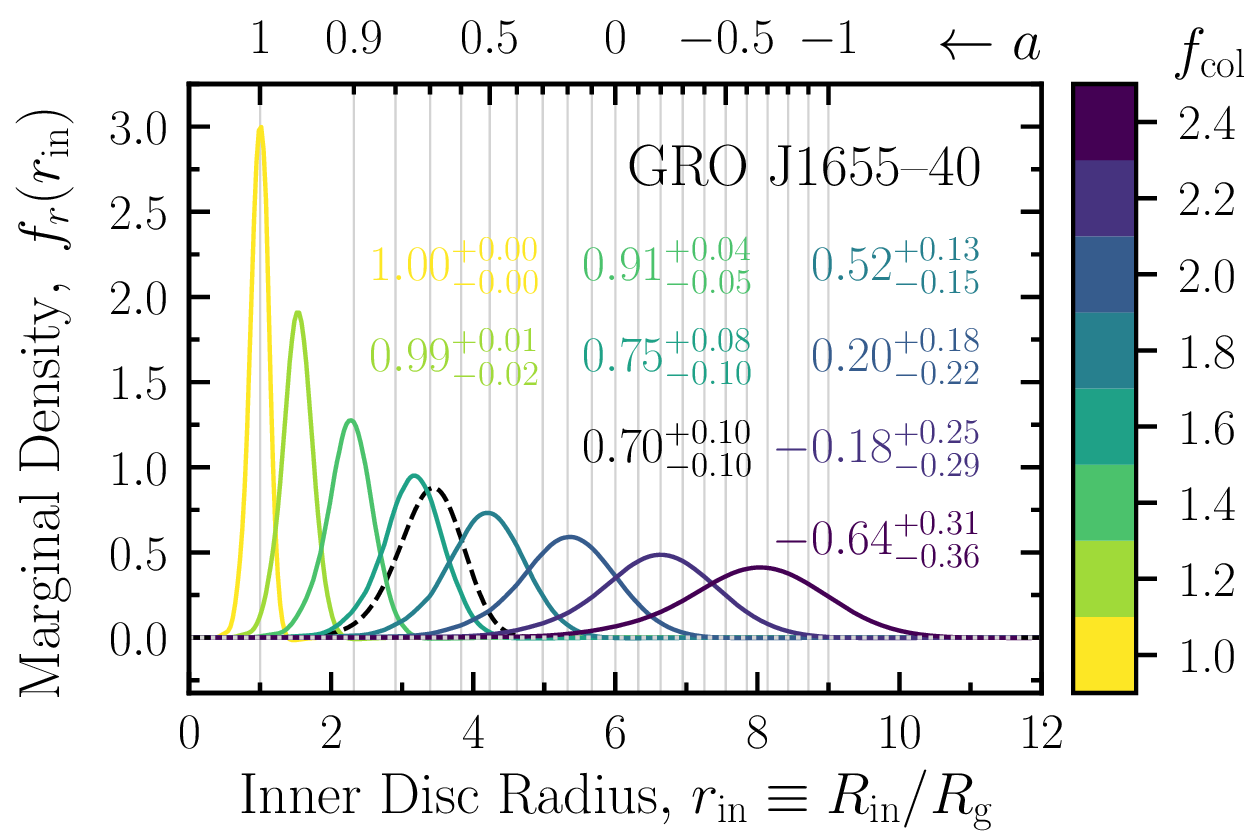}
     \end{center}
    \vspace{-4mm}
    \caption{Inner disc radius marginal density $f_{r}( r_{\rm in} )$ for different choices of the colour correction factor $f_{\rm col}$, using GRO J1655--40 as an example. The \textit{top} $x$-axis shows the black hole spin parameter $a$ corresponding to $r_{\rm in} = r_{\rm ISCO}$ and the \textit{dashed black line} shows the original disc continuum fitting measurement. Increasing $f_{\rm col}$ causes the measured black hole spin to decrease and broadens its marginal density. The colour-coded \textit{legend} gives each black hole spin measurement, which changes substantially in response to modest changes in $f_{\rm col}$, especially for low-to-moderate spins.}
    \label{fig:fr_fcolVal}
\end{figure}

Calculating $f_{r}( r_{\rm in} )$ from equation \eqref{eqn:fr_fK_fcol} with different choices of $f_{\rm col}$ ranging from 1.0 to 2.4 in increments of 0.2, Figure \ref{fig:fr_fcolVal} shows the corresponding black hole spin measurements that would be obtained for GRO J1655--40, as a representative example. Choosing a larger $f_{\rm col}$ leads to a lower black hole spin measurement because the associated increase in $r_{\rm in} \propto f_{\rm col}^{2}$ (see equation \ref{eqn:Kflux_body}) maps to a decrease in $a$, under the assumption that $r_{\rm in}$ coincides with the ISCO (see Figure \ref{fig:rISCO_spin}). This strong degeneracy between $f_{\rm col}$ and $a$ is well-known \citep[e.g.,][]{Nowak2008}, but often goes under-appreciated, which compelled us to explicitly show how an erroneous choice for $f_{\rm col}$ impacts the existing spin measurements.

\begin{figure}
    \begin{center}
        \includegraphics[width=0.495\textwidth]{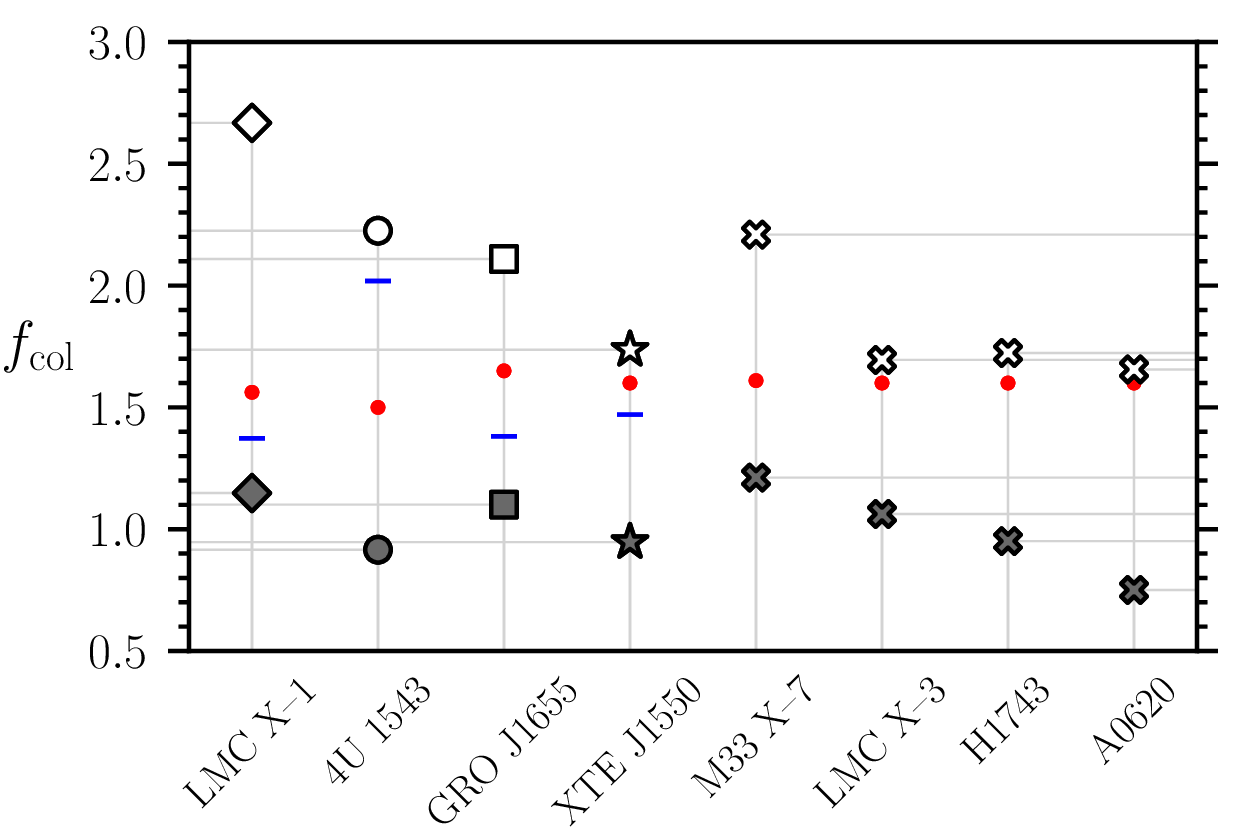}
     \end{center}
    \vspace{-4mm}
    \caption{Colour correction factor $f_{\rm col}$ for each X-ray binary in Table \ref{tab:obs} that would yield a black hole spin measurement of $a = 0$ (\textit{white symbols}) and $a = 0.998$ (\textit{dark symbols}). Measuring $a = 0.998$ is not possible in sources that require $f_{\rm col} < 1$, which is unphysical. \textit{Vertical} and \textit{horizontal lines} help guide the eye. In most cases, $0.998 > a > 0$ is possible with $1.0 < f_{\rm col} \lesssim 2.2$. The \textit{red dots} mark the $f_{\rm col}^{\rm CF}$ values from Table \ref{tab:obs} and the \textit{blue dashes} mark the $f_{\rm col}$ values needed to make $a$ consistent with the iron line spin measurements.}
    \label{fig:fcol_a0}
\end{figure}

Although not shown, we repeat this exercise of varying $f_{\rm col}$ and calculating $f_{r}( r_{\rm in} )$ for the eight X-ray binaries we consider. Figure \ref{fig:fcol_a0} summarizes the results, showing the $f_{\rm col}$ values that cause the measured black hole to be maximally-spinning ($a = 0.998$; \textit{black symbols}) and non-spinning ($a = 0$; \textit{white symbols}). The spin measurements for 4U 1543--47, XTE J1550--564, H1743--322, and A0620--00 cannot be made maximally-spinning by changing $f_{\rm col}$ alone because this would require $f_{\rm col} < 1$, which is unphysical. The moderate spin measurements in 4U 1543--47, GRO J1655-40, and M33 X--7 can be made non-spinning by adopting $f_{\rm col} \simeq 2.2$. This result is seemingly inconsistent with $f_{\rm col} \gtrsim 3$ \citep[e.g.,][]{BegelmanPringle2007}, unless retrograde discs are common.

We also determined the $f_{\rm col}$ values needed to bring the disc continuum fitting black hole spin measurements into agreement with those from the iron line method. These are: $f_{\rm col} = 1.37$, 2.02, 1.38, 1.47 for LMC X--1, 4U 1543--47, GRO J1655--40, XTE J1550--564, respectively. Notably, the three systems with discrepant spin measurements cannot all be brought into agreement by applying a systematic shift in $f_{\rm col}$. Taken at face value, this suggests that different disc atmospheric physics might be operating in these systems to drive $f_{\rm col}$ in different directions. Alternatively, other uncertainties inherent to the disc continuum fitting method (e.g., does $i_{\rm disc} = i_{\rm orb}$?), or the iron line method, or both as is often the case, could be responsible for the mismatched spins. Regardless, given the uncertainties in the continuum fitting method and the system parameters, Figure \ref{fig:fcol_a0} is plausibly consistent with the relatively narrow range of $f_{\rm col} \simeq 1.4$--1.8 predicted by spectral models \citep{Davis2005, DavisElAbd2019}.

\section{Discussion}
\label{sec:disc}
Upon the inception of the disc continuum fitting concept for measuring black hole spin, \citet{Zhang1997} recognized that, ``The largest theoretical uncertainty in [measuring the inner disc radius] is from $f_{\rm col}$.'' Over two decades later, we think this still describes the situation today. The modern technique employs the gold standard in disc atmosphere models, which is commendable, but still uncertain, as discussed below.

\subsection{Uncertainties in the Colour Correction Factor $f_{\rm col}$}
\label{sec:uncertain}
Our discussion will focus on the degree to which disc atmospheric physics and its $f_{\rm col}$ parametrization remains not fully understood, but first we point out that several other uncertainties abound with disc continuum fitting measurements of black hole spin. These include: reliance on an $\alpha$-disc model, whereas other disc models might also provide good fits to the data; whether or not the inner disc truncates at the ISCO with a zero-torque boundary condition \citep[e.g.,][]{ReynoldsArmitage2001}; whether or not the inner disc inclination is interchangeable with the binary orbital or jet axes inclinations \citep[e.g.,][]{SalvesenPokawanvit2020}; contamination of the disc continuum Wien tail by a high-energy, non-thermal spectral component that is always present at some level \citep[e.g.,][]{Kubota2001}; the role of bulk Comptonization of thermal disc photons by turbulent motions \citep[e.g.,][]{Socrates2004}; and the role of heat transport mechanisms other than radiative diffusion, such as convection \citep[e.g.,][]{Coleman2017} or magnetic reconnection / flares \citep[e.g.,][]{Beloborodov1999}.

To its credit, the modern disc continuum fitting technique makes an educated $f_{\rm col}$ choice for each spectral fit. This is done by tabulating $f_{\rm col}$ on a grid of the two free parameters $( \dot{M}, a )$ by fitting the colour-corrected \texttt{kerrbb} model to the physically-motivated \texttt{bhspec} model (see \S\ref{sec:fcol}). The statistical uncertainty on $f_{\rm col}$ from this fitting process can be roughly $\pm0.03$, but is not considered, and choosing an effective viscosity $\alpha = 0.01$ can systematically lower $f_{\rm col}$ compared to $\alpha = 0.1$ \citep[see Figure 3 of][]{DoneDavis2008}. In most cases, practitioners perform two identical analyses with different $f_{\rm col}$ grids, one tabulated with $\alpha = 0.01$ and the other with $\alpha = 0.1$, and then combine the two results into the reported black hole spin error budget. Figure 3 of \citet{DoneDavis2008} shows that this practice amounts to including a systematic uncertainty on $f_{\rm col}$, but at a level of perhaps $\pm0.02$ for $\dot{m} \lesssim 0.1$, which is the relevant $\dot{m}$ for most cases, and as much as about $\pm0.05$ for $\dot{m} \simeq 0.3$. To be clear, our analysis assumed that no $f_{\rm col}$ uncertainties went into the black hole spin error budget. But to be fair, $f_{\rm col}$ uncertainties are probably included at about the $\pm0.02$-level, but are not directly reported.

The salient point we want to make is that disc continuum fitting spin measurements effectively ignore $f_{\rm col}$ uncertainties, by which we mean there are neglected systematics due to the model-dependence of \texttt{bhspec} that are potentially much larger than the nominal $\pm0.02$-level. We now turn to the literature to demonstrate that $f_{\rm col}$ uncertainties at the $\pm$0.2--0.3-level are reasonable, which we showed are generally enough to dominate the black hole spin error budget (see Figure \ref{fig:spinPDFs}).

Hints that there may be missing physics in the \texttt{bhspec} model come from the exquisite thermal component in LMC X--3, which cannot be reconciled with the predicted relativistically broadened absorption edges below $\sim1~{\rm keV}$ \citep{Kubota2010}.\footnote{For observations lacking soft X-ray coverage, \texttt{bhspec} sometimes fits the data better than colour-corrected disc models like \texttt{kerrbb} \citep{Davis2006}.} The missing ingredient from \texttt{bhspec} might be a contribution from magnetic pressure support to the disc's vertical hydrostatic balance, which reduces the strength of these edges as follows \citep[see][]{Blaes2006}: Increased magnetic pressure support alleviates the need for steep gas pressure gradients in combating the large magnitude of the vertical tidal gravitational acceleration at high disc altitudes. Consequently, magnetically supported discs have more vertically extended atmospheres with larger density scale heights and \textit{lower} gas densities at the effective photosphere.\footnote{See Appendix \ref{app:physics} for a discussion of the effective photosphere.} Note that the ratio of the ionization rate to the recombination rate scales as $N n_{i} / ( n_{i} n_{e} )$, where $N$ is the number of ionizing photons and $n_{i} \simeq n_{e}$ are the ion and electron number densities. Therefore, a reduced gas density at the effective photosphere due to magnetic support corresponds to an increased ionization state of metals; thus, reducing the strength of absorption edges.

This same decrease in gas density at the effective photosphere $\rho_{\ast}$, combined with the increase in gas temperature $T_{\ast}$, associated with the inclusion of magnetic support also tends to harden the disc continuum, as follows. The blackbody spectrum produced at the effective photosphere emerges modified (see Appendix \ref{app:physics}), with the effective temperature hardened by an amount $f_{\rm col} \sim ( \kappa_{\rm es}^{\ast} / \kappa_{\rm th}^{\ast} )^{1/8}$, where $\kappa_{\rm es}^{\ast} / \kappa_{\rm th}^{\ast} \propto \rho_{\ast}^{-1} T_{\ast}^{7/2}$ is the scaling of the scattering-to-absorption grey opacities, assuming Kramers's opacity law. Applying the same radiative transfer code used in the \texttt{bhspec} model to the time- and horizontally-averaged vertical structure of a \textit{weakly magnetized} radiation magnetohydrodynamic (MHD) shearing box simulation, \citet{Blaes2006} found that including magnetic support indeed hardened the disc-integrated spectrum, with a best-fit $f_{\rm col} = 1.74$, as compared to $f_{\rm col} = 1.48$ without magnetic support.

On the other hand, magnetically supported disc surface layers are prone to Parker instabilities, which leads to large gas density
inhomogeneities at the scattering and effective photospheres \citep{Blaes2007}. Interestingly, in the weakly magnetized regime accessible to zero-net vertical flux MHD simulations, the characteristic $\sim10\%$ increase in $f_{\rm col}$ from magnetic pressure support is compensated by a $\sim10\%$ decrease from density inhomogeneities \citep{Davis2009}. This softening results from the denser regions enhancing the overall emissivity (because $\epsilon_{\nu}^{\rm ff} \propto \rho^{2}$), as follows. The total liberated gravitational energy gets partitioned into more photons that individually carry less energy, as compared to a homogeneous medium with a lower emissivity and hence fewer photons to carry the same total energy. Consequently, in an inhomogeneous medium the radiation temperature is lower, which means the spectrum is softer. A typical photon produced at the effective photosphere is then able to scatter to a low density region (because $\kappa_{\rm es} \gg \kappa_{\rm th}$) from which it is more likely to escape before being absorbed.

The aforementioned studies of how magnetic fields affect the emergent disc spectrum used disc structures from shearing box simulations with weak magnetization; that is, the ratio of gas-to-magnetic pressures $\beta \equiv P_{\rm gas} / P_{\rm mag} \ll 1$ at the disc mid-plane. It remains to be seen whether or not the hardening/softening balance between magnetic support and density inhomogeneities extends to more strongly magnetized regimes, which dramatically influence the disc structure \citep[e.g.,][]{Salvesen2016a, Mishra2020}.

Crude estimates suggest that $f_{\rm col}$ might achieve extreme values for strong disc magnetizations. Extending the $\alpha$-disc model to the magnetic pressure-dominated regime \citep{Pariev2003}, then assuming LTE inside the disc and that radiative diffusion transports the liberated gravitational energy outward, \citet{BegelmanPringle2007} found the scaling of the lower limit on the colour correction factor to be
\begin{equation}
f_{\rm col} \gtrsim 3.4~\left( \frac{\alpha}{1} \right)^{5/27} \left( \frac{m}{10} \right)^{-1/108} \left( \frac{\dot{m}}{0.1} \right)^{5/108} \left( \frac{x}{20} \right)^{-25/108} \nonumber
\end{equation}
in the maximum field strength limit \citep{PessahPsaltis2005}, which suggests that strongly magnetized discs have very large colour temperatures. Here, $m \equiv M / M_{\odot}$, $\dot{m} \equiv \dot{M} / \dot{M}_{\rm Edd}$, and $x \equiv R / R_{\rm g}$. Importantly, we include the radiative efficiency factor (adopting $\eta = 0.057$) in the Eddington mass accretion rate $\dot{M}_{\rm Edd} = L_{\rm Edd} / (\eta c^{2})$, where $L_{\rm Edd} = 4 \pi G M m_{\rm p} c / \sigma_{\rm T}$ is the Eddington luminosity. This way, $\dot{m}$ is equivalent to $l \equiv L_{\rm disc} / L_{\rm Edd}$, where $L_{\rm disc} = \eta \dot{M} c^{2}$ is the disc luminosity.\footnote{Disc structure models sometimes omit $\eta$ in their definition of $\dot{m}$ \citep[e.g.,][]{SvenssonZdziarski1994, BegelmanPringle2007}.} We chose $\alpha = 1$ as representative based on results from magnetic pressure-dominated shearing box simulations \citep{Salvesen2016a}. For comparison, we find that the same analysis for a non-magnetized, radiation pressure-dominated disc structure gives the much lower lower limit \citep{ShakuraSunyaev1973, SvenssonZdziarski1994}
\begin{equation}
f_{\rm col} \gtrsim 1.7~\left( \frac{\alpha}{0.1} \right)^{2/9} \left( \frac{m}{10} \right)^{1/36} \left( \frac{\dot{m}}{0.1} \right)^{23/36} \left( \frac{x}{20} \right)^{-11/12}. \nonumber
\end{equation}

By neglecting the effect of thermal Comptonization to down-scatter escaping photons into a Wien-like spectrum, \citet{BegelmanPringle2007} overestimate $f_{\rm col}$ for the magnetically dominated regime they consider. One expects Comptonization to be important for large $f_{\rm col}$ because the number of scatterings experienced by a photon escaping from the effective photosphere is approximately $N \sim \tau_{\ast}^{2} \sim f_{\rm col}^{8}$. This strong $f_{\rm col}$ scaling comes from $f_{\rm col} \sim T_{\ast} / T_{\rm eff} \sim \tau_{\ast}^{1/4}$ (see Appendix \ref{app:physics}). Following a similar analysis to \citet{BegelmanPringle2007} in estimating $f_{\rm col}$, we estimate the Compton $y$-parameter $y_{\ast} \equiv 4 k_{\rm B} T_{\ast} / ( m_{\rm e} c^{2} ) \max( \tau_{\ast}^{2}, \tau_{\ast})$ at the optically thick ($\tau_{\ast} \gg 1$) effective photosphere to be
\begin{equation}
y_{\ast} = \frac{4 k_{\rm B} T}{m_{\rm e} c^{2}} \tau_{\rm es}^{-2} \left( \frac{\kappa_{\rm bf}}{\kappa_{\rm es}} \right)^{-2} \left( \frac{\rho_{\ast}}{\rho} \right)^{-2}, \nonumber
\end{equation}
where $T$, $\rho$, and $\tau_{\rm es}$ are the gas temperature, gas density, and electron scattering optical depth, all at the disc mid-plane. Evaluating the Kramers's bound-free opacity for solar abundances, $\kappa_{\rm bf} = 1.6 \times 10^{24} \rho T^{-7/2}~{\rm cm^{2}~g^{-1}}$, at the disc mid-plane and adopting an electron scattering opacity $\kappa_{\rm es} = 0.4~{\rm cm^{2}~g^{-1}}$, we find that in the maximally magnetized limit \citep{BegelmanPringle2007},
\begin{equation}
y_{\ast} \gtrsim 1.4 \times 10^{2}~\left( \frac{\alpha}{1} \right)^{5/3} \left( \frac{m}{10} \right)^{-1/3} \left( \frac{\dot{m}}{0.1} \right)^{2/3} \left( \frac{x}{20} \right)^{-17/6}, \nonumber
\end{equation}
with the lower limit being because $\rho_{\ast} / \rho < 1$. Figures 1b, 2b, and 3b in \citet{ShimuraTakahara1995b} compare the $y_{\ast} \gg 1$ inner disc annuli to the $y_{\ast} \ll 1$ outer disc annuli, finding only a mild $f_{\rm col}$ reduction attributed to thermal Comptonization. The extent to which Comptonization limits large $f_{\rm col}$ in the magnetically dominated regime may be more severe, but understanding the degree to which thermal Comptonization reduces $f_{\rm col}$ requires knowledge of the detailed vertical disc structure (e.g., the gas temperature at the location where $y = 1$).

The low gas density $\rho_{\ast}$ at the effective photosphere is a principle driver of the large $f_{\rm col}$ estimates from magnetically dominated discs. A related quantity is the mid-plane column mass $m_{0}$, with the \citet{BegelmanPringle2007} model predicting
\begin{equation}
m_{0} \sim 1.2 \times 10^{2}~\left( \frac{\alpha}{1} \right)^{-8/9} \left( \frac{m}{10} \right)^{1/9} \left( \frac{\dot{m}}{0.1} \right)^{7/9} \left( \frac{x}{20} \right)^{-5/9}~{\rm g~cm^{-2}}. \nonumber
\end{equation}
Like $y_{\ast}$, the expected $m_{0}$ is sensitive to $\alpha$. Although the existing disc atmosphere models lack magnetic pressure support, they do span a large range of $m_{0}$ down to  $10^{2.5}~{\rm g~cm^{-2}}$ and find that $f_{\rm col} \lesssim 2.0$ for $\dot{m} \sim 0.1$ \citep[e.g.,][]{DavisElAbd2019}. Based on this $m_{0}$ parameter study, the very large $f_{\rm col}$ values predicted for magnetically dominated discs might not be plausible.

Besides magnetization, several other pieces of ``missing physics'' from \texttt{bhspec} can affect $f_{\rm col}$, with different degrees of importance. Encouragingly for disc continuum fitting applications, when most of the accretion power gets dissipated below the effective photosphere, the vertical dissipation profile has little impact on $f_{\rm col}$ \citep{Davis2005, Blaes2006}. If instead some fraction $\chi$ of the accretion energy is dissipated in a corona, then the disc becomes colder, denser, and optically thicker \citep{SvenssonZdziarski1994}. These effects conspire to produce a softer disc spectrum, where an example calculation showed that depositing 80\% of the accretion energy into the corona, but neglecting its irradiation on the disc, decreased $f_{\rm col}$ from 1.62 to 1.43 \citep{DavisElAbd2019}.\footnote{\citet{DavisElAbd2019} offer plausible explanations for why \citet{Merloni2000} reported the opposite trend of increasing $f_{\rm col}$ with increasing $\chi$.} Less explored are the effects on the disc continuum from non-solar metallicity, bound-bound opacities, and irradiation from a corona and/or returning radiation \citep[e.g.,][]{Cunningham1976}. Finally, current models cannot simultaneously consider the three-dimensional (3D) local disc structure and treat the radiative transfer self-consistently: \texttt{bhspec} is self-consistent, at the expense of only considering the vertical disc structure \citep{DavisHubeny2006}; while Monto Carlo radiative transfer post-processing of 3D shearing box data sacrifices self-consistency \citep{Davis2009}.

\subsection{Path Forward: The Case for Disc Magnetization}
\label{sec:path}
The discussion above harps on the need to better understand how various physics affect $f_{\rm col}$. Perhaps more difficult is deciding what physics is most relevant to real black hole X-ray binary systems in the high/soft state, where disc continuum fitting is appropriate. Here, we offer several pieces of evidence that warrant future studies on how magnetic fields influence the disc spectrum.

Disc accretion in X-ray binaries is a fundamentally magnetic process; that is, the magneto-rotational instability (MRI) generates turbulence in the disc and mediates angular momentum transport that drives accretion \citep{BalbusHawley1991}. The MRI establishes the critical role of magnetic fields in X-ray binaries and its saturation sets toroidal field strength extrema \citep[e.g.,][]{Hawley1995, Das2018}. Beyond that, magnetization remains perhaps the least understood aspect of disc physics, with the field strength, topology, and dynamical importance being open areas of active study.

A compelling case for dynamically important magnetic fields is made by the highly-ionized, blue-shifted absorption lines from the GRO J1655-40 disc wind, which cannot be explained without invoking a magnetic-driving mechanism \citep{Miller2006a, Balakrishnan2020}. Strong magnetic fields ($B \sim 10^{3}$--$10^{5}$ G) are also required to drive the GRS 1915+105 disc wind in the high/soft state when the mass accretion rate $\dot{m}$ is large \citep{Miller2016}, but a ``failed'' disc wind is observed at lower $\dot{m}$ \citep{Miller2020}. Speculatively, this variable disc wind may be consistent with the idea that $\dot{m}$ regulates the field strength \citep[e.g.,][]{BegelmanArmitage2014}, such that a decrease in $\dot{m}$ diminishes inward advection and the resulting weaker field cannot accelerate a wind to escape from the inner disc regions.

Light curve modeling of black hole X-ray binary outbursts favor a large effective viscosity $\alpha \simeq 0.2$ \citep{Tetarenko2018}, which large-scale poloidal magnetic fields threading the disc naturally provides \citep[e.g.,][]{Hawley1995, Salvesen2016a}. Another attractive property of magnetic pressure-dominated discs is that they are not prone to thermal or viscous instabilities \citep[e.g.,][]{BegelmanPringle2007, Jiang2019}, unlike radiation pressure-dominated discs \citep{ShakuraSunyaev1976, LightmanEardley1974}, which is consistent with the steady disc emission (as opposed to limit cycle behaviour) observed in the high/soft state.

\bigskip
The goal of this discussion was to show that an $f_{\rm col}$ uncertainty of $\pm$0.2--0.3 is a plausible level to include in the black hole spin error budget. For the disc accretion regimes relevant to X-ray binaries in the high/soft state, the current best theoretical understanding is that $f_{\rm col}$ can span $\simeq 1.4$--2.0 \citep[e.g.,][]{Davis2005, DavisElAbd2019}. Observational attempts to measure $f_{\rm col}$ are consistent with this range, but very uncertain (see \S\ref{sec:intro}). The case for larger $f_{\rm col}$ is much more speculative.

Theoretically, $f_{\rm col} > 2.0$ either requires strong magnetization \citep{BegelmanPringle2007}; or $\alpha \sim 1$ \citep{DavisElAbd2019}, as seen in magnetic pressure-dominated shearing box simulations \citep{Salvesen2016a}; or the disc surface density to be low enough to make it ``photon starved,'' meaning that the resulting lowered emissivity cannot radiate away the liberated gravitational energy and the radiation temperature must increase to compensate \citep{DavisElAbd2019}. Observationally, a static $f_{\rm col}$ is \textit{not} consistent with an inner disc edge truncated out to $\gtrsim100~R_{\rm g}$ that migrates inwards as the source softens \citep{ReynoldsMiller2013}. Instead, the alternative interpretation that the inner disc edge remains near the ISCO during state transitions implies a dynamic $f_{\rm col}$ that varies by a factor of $\sim$2.0--3.5 \citep{Salvesen2013}. 

Regardless of whether $f_{\rm col}$ can reach such extremes, the discussion above justifies our representative $f_{\rm col}$ uncertainties of $\pm$0.2--0.3, which are enough to dominate the uncertainty in black hole spin measurements made with the disc continuum fitting technique (see Figure \ref{fig:spinPDFs}).

\section{Summary and Conclusions}
\label{sec:sumconc}
The disc continuum fitting community reports black hole spin measurements that meet certain quality criteria for 10 different X-ray binaries \citep{McClintock2014}, but the reported spins effectively do not marginalize over the colour correction factor $f_{\rm col}$ (see \S\ref{sec:fcol}). By re-parametrizing the disc continuum model in terms of a disc flux normalization $K_{\rm flux}$ (see \S\ref{sec:meth}; Appendix \ref{app:MCD}), we reverse-engineered each of the eight non-maximal black hole spin marginal densities to obtain the marginal density $f_{K}( K_{\rm flux} )$ (see \S\ref{sec:analres}; Appendix \ref{app:meth_PDFs}), which we treat as the observable. Combining uncertainties on all parameters in the disc continuum model, \textit{including} $f_{\rm col}$, we presented revised black hole spin measurements with different choices for either the $f_{\rm col}$ uncertainty level (see \S\ref{sec:fcolErr}) or the adopted $f_{\rm col}$ value (see \S\ref{sec:fcolVal}). From this analysis, our conclusions are that:
\begin{itemize}
\item Adopting systematic uncertainties on $f_{\rm col}$ at the level of $\pm0.2$--0.3 is reasonable (see \S\ref{sec:disc}) and enough to dominate the error budget of most black hole spin measurements from disc continuum fitting (see Figure \ref{fig:spinPDFs}).
\item The measured black hole spin decreases precipitously with increasing $f_{\rm col}$ (see Figure \ref{fig:fr_fcolVal}). A systematic offset between the adopted $f_{\rm col}$ and its ``actual value'' will lead to an erroneous black hole spin measurement, especially for low-to-moderate intrinsic spins.
\item Of the six X-ray binaries with vetted black hole spin measurements from two independent techniques, the three non-extremal cases (i.e., $a < 0.9$) are discrepant at the $68\%$-level (see Figure \ref{fig:bhspins}). Incorporating $f_{\rm col}$ uncertainties may help alleviate this tension, as the discrepant sources could be brought into agreement with $f_{\rm col}$ values in the range $1.4 \le f_{\rm col} \le 2.0$.
\end{itemize}

The widespread success of disc continuum fitting applications to X-ray binaries in the high/soft state is a testament to the applicability of a colour-corrected, multi-temperature, disc blackbody model in this accretion regime. However, using this disc model to measure physical parameters of the system, like the black hole spin, requires a firm understanding of its phenomenological parametrization in terms of $f_{\rm col}$. Reporting accurate measurements requires knowledge and inclusion of the $f_{\rm col}$ uncertainties, both statistical and systematic. Statistical uncertainties refer to the validity of a colour-corrected approximation to the true disc continuum (e.g., \texttt{kerrbb} fits to \texttt{bhspec} models are not perfect). Systematic uncertainties refer to the possibility that the physical disc atmosphere model (e.g., \texttt{bhspec}) being used to inform the choices for $f_{\rm col}$ is missing potentially relevant physics, such as the effects of magnetization. To better understand how disc atmospheric physics affects the observed spectrum, the path forward might involve local and global numerical simulations of thin discs that self-consistently treat magnetohydrodynamics and frequency-dependent radiative transfer, which may soon be possible with state-of-the art codes like \texttt{MOCMC} \citep{RyanDolence2020}.

\section*{Acknowledgements}
We are grateful to the anonymous referee for their constructive feedback. GS appreciates the helpful comments and suggestions from an anonymous referee on a withdrawn draft. We thank Mitch Begelman, Omer Blaes, Eric Coughlin, Josh Dolence, Brooks Kinch, Tom Maccarone, Jon Miller, Jordan Mirocha, and Ben Ryan for productive discussions, along with Robert Feldmann for helpful e-mail exchanges. GS acknowledges support through an NSF Astronomy \& Astrophysics Postdoctoral Fellowship under award AST-1602169 and the NASA Earth and Space Science Graduate Fellowship program.

This work was supported by the US Department of Energy through the Los Alamos National Laboratory (LANL). Additional funding was provided by the Laboratory Directed Research and Development Program and the Center for Space and Earth Science (CSES) at LANL under project numbers 20190021DR and 20180475DR (TS). This research used resources provided by the LANL Institutional Computing Program. LANL is operated by Triad National Security, LLC, for the National Nuclear Security Administration of U.S. Department of Energy under Contract No. 89233218CNA000001. LANL assigned this article the document release number LA-UR-20-28487.

\textit{Software}: We are grateful to the countless developers contributing to open source projects on which this work relied, including \texttt{Python} \citep{rossumPythonWhitePaper}, \texttt{NumPy} and \texttt{SciPy} \citep{numpy,scipy}, \texttt{Matplotlib} \citep{matplotlib}, \texttt{HDF5} \citep{hdf5}, and \texttt{DataThief} \citep{DataThief}.

\textit{Data Availability}: The data and code used in this paper are available as a \texttt{GitHub} repository: \href{https://github.com/gregsalvesen/bhspinf}{https://github.com/gregsalvesen/bhspinf}. If this link goes defunct, contact the corresponding author to request access to the data and code. LANL approved the data and code for open source distribution under data release number LA-UR-20-28697 and the BSD-3 License, respectively.

\bibliographystyle{aasjournal}
\bibliography{SalvesenMiller2020}

\appendix

\section{Physics of Disc Continuum Spectral Hardening}
\label{app:physics}
The standard $\alpha$-disc model for an X-ray binary predicts a sufficiently high gas temperature ($\sim10^{7}~{\rm K}$) and low gas density ($\sim10^{-3}~{\rm g~cm^{-3}}$) at the surface of the radiation pressure-dominated inner disc regions, such that electron scattering opacities $\kappa_{\nu}^{\rm es}$ dominate over absorptive opacities $\kappa_{\nu}^{\rm th}$ for typical photon energies \citep[$\sim1~{\rm keV}$; e.g.,][]{ShakuraSunyaev1973}. As explained below, it is primarily this importance of electron scattering that causes the local disc continuum to deviate from a blackbody.

Introducing electron scattering opacities into an atmosphere that contains emissive/absorptive opacities necessitates introducing the concept of an ``effective photosphere,'' which corresponds to the location where a typical photon experiences its last absorption event before embarking on a scattering random walk out of the atmosphere. The location $z_{\nu}^{\ast}$ of this effective photosphere is set by where the effective ``optical depth'' $\tau_{\nu}^{\rm eff} \equiv - \int_{z_{\nu}^{\ast}}^{0} [ 3 \kappa_{\nu}^{\rm th} ( \kappa_{\nu}^{\rm th} + \kappa_{\nu}^{\rm es} ) ]^{1/2} \rho dz = 1$ \citep[e.g.,][]{DavisonSykes1957}, which is not an optical depth in the usual sense. Rather, the atmospheric properties (e.g., gas density, gas temperature, ionization state) at $z_{\nu}^{\ast}$ dictate the formation of the emergent spectrum. The actual optical depth down to the effective photosphere is called the depth of formation $\tau_{\nu}^{\ast} = - \int_{z_{\nu}^{\ast}}^{0} ( \kappa_{\nu}^{\rm th} + \kappa_{\nu}^{\rm es} ) \rho dz$. Because $\tau_{\nu}^{\ast} \gg 1$ when $\kappa_{\nu}^{\rm es} \gg \kappa_{\nu}^{\rm th}$, the emergent spectrum is formed deeper in the atmosphere --- where the density and temperature are higher --- than the usual $\tau_{\nu} = 1$ photosphere definition. The gas temperature at the depth of formation is $T_{\nu}^{\ast}$ and an asterisk on a quantity denotes its evaluation at the effective photosphere.\footnote{The temperature at the effective photosphere is $T_{\nu}^{\ast}$, \textit{not} the effective temperature $T_{\rm eff}$ defined by the outgoing flux.}

Figure \ref{fig:fcol} shows how an atmosphere that is optically thick ($\tau_{\nu}^{\ast} \gg 1$) and electron scattering-dominated ($\kappa_{\nu}^{\rm es} \gg \kappa_{\nu}^{\rm th}$) produces an emergent specific intensity
\begin{equation}
I_{\nu} \sim \sqrt{\frac{\kappa_{\nu}^{\rm th \ast}}{\kappa_{\nu}^{\rm es \ast}}} B_{\nu}\left( T_{\nu}^{\ast} \right), \label{eqn:modbbody}
\end{equation}
called a modified blackbody \citep[e.g.,][]{Schuster1905, Stromgren1951, FeltenRees1972}.\footnote{The modified blackbody can also be derived by solving the radiative transfer equation using the Eddington and two-stream approximations \citep[e.g.,][]{Madej1974, RybickiLightman1979}.} Here, $I_{\nu}$ is a blackbody $B_{\nu}$ at the temperature $T_{\nu}^{\ast}$, but diluted in amplitude by the factor $\sqrt{\kappa_{\nu}^{\rm th \ast} / \kappa_{\nu}^{\rm es \ast}}$ and hardened relative to a blackbody with effective temperature $T_{\rm eff} < T_{\nu}^{\ast}$, which can be understood for a grey atmosphere as follows:

\begin{figure}[b!]
    \vspace{0mm}
    \begin{center}
        \includegraphics[width=0.495\textwidth]{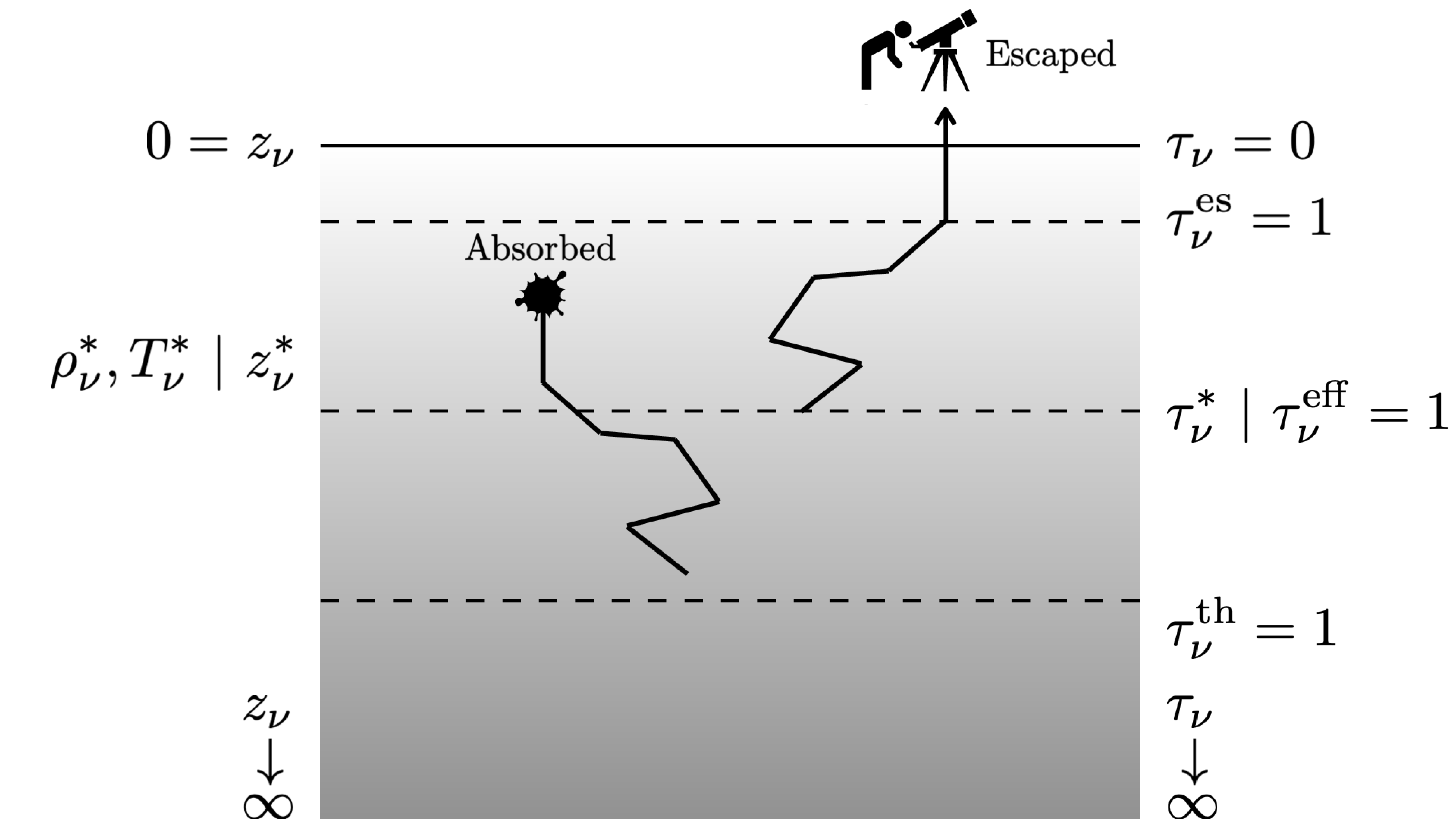}
    \end{center}
    \vspace{-2mm}
    \caption{Cartoon explaining the modified blackbody spectrum emerging from an optically thick, semi-infinite, plane-parallel, emitting atmosphere with total absorptive opacity $\kappa_{\nu}^{\rm th}$, but dominated by isotropic, coherent electron scattering opacities $\kappa_{\nu}^{\rm es} \gg \kappa_{\nu}^{\rm th}$ in the non-relativistic limit ($k T_{\rm e} \ll m_{\rm e} c^{2}$; $h\nu \ll m_{\rm e} c^{2}$). At the altitude $z_{\nu}^{\ast}$ of the effective photosphere, defined by where the effective ``optical depth'' $\tau_{\nu}^{\rm eff} \equiv - \int_{z_{\nu}^{\ast}}^{0} [ 3 \kappa_{\nu}^{\rm th} ( \kappa_{\nu}^{\rm th} + \kappa_{\nu}^{\rm es} ) ]^{1/2} \rho dz = 1$, the radiation gets thermalized with the gas temperature $T_{\nu}^{\ast}$. That is, a typical photon experiences its last absorption event at $z_{\nu}^{\ast}$. Above $z_{\nu}^{\ast}$, the radiation performs a scattering random walk to escape the atmosphere. The actual optical depth to the effective photosphere is $\tau_{\nu}^{\ast} = - \int_{z_{\nu}^{\ast}}^{0} ( \kappa_{\nu}^{\rm th} + \kappa_{\nu}^{\rm es} ) \rho dz \gg 1$, so we assume thermal equilibrium there and Kirchoff's law gives the emission coefficient at the effective photosphere, $j_{\nu}^{{\rm th} \ast} = \kappa_{\nu}^{{\rm th} \ast} \rho_{\nu}^{\ast} B_{\nu}( T_{\nu}^{\ast} )$. Absorption is not important above $z_{\nu}^{\ast}$, so the radiative transfer equation for emission only gives the approximate emergent specific intensity $I_{\nu} \sim j_{\nu}^{{\rm th} \ast} z_{\nu}^{\ast}$. The modified blackbody of equation \eqref{eqn:modbbody} then follows, given $j_{\nu}^{{\rm th} \ast}$ along with $\rho_{\nu}^{\ast} z_{\nu}^{\ast}$ from solving the approximation $\tau_{\nu}^{\rm eff} = 1 \sim \sqrt{ \kappa_{\nu}^{{\rm th} \ast} \kappa_{\nu}^{{\rm es} \ast} } \rho_{\nu}^{\ast} z_{\nu}^{\ast}$. The cartoon also marks the locations of the scattering-only photosphere $\tau_{\nu}^{\rm es} = 1$ and the absorption-only photosphere $\tau_{\nu}^{\rm th} = 1$. If Comptonization is saturated at the effective photosphere ($y_{\nu}^{\ast} \gg 1$), then thermalization occurs at the location where $y_{\nu} = 1$ rather than where $\tau_{\nu}^{\rm eff} = 1$.}
    \label{fig:fcol}
\end{figure}

Considering radiative diffusion as the only heat transport mechanism, the outgoing bolometric radiative flux is $F = c (dP_{\rm rad} / d\tau)$. We assume that any flux added above the effective photosphere is small compared to the flux at the effective photosphere (i.e., $F \simeq {\rm constant}$). This amounts to the assumption that the energy dissipation rate per unit volume $\epsilon = dF / dz$, or the volume-specific dissipation of accretion power, is negligible above the effective photosphere compared to all greater depths.\footnote{Turbulence generated by the magneto-rotational instability, and presumably its associated dissipation, peaks above the mid-plane in the disc surface layers \citep[e.g.,][]{ZhuStone2018}, which if above the effective photosphere would nullify this $F \simeq {\rm constant}$ assumption.} Next, assuming local thermodynamic equilibrium (LTE) means the radiation pressure $P_{\rm rad}^{\ast} = (4 \sigma / 3 c) T_{\ast}^{4}$ at the effective photosphere, and integrating the radiative diffusion equation from $\tau = 0$ down to the depth of formation $\tau = \tau_{\ast}$ gives $F \sim \sigma T_{\ast}^{4} /  \tau_{\ast}$. Equating this expression to the definition of the effective temperature $F \equiv \sigma T_{\rm eff}^{4}$ (recall $\tau_{\ast} \gg 1$) gives $T_{\ast} \sim \tau_{\ast}^{1/4} T_{\rm eff}$. Substituting\footnote{This comes from electron scattering being the dominant opacity source, such that $\tau_{\ast} \simeq \tau_{\rm es}^{\ast} \sim \kappa_{\rm es}^{\ast} \rho_{\ast} z_{\ast}$, combined with the approximation $\tau_{\rm eff} = 1 \sim \sqrt{ \kappa_{\rm th}^{\ast} \kappa_{\rm es}^{\ast} } \rho_{\ast} z_{\ast}$ (see Figure \ref{fig:fcol} caption).} $\tau_{\ast} \sim \sqrt{\kappa_{\rm es}^{\ast} / \kappa_{\rm th}^{\ast}}$ and recalling that $\kappa_{\rm es}^{\ast} \gg \kappa_{\rm th}^{\ast}$, we find the temperature at the effective photosphere $T_{\ast}$ is $\sim( \kappa_{\rm es}^{\ast} / \kappa_{\rm th}^{\ast} )^{1/8}$ times hotter than the effective temperature $T_{\rm eff}$, which means $B_{\nu}( T_{\ast} )$ in equation \eqref{eqn:modbbody} is spectrally harder than $B_{\nu}( T_{\rm eff} )$.

Noticing that the (grey) dilution factor $\sqrt{\kappa_{\rm th}^{\ast} / \kappa_{\rm es}^{\ast}}$ is the minus fourth power of the hardening factor $( \kappa_{\rm es}^{\ast} / \kappa_{\rm th}^{\ast} )^{1/8}$ motivates re-expressing equation \eqref{eqn:modbbody} as a colour-corrected blackbody \citep[e.g.,][]{ShimuraTakahara1995b}
\begin{equation}
I_{\nu} = \frac{1}{f_{\rm col}^{4}} B_{\nu}\left( f_{\rm col} T_{\rm eff} \right), \label{eqn:ccbbody}
\end{equation}
where the colour correction factor $f_{\rm col} \sim ( \kappa_{\rm es}^{\ast} / \kappa_{\rm th}^{\ast} )^{1/8}$ phenomenologically relates the theoretically predicted effective temperature to the observed colour temperature $T_{\rm col} = f_{\rm col} T_{\rm eff}$. The interpretation of $f_{\rm col}$ is the multiplicative shift factor of spectral features to higher energies, while the $f_{\rm col}^{-4}$ factor out front keeps the $\nu I_{\nu}$ amplitude, and therefore the frequency-integrated intensity, constant \citep[e.g.,][]{DavisElAbd2019}. The colour-corrected blackbody approximation of equation \eqref{eqn:ccbbody} neglects the frequency-dependence of thermalization. In general, the frequency-dependence of the depth of formation $\tau_{\nu}^{\ast}$ and the location where the Compton $y$-parameter $y_{\nu} = 1$ means that different frequencies probe different atmospheric depths and therefore different temperatures. Assessing the applicability of the colour-corrected blackbody approximation to accretion disc spectra requires sophisticated modeling.

The modified blackbody arises when electron scattering is an important opacity source, but spectral hardening is also sensitive to the profile of gas temperature, which generally decreases at higher disc altitudes. When the effective photosphere (i.e., where escaping radiation gets thermalized with the gas) is deeper in the atmosphere, the thermalization temperature $T_{\ast}$ is hotter and so the emergent spectrum is harder. Adding bound-free opacities increases the total absorptive opacity $\kappa_{\rm th}$, which pushes the effective photosphere to higher disc altitudes where $T_{\ast}$ is colder, making the emergent modified blackbody softer. Adding bound-free opacities reduces the ratio of scattering-to-absorption opacities $\kappa_{\rm es} / \kappa_{\rm th}$, which also softens the emergent spectrum. These two effects might explain why $f_{\rm col}$ is slightly lower when bound-free absorption opacities are included \citep{Davis2005}, compared to considering only free-free absorption opacities \citep{ShimuraTakahara1995b}.

Finally, there is the question of what physical process is responsible for thermalizing the radiation? The preceding discussion supposes that absorption dictates thermalization, but thermalization with the non-relativistic electrons is possible for saturated Comptonization; that is, when the Compton $y$-parameter at the effective photosphere $y_{\nu}^{\ast} \equiv 4 k_{\rm B} T_{\nu}^{\ast} / (m_{\rm e} c^{2}) \max( \tau_{\nu}^{\ast 2}, \tau_{\nu}^{\ast})$ is large ($y_{\nu}^{\ast} \gg 1$). In this scenario, the blackbody spectrum $B_{\nu}( T_{\nu}^{\ast} )$ produced at the effective photosphere diffuses outward and continually experiences incoherent Compton (down)scattering, ultimately emerging as a softer, Wien-like spectrum with a temperature corresponding to the altitude where $y \simeq 1$ \citep{Ross1992}. Early models neglecting bound-free opacities showed saturated Comptonization ($y_{\nu}^{\ast} \gg 1$) and found that a colour-corrected disc blackbody well-approximated the emergent disc continuum for luminosities $l \gtrsim 0.01$ \citep{ShimuraTakahara1995b}. These models found that disc annuli with $y_{\nu}^{\ast} \gg 1$ were better approximated by a colour-corrected blackbody than their $y_{\nu}^{\ast} \ll 1$ counterparts, and attributed this improvement to Comptonization providing enhanced thermalization of the radiation with the gas. More recent models including bound-free opacities showed Comptonization to be much less important ($y_{\nu}^{\ast} \lesssim 1$), but still found that a colour-corrected disc blackbody well-approximated the emergent disc continuum \citep{Davis2005}. This success was attributed to the relative frequency-independence of $\tau_{\nu}^{\ast}$, such that photons with different frequencies probe similar temperatures because thermalization occurs over a confined range of atmospheric depths.

\section{Accretion Disc Continuum Model}
\label{app:MCD}
Our starting point is the observed specific disc flux calculated from the \texttt{kerrbb} model \citep[equation E2 of][]{Li2005},
\begin{equation}
F_{E_{\rm obs}} = \int g^{3} I_{E_{\rm em}} d\Omega_{\rm obs}, \label{eqn:FEobs_E2}
\end{equation}
which accounts for relativistic effects on photon propagation from the disc surface to the observer by using a ray-tracing technique. Here, $I_{E_{\rm em}}$ is the specific intensity emitted from the disc surface, $E_{\rm obs}$ is the observed photon energy, and $g \equiv E_{\rm obs} / E_{\rm em}$ is the redshift factor relating the observed and emitted photon energies. The solid angle element of the disc image plane seen by the observer, but written in terms of polar coordinates in the disc equatorial plane, is $d\Omega_{\rm obs} = R~dR~d\phi~\cos( i _{\rm disc} ) / D^{2}$. Assuming $I_{E_{\rm em}}$ obeys a colour-corrected blackbody spectrum according to the local disc colour temperature $T_{\rm col}( R ) = f_{\rm col} T_{\rm eff}( R )$, equation \eqref{eqn:FEobs_E2} becomes \citep[equation E7 of][]{Li2005}
\begin{equation}
F_{E_{\rm obs}} = \frac{1}{f_{\rm col}^{4}} \frac{2 E_{\rm obs}^{3}}{h^{3} c^{2}} \int \left[ \exp\left( \frac{E_{\rm obs}}{g k_{\rm B} T_{\rm col}\left( R \right)}\right) -1 \right]^{-1} \Upsilon\left( \theta \right) d\Omega_{\rm obs}, \label{eqn:FEobs_kerrbb}
\end{equation}
where the colour correction factor $f_{\rm col}$ (assumed to be constant) relates the colour and effective temperatures. The \texttt{kerrbb} model obtains $T_{\rm eff}( R ) = [ F_{\rm out}( R ) / \sigma_{\rm SB}]^{1/4}$ by calculating the outgoing flux $F_{\rm out}$ through an iterative procedure that uses the aforementioned ray-tracing technique. The limb-darkening law describing the angular distribution of $I_{E_{\rm em}}$ is taken to be that of a plane-parallel, semi-infinite, electron scattering atmosphere \citep{Chandrasekhar1960, FukueAkizuki2006},
\begin{equation}
\Upsilon\left( \theta \right) = \frac{1}{2} + \frac{3}{4} \cos\left( \theta \right), \label{eqn:limbdark}
\end{equation}
where $\theta$ is the polar angle measured from the disc surface normal to the wavevector of the emitted photon. Defining $\tilde{r} \equiv R / R_{\rm in}$ as the radial coordinate in units of the inner disc radius $R_{\rm in}$ and performing the change of variables $R \rightarrow \tilde{r}$, equation \eqref{eqn:FEobs_kerrbb} becomes
\begin{equation}
F_{E_{\rm obs}} = \frac{r_{\rm in}^{2}}{f_{\rm col}^{4}} \left( \frac{G M / c^{2}}{D} \right)^{2} \cos\left( i_{\rm disc} \right) \frac{2 E_{\rm obs}^{3}}{h^{3} c^{2}} \int\limits_{0}^{2 \pi} \int\limits_{1}^{\tilde{r}_{\rm out}} \left[ \exp\left( \frac{E_{\rm obs}}{g k_{\rm B} T_{\rm col}\left( \tilde{r} \right)}\right) -1 \right]^{-1} \Upsilon\left( \theta \right) \tilde{r}~d\tilde{r}~d\phi, \label{eqn:FEobs}
\end{equation}
where (no tilde) $r \equiv R / R_{\rm g}$ is the radial coordinate in units of the gravitational radius $R_{\rm g} = G M / c^{2}$. The dependence of the disc continuum model on the system parameters $\{ r_{\rm in}, f_{\rm col}, M, D, i_{\rm disc} \}$ is now apparent and a dependence on the mass accretion rate $\dot{M}$ is contained within $T_{\rm col}$. Integrating $F_{E_{\rm obs}}$ over $E_{\rm obs}$ then gives the total observed disc flux,
\begin{equation}
F_{\rm obs} = \int\limits_{0}^{\infty} F_{E_{\rm obs}} dE_{\rm obs}. \label{eqn:Fobs}
\end{equation}

Our first objective is to find an equivalent expression for $F_{\rm obs}$ that replaces the integral in equation \eqref{eqn:FEobs} with something more tractable (i.e., that does not require ray tracing). We will achieve this by isolating the relativistic effects on photon propagation into ``disc flux correction factors,''  as follows. Ignoring relativistic effects on photon propagation (as indicated by the $\widehat{\;}$ symbol below) means that the outgoing flux is axisymmetric, the photons experience no redshift ($g = 1$), and only the photons emitted from the disc surface with polar angle $\theta = i_{\rm disc}$ reach the observer. The total observed disc flux that results from this simplification is \citep[e.g.,][]{Mitsuda1984, Zimmerman2005}
\begin{equation}
\widehat{F}_{\rm obs} = \frac{r_{\rm in}^{2}}{f_{\rm col}^{4}} \left( \frac{G M / c^{2}}{D} \right)^{2} \cos\left( i_{\rm disc} \right) \Upsilon\left( i_{\rm disc} \right) \int\limits_{0}^{\infty} \frac{4 \pi E_{\rm obs}^{3}}{h^{3} c^{2}} \int\limits_{1}^{\tilde{r}_{\rm out}} \left[ \exp\left( \frac{E_{\rm obs}}{k_{\rm B} T_{\rm col}\left( \tilde{r} \right)}\right) -1 \right]^{-1} \tilde{r}~d\tilde{r}~dE_{\rm obs}. \label{eqn:Fobs_noGR}
\end{equation}
To isolate the effect of neglecting relativistic photon propagation on the total observed disc flux, we must adopt the same underlying disc physics used by the \texttt{kerrbb} model; that is, the standard relativistic ``$\alpha$-disc'' model for a geometrically thin, optically thick accretion disc. The $\alpha$-disc model predicts the radial profile of the colour temperature at the disc mid-plane to be \citep{ShakuraSunyaev1973}
\begin{equation}
T_{\rm col}(x) = f_{\rm col} \left[ \frac{3 G M \dot{M}}{8 \pi \sigma_{\rm SB} R_{\rm g}^{3}} f\left( x \right) \right]^{1/4}, \label{eqn:Tcol_x}
\end{equation}
where the radial coordinate is parametrized as $x \equiv ( R / R_{\rm g} )^{1/2} = ( \tilde{r} R_{\rm in} / R_{\rm g} )^{1/2}$ and $f( x )$ describes the radial profile of gravitational energy released in the form of a radiation flux through the disc surfaces. The time-steady mass accretion rate $\dot{M}$ through the disc is taken to be radially independent and one generally assumes a zero-torque boundary condition on the inner disc edge. With these assumptions, the relativistic radial disc structure adopted by \texttt{kerrbb} gives $f( x )$ as \citep{NovikovThorne1973, PageThorne1974}
\begin{align}
f_{\rm NT}\left( x \right) &= \frac{1}{x^{4} \left( x^{3} - 3 x + 2 a \right)} \left[ x - x_{0} - \frac{3}{2} a \ln\left( \frac{x}{x_{0}} \right) - \frac{3 \left( x_{1} - a \right)^{2}}{x_{1} \left( x_{1} - x_{2} \right) \left( x_{1} - x_{3} \right)} \ln\left( \frac{x - x_{1}}{x_{0} - x_{1}} \right) \right. \nonumber \\
&\left.- \frac{3 \left( x_{2} - a \right)^{2}}{x_{2} \left( x_{2} - x_{1} \right) \left( x_{2} - x_{3} \right)} \ln\left( \frac{x - x_{2}}{x_{0} - x_{2}} \right) - \frac{3 \left( x_{3} - a \right)^{2}}{x_{3} \left( x_{3} - x_{1} \right) \left( x_{3} - x_{2} \right)} \ln\left( \frac{x - x_{3}}{x_{0} - x_{3}} \right) \right], \label{eqn:fx_NT}
\end{align}
where,
\begin{equation}
x_{0} \equiv \left( \frac{R_{\rm in}}{R_{\rm g}} \right)^{1/2}, \qquad
x_{1} \equiv 2 \cos\left( \frac{1}{3} \cos^{-1}\left( a \right) - \frac{\pi}{3} \right), \qquad
x_{2} \equiv 2 \cos\left( \frac{1}{3} \cos^{-1}\left( a \right) + \frac{\pi}{3} \right), \qquad
x_{3} \equiv -2 \cos\left( \frac{1}{3} \cos^{-1}\left( a \right) \right),
\end{equation}
and $a = J / ( R_{\rm g} M c )$ is the black hole spin parameter, where $J$ is the black hole angular momentum. The black hole spin is interchangeable with the inner disc radius under the standard assumption that $r_{\rm in}$ coincides with the location of the innermost stable circular orbit (ISCO), which depends on $a$ monotonically \citep{Bardeen1972},
\begin{equation}
r_{\rm ISCO} = \left\{
    \begin{array}{@{\hspace{0mm}}l@{\hspace{1mm}}l@{\hspace{0mm}}}
        3 + Z_{2} - \sqrt{\left( 3 - Z_{1} \right) \left( 3 + Z_{1} + 2 Z_{2} \right)} &,~\mathrm{for}~ 0 \le a \le +1 \\ \\
        3 + Z_{2} + \sqrt{\left( 3 - Z_{1} \right) \left( 3 + Z_{1} + 2 Z_{2} \right)} &,~\mathrm{for}~ -1 \le a < 0,
    \end{array}
    \right. \label{eqn:risco}
\end{equation}
where,
\begin{equation}
Z_{1} \equiv 1 + \left( 1 - a^{2} \right)^{1/3} \left[ \left( 1 + a \right)^{1/3} + \left( 1 - a\right)^{1/3} \right], \qquad
Z_{2} \equiv \sqrt{3 a^{2} + Z_{1}^{2}}. \label{eqn:Z1Z2}
\end{equation}

For a set of model parameters $\{ r_{\rm in}, f_{\rm col}, M, D, i_{\rm disc}, \dot{M}\}$, we can calculate $\widehat{F}_{\rm obs}$ from equation \eqref{eqn:Fobs_noGR}, given $T_{\rm col}( \tilde{r} )$ as determined by inserting an explicit expression for $f( x )$ in equation \eqref{eqn:Tcol_x}. Choosing the relativistic radial disc structure $f_{\rm NT}( x )$ from equation \eqref{eqn:fx_NT} defines the radial profile of the colour temperature $T_{\rm col}^{\rm NT}( x )$, which we insert into equation \eqref{eqn:Fobs_noGR} to obtain the corresponding total observed disc flux $\widehat{F}_{\rm obs}^{\rm NT}$. Because $\widehat{F}_{\rm obs}^{\rm NT}$ and $F_{\rm obs}$ have identical temperature profiles, they only differ in their treatment of relativistic effects on photon propagation, which $\widehat{F}_{\rm obs}^{\rm NT}$ ignores and $F_{\rm obs}$ includes. To isolate these effects on the total observed disc flux, we define the disc flux correction factor \citep[e.g.,][]{Cunningham1975, Zhang1997}
\begin{equation}
g_{\rm GR}\left( r_{\rm in}, i_{\rm disc} \right) = \frac{F_{\rm obs}}{\widehat{F}_{\rm obs}^{\rm NT}}, \label{eqn:gGR}
\end{equation}
which we tabulate in \S\ref{sec:gGR_gNT}. Having condensed the relativistic effects on the propagation of photons from the disc surface to the observer into the correction factor $g_{\rm GR}( r_{\rm in}, i_{\rm disc} )$, equation \eqref{eqn:Fobs} becomes
\begin{equation}
F_{\rm obs} = \frac{r_{\rm in}^{2}}{f_{\rm col}^{4}} \left( \frac{G M / c^{2}}{D} \right)^{2} \cos\left( i_{\rm disc} \right) \Upsilon\left( i_{\rm disc} \right) g_{\rm GR}\left( r_{\rm in}, i_{\rm disc} \right) \int\limits_{0}^{\infty} \frac{4 \pi E_{\rm obs}^{3}}{h^{3} c^{2}} \int\limits_{1}^{\tilde{r}_{\rm out}} \left[ \exp\left( \frac{E_{\rm obs}}{k_{\rm B} T_{\rm col}^{\rm NT}\left( \tilde{r} \right)}\right) -1 \right]^{-1} \tilde{r}~d\tilde{r}~dE_{\rm obs}, \label{eqn:Fobs_gGR}
\end{equation}
which is equivalent to $F_{\rm obs}$ calculated using \texttt{kerrbb}. Notably, we took advantage of the \texttt{kerrbb} ray-tracing calculations by pre-computing $g_{\rm GR}$, such that we can calculate $F_{\rm obs}$ from equation \eqref{eqn:Fobs_gGR} with straightforward integration.

Our next objective is to reduce equation \eqref{eqn:Fobs_gGR} to a two-parameter disc model described by only a disc flux normalization and a characteristic colour temperature, just like the \texttt{diskbb} family of disc models \citep{Mitsuda1984}. As equation \eqref{eqn:Fobs_gGR} currently stands, the factors outside the integrals could be rolled into a single ``normalization'' model parameter, but parametrizing the colour temperature profile is more complicated. Notably, in the non-relativistic limit for the radial disc structure, we have $f( x ) \rightarrow f_{\rm SS}( x ) = ( x - x_{0} ) / x^{7}$ and the corresponding colour temperature profile \citep{ShakuraSunyaev1973}
\begin{equation}
T_{\rm col}^{\rm SS}\left( \tilde{r} \right) = T_{\rm col}^{\ast} \tilde{r}^{-3/4} \left( 1 - \tilde{r}^{-1/2} \right)^{1/4} \label{eqn:Tcol_SS}
\end{equation}
is completely specified by the characteristic colour temperature \citep[e.g.,][]{FKR2002}
\begin{equation}
T_{\rm col}^{\ast} \equiv f_{\rm col} \left( \frac{3 G M \dot{M}}{8 \pi \sigma_{\rm SB} R_{\rm in}^{3}} \right)^{1/4}, \label{eqn:Tcol_star}
\end{equation}
which allows equation \eqref{eqn:Fobs_gGR} to be reduced to a two-parameter model. However, unlike $T_{\rm col}^{\rm SS}( \tilde{r} )$, the colour temperature profile $T_{\rm col}^{\rm NT}( x )$ associated with the relativistic radial disc structure used by \texttt{kerrbb} cannot be parametrized in terms of only $T_{\rm col}^{\ast}$, but also requires specifying $r_{\rm in}$ because this sets the black hole spin $a$ (assuming $r_{\rm in} = r_{\rm ISCO}$) as needed by $f_{\rm NT}( x )$. Analogously to our calculation of $g_{\rm GR}$ above, we can isolate the effects of a relativistic radial disc structure by defining another disc flux correction factor (tabulated in \S\ref{sec:gGR_gNT})
\begin{equation}
g_{\rm NT}\left( r_{\rm in} \right) = \frac{\widehat{F}_{\rm obs}^{\rm NT}}{\widehat{F}_{\rm obs}^{\rm SS}}, \label{eqn:gNT}
\end{equation}
where we calculate $\widehat{F}_{\rm obs}^{\rm SS}$ using equation \eqref{eqn:Fobs_noGR}, with $T_{\rm col}^{\rm SS}( \tilde{r} )$ from equation \eqref{eqn:Tcol_SS} inserted for $T_{\rm col}( \tilde{r} )$. Condensing the relativistic effects on the radial profile of the colour temperature into the correction factor $g_{\rm NT}( r_{\rm in} )$, we can write equation \eqref{eqn:Fobs_gGR} as a two-parameter --- $K_{\rm flux}$ and $T_{\rm col}^{\ast}$ --- model for the total observed disc flux
\begin{equation}
F_{\rm obs} = K_{\rm flux} \int\limits_{0}^{\infty} \frac{4 \pi E_{\rm obs}^{3}}{h^{3} c^{2}} \int\limits_{1}^{\tilde{r}_{\rm out}} \left[ \exp\left( \frac{E_{\rm obs}}{k_{\rm B} T_{\rm col}^{\ast} \tilde{r}^{-3/4} \left( 1 - \tilde{r}^{-1/2} \right)^{1/4}}\right) -1 \right]^{-1} \tilde{r}~d\tilde{r}~dE_{\rm obs}, \label{eqn:Fobs_gGR_gNT}
\end{equation}
where we combined $\{ r_{\rm in}, f_{\rm col}, M, D, i_{\rm disc} \}$ into a single model parameter called the ``disc flux normalization''
\begin{equation}
K_{\rm flux} \equiv \frac{r_{\rm in}^{2}}{f_{\rm col}^{4}} \left( \frac{G M / c^{2}}{D} \right)^{2} \cos\left( i_{\rm disc} \right) \Upsilon\left( i_{\rm disc} \right) g_{\rm GR}\left( r_{\rm in}, i_{\rm disc} \right) g_{\rm NT}\left( r_{\rm in} \right). \label{eqn:Kflux}
\end{equation}
For the same set of input parameters $\{ r_{\rm in}, f_{\rm col}, M, D, i_{\rm disc}, \dot{M}\}$, the total observed disc flux $F_{\rm obs}$ calculated from equation \eqref{eqn:Fobs_gGR_gNT} is identical to that calculated from the \texttt{kerrbb} model.

\begin{figure}
    \begin{center}
        \includegraphics[width=0.495\textwidth]{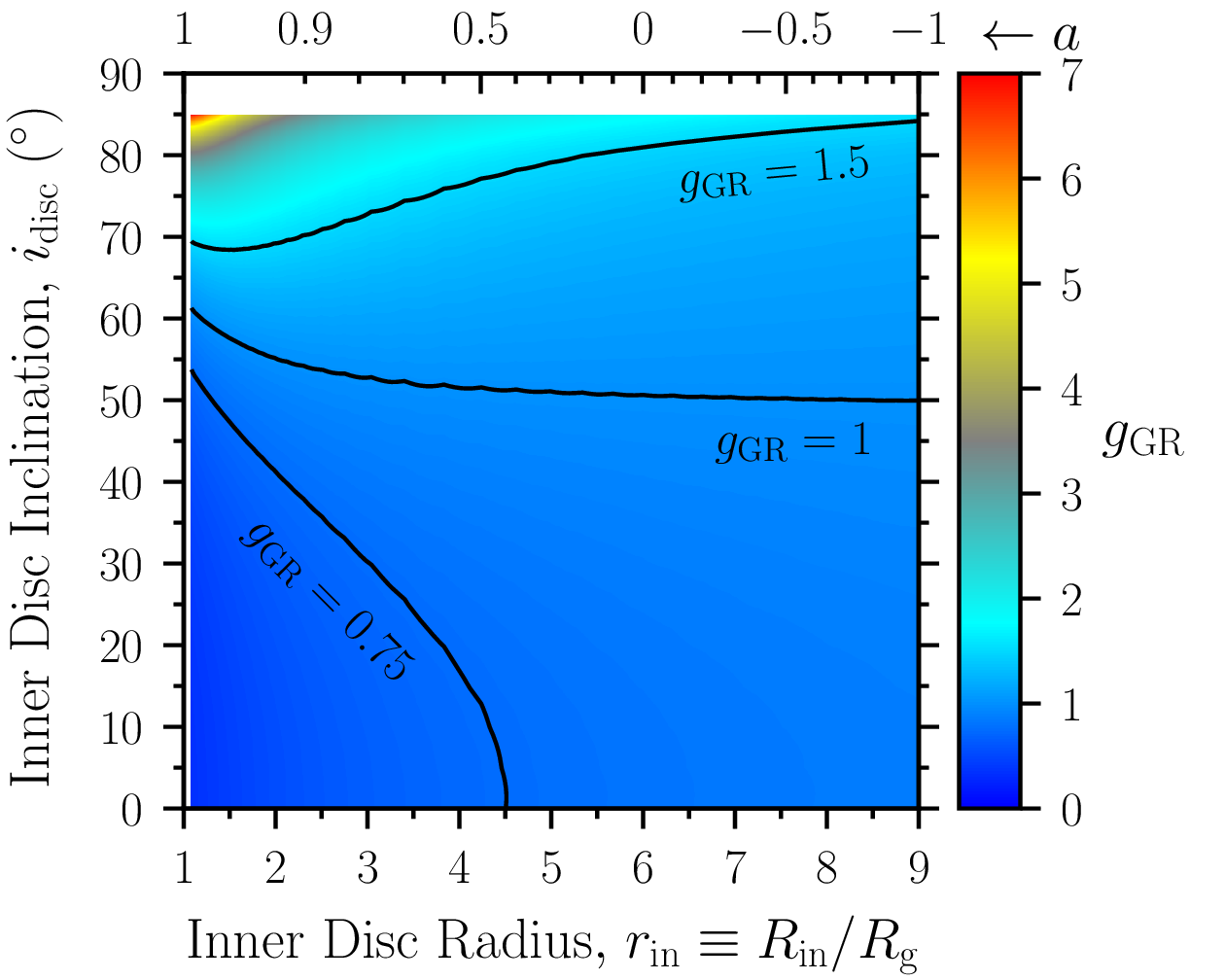}
        \hfill
        \includegraphics[width=0.495\textwidth]{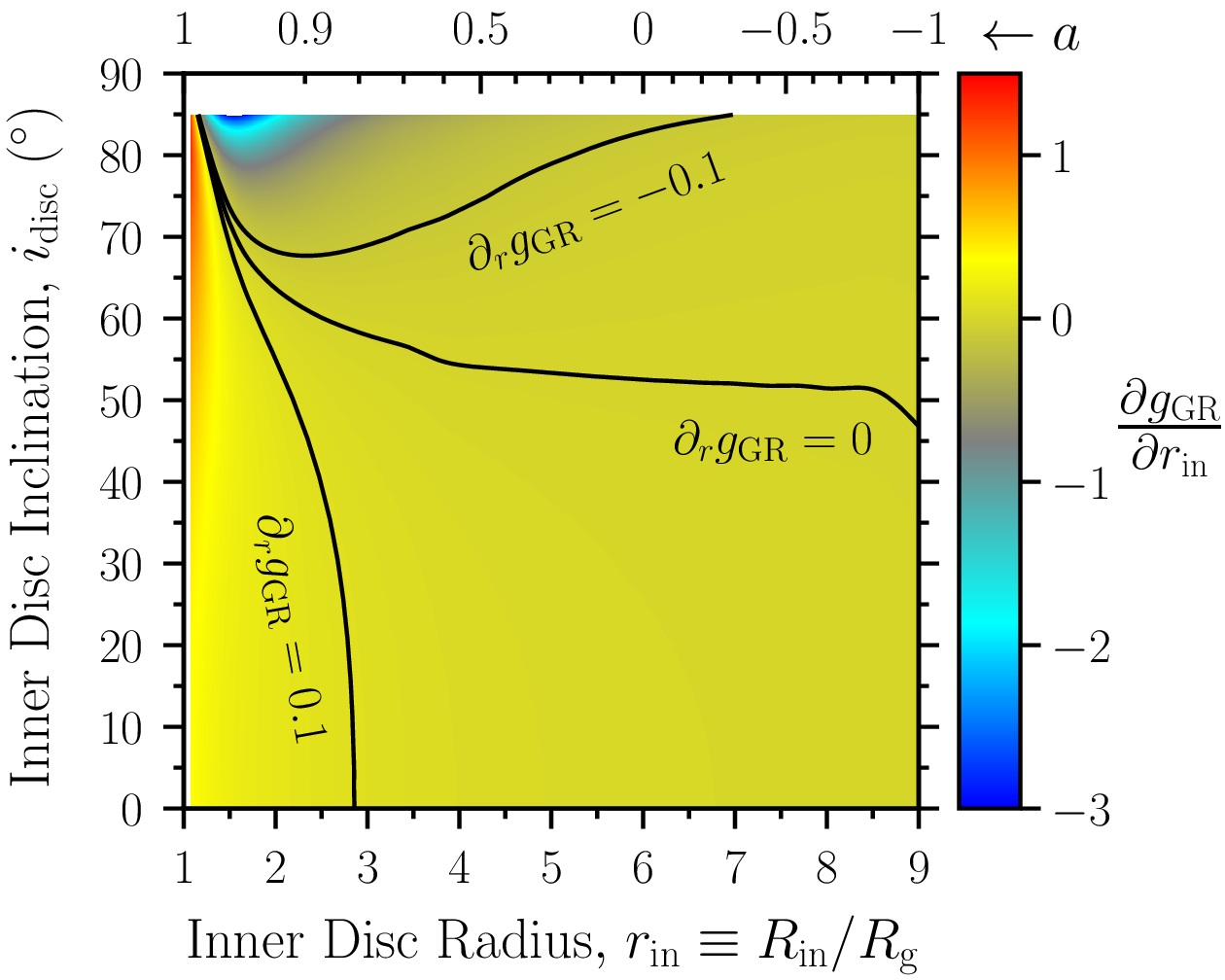}
     \end{center}
    \vspace{-2mm}
    \caption{\textit{Left}: Disc flux correction factor $g_{\rm GR}( r_{\rm in}, i_{\rm disc} )$, which accounts for relativistic effects on photon propagation, on a grid of inner disc radius $r_{\rm in}$ and inner disc inclination angle $i_{\rm disc}$. The \textit{top} $x$-axis gives the black hole spin parameter $a$ corresponding to $r_{\rm in}$, assuming $r_{\rm in} = r_{\rm ISCO}$. Significant departures of $g_{\rm GR}$ away from unity occur for moderate-to-extreme prograde black hole spins and/or more edge-on disc inclinations. For a non-spinning black hole, $g_{\rm GR} = 1$ when $i_{\rm disc} = 50\fdg7$. The $g_{\rm GR}$ extrema are $[ 0.33, 6.9 ]$. {\it White regions} show the parameter space inaccessible to \texttt{kerrbb}. \textit{Right}: Inner disc radius partial derivative of the disc flux correction factor $\partial g_{\rm GR} / \partial r_{\rm in}$, which is needed to compute the Jacobian determinant of equation \eqref{eqn:J_Kflux2rin}. The $g_{\rm GR}$ contour lines show artifacts (resembling ocean surface waves) that get amplified to large oscillations in $\partial g_{\rm GR} / \partial r_{\rm in}$, so we smoothed $\partial g_{\rm GR} / \partial r_{\rm in}$ along the $r_{\rm in}$-dimension using a Gaussian filter, whose kernel required a standard deviation of $25~R_{\rm g}$ to yield the smooth, non-oscillatory contour lines shown in the plot. After smoothing, the $\partial g_{\rm GR} / \partial r_{\rm in}$ extrema are $[ -3.1, 1.3 ]$.}
    \label{fig:gGR}
\end{figure}

\begin{figure}
    \begin{center}
        \includegraphics[width=0.495\textwidth]{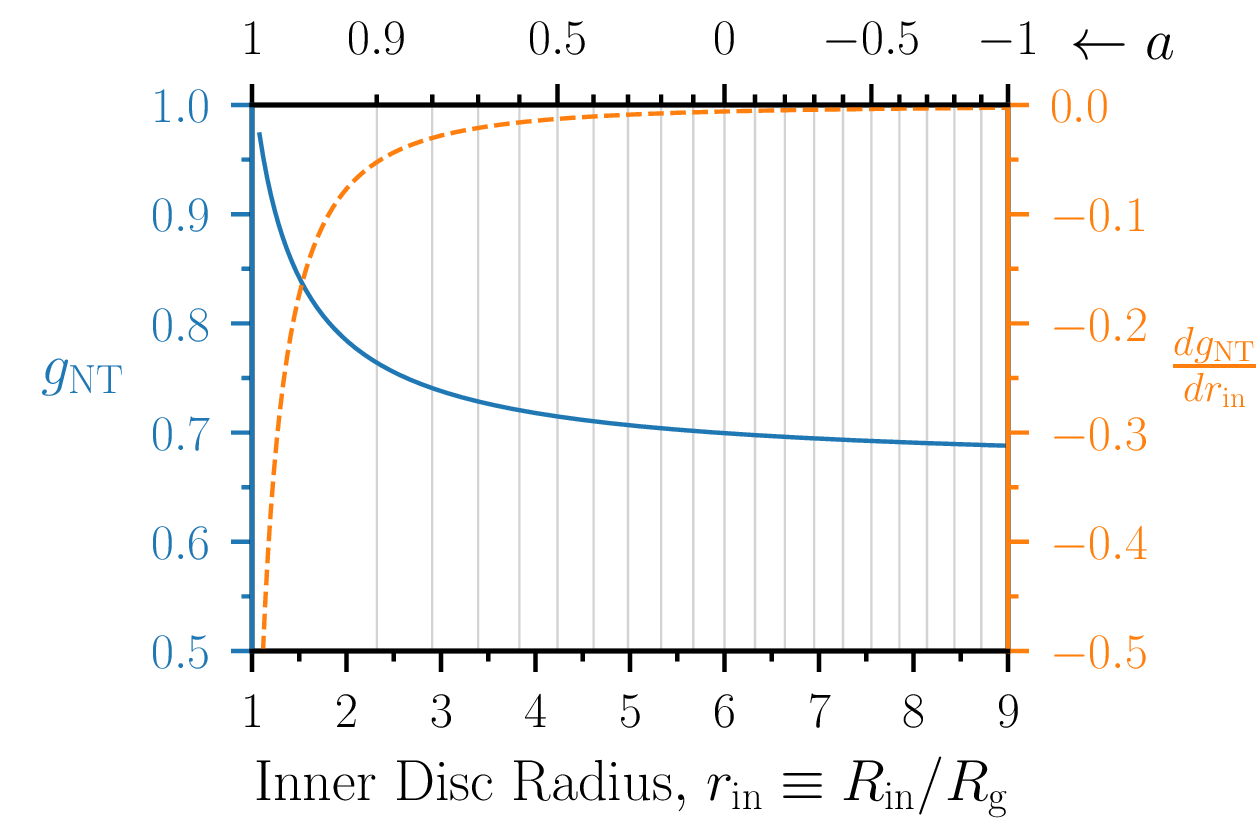}
    \end{center}
    \vspace{-2mm}
    \caption{Disc flux correction factor $g_{\rm NT}(r_{\rm in})$ (\textit{left axis}; \textit{solid blue line}), which accounts for relativistic effects on the radial disc structure, and its derivative $d g_{\rm NT} / d r_{\rm in}$ (\textit{right axis}; \textit{dashed orange line}), which is needed to compute the Jacobian determinant of equation \eqref{eqn:J_Kflux2rin}. The \textit{top} $x$-axis gives the black hole spin parameter $a$ corresponding to $r_{\rm in}$, assuming $r_{\rm in} = r_{\rm ISCO}$.}
    \label{fig:gNT}
\end{figure}

\subsection{Relativistic Disc Flux Correction Factors $g_{\rm GR}( r_{\rm in}, i_{\rm disc} )$ and $g_{\rm NT}( r_{\rm in} )$}
\label{sec:gGR_gNT}
The disc flux correction factor $g_{\rm GR}( r_{\rm in}, i_{\rm disc} ) = F_{\rm obs} / \widehat{F}_{\rm obs}^{\rm NT}$ is the ratio of the total observed disc flux including relativistic effects on photon propagation to that without such effects. We calculate $\widehat{F}_{\rm obs}^{\rm NT}$ using equation \eqref{eqn:Fobs_noGR} with $T_{\rm col}( \tilde{r} ) \rightarrow T_{\rm col}^{\rm NT}( \tilde{r} )$, which follows from inserting equation \eqref{eqn:Tcol_x} into \eqref{eqn:fx_NT}. The integral over $\tilde{r}$ becomes numerically tractable by assuming a large outer disc radius $R_{\rm out} = 10^{6} R_{\rm g}$ (the same as \texttt{kerrbb}) and changing variables $\tilde{r} \rightarrow \tilde{\mathscr{R}} \equiv \log_{10}( \tilde{r} )$ to accommodate the large range in $\tilde{r}$ spanned by the integration bounds. We evaluate this integral using 600 bins linearly-spaced in $\tilde{\mathscr{R}}$. We numerically evaluate the $E_{\rm obs}$ integral over the range $\left[ E_{\rm obs}^{\rm min}, E_{\rm obs}^{\rm max} \right] = \left[ 0.001, 100 \right]~{\rm keV}$ with 500 logarithmically-spaced bins. To get $F_{\rm obs}$, we first use \texttt{kerrbb} to calculate the specific observed disc flux $F_{{E}_{\rm obs}}$ with the flags for limb-darkening and self-irradiation turned on and the inner disc torque set to zero, as is always done by the continuum fitting community. Then we integrate $F_{{E}_{\rm obs}}$ over $E_{\rm obs}$ using the \texttt{flux} command in \texttt{XSPEC} \citep[version 12.10.1f;][]{Arnaud1996} and the same $E_{\rm obs}$ grid used to calculate $\widehat{F}_{\rm obs}^{\rm NT}$.

The disc flux correction factor $g_{\rm NT}( r_{\rm in} ) = \widehat{F}_{\rm obs}^{\rm NT} / \widehat{F}_{\rm obs}^{\rm SS}$ is the ratio of the total observed disc flux including relativistic effects on the radial temperature profile to that without such effects (both fluxes ignore relativistic effects on photon propagation). We calculate $\widehat{F}_{\rm obs}^{\rm NT}$ as described in the preceding paragraph and $\widehat{F}_{\rm obs}^{\rm SS}$ similarly follows from choosing $T_{\rm col}( \tilde{r} ) \rightarrow T_{\rm col}^{\rm SS}( \tilde{r} )$, as defined in equations \eqref{eqn:Tcol_SS}--\eqref{eqn:Tcol_star}.

For each of the eight black hole X-ray binaries in Table \ref{tab:obs} with non-maximal spin measurements, we compute grids of $g_{\rm GR}( r_{\rm in}, i_{\rm disc} )$ and $g_{\rm NT}( r_{\rm in} )$ over the parameter ranges $[ r_{\rm in}^{\rm min}, r_{\rm in}^{\rm max} ] = [ 1.08, 9 ]$ and $[ i_{\rm disc}^{\rm min}, i_{\rm disc}^{\rm max} ] = [ 0^{\circ}, 85^{\circ} ]$ in increments of $\Delta r_{\rm in} = 0.01$ and $\Delta i_{\rm disc} = 0\fdg1$. Adopting the usual $r_{\rm in} = r_{\rm ISCO}$ assumption makes $r_{\rm in}$ synonymous with the black hole spin $a$, which \texttt{kerrbb} takes as an input. Starting with LMC X--1 in Table \ref{tab:obs} and considering each source from \textit{top} to \textit{bottom}, we adopt the listed input parameters $\{ f_{\rm col}, M, D \}$ and choose an $\dot{M}$ based on the quoted values in the $a^{\rm CF}$ reference paper:\footnote{The Eddington mass accretion rate is $\dot{M}_{\rm Edd} = L_{\rm Edd} / ( \eta c^{2} )$, where the Eddington luminosity is $L_{\rm Edd} = 4 \pi G M m_{\rm p} c / \sigma_{\rm T}$ and the radiative efficiency is $\eta = 1 - [ 1 - 2 R_{\rm g} / ( 3 R_{\rm ISCO} ) ]^{1/2}$.} $\dot{m} \equiv \dot{M} / \dot{M}_{\rm Edd} = 0.16$, 0.1, 0.095, 0.12, 0.082, 0.2, 0.2, 0.11. For the representative system GRO J1655--40, Figure \ref{fig:gGR} shows $g_{\rm GR}( r_{\rm in}, i_{\rm disc} )$ and its $r_{\rm in}$ partial derivative, while Figure \ref{fig:gNT} shows $g_{\rm NT}( r_{\rm in} )$ and its derivative. Although we tailor the $g_{\rm GR}$ and $g_{\rm NT}$ grids to each individual system by using the appropriate input parameters $\{ f_{\rm col}, M, D, \dot{M} \}$ for that system, it turns out these customized grids are practically identical to each other, confirming that $g_{\rm GR} = g_{\rm GR}( r_{\rm in}, i_{\rm disc} )$ and $g_{\rm NT} = g_{\rm NT}( r_{\rm in} )$ have no appreciable $\{ f_{\rm col}, M, D, \dot{M} \}$-dependence.

\section{Methodology for Revising Black Hole Spin Measurements}
\label{app:meth_PDFs}
Here, we present the step-by-step methodology for incorporating uncertainties in the colour correction factor $f_{\rm col}$ into measurements of the black hole spin parameter $a$. In summary, the published disc continuum fitting literature provides a measured marginal density $f_{a}^{\rm CF}( a )$ for the black hole spin, which effectively does not include $f_{\rm col}$ uncertainties. We transform this $f_{a}^{\rm CF}( a )$ into a marginal density $f_{\circ}^{\rm CF}( r_{\rm ISCO} )$ for the innermost stable circular orbit (ISCO) by a change of variables (see \S\ref{app:a2risco}) and make the standard assumption that $f_{\circ}^{\rm CF}( r_{\rm ISCO} ) = f_{r}^{\rm CF}( r_{\rm in} )$, the marginal density for the inner disc radius. Next, we use $f_{r}^{\rm CF}( r_{\rm in} )$ to reverse-engineer the marginal density $f_{K}( K_{\rm flux} )$ for the disc flux normalization associated with the published black hole spin measurement. We accomplish this by taking advantage of the variables $\{ K_{\rm flux}, f_{\rm col}, M, D, i_{\rm disc} \}$ being independent (see \S\ref{app:rin2Kflux}), then inverting an integral transform equation using properties of orthonormal basis functions (see \S\ref{app:basisfunc}) and imposing constraints (see \S\ref{app:constopt}). With the original continuum fitting observable $f_{K}( K_{\rm flux} )$ in hand, we measure a revised $f_{r}( r_{\rm in} )$ --- this time accounting for $f_{\rm col}$ uncertainty --- by adopting an equivalent approach to the modern continuum fitting method (see \S\ref{app:Kflux2rin}). To come full circle, we assume this updated $f_{r}( r_{\rm in} )$ is interchangeable with $f_{\circ}( r_{\rm ISCO} )$, which we transform to a revised black hole spin measurement $f_{a}( a )$.

We use the superscript ``CF'' to uniquely associate a probability density or a variable with the disc continuum fitting literature. For example, $f_{r}^{\rm CF}( r_{\rm in} )$ is the measured marginal density for the inner disc radius reported in a continuum fitting paper and derived assuming the marginal density for the colour correction factor $f_{f}^{\rm CF}( f_{\rm col} ) = \delta( f_{\rm col} - f_{\rm col}^{\rm CF} )$ is a Dirac delta function.\footnote{Understanding that $f_{f}^{\rm CF}( f_{\rm col} ) = \delta( f_{\rm col} - f_{\rm col}^{\rm CF} )$, the marginal density $f_{r}^{\rm CF}( r_{\rm in} )$ is equivalent to the conditional density $f_{r | f}^{\rm CF}( r_{\rm in} | f_{\rm col}^{\rm CF} )$.} In what follows, we drive toward calculating $f_{r}( r_{\rm in} )$, which revises $f_{r}^{\rm CF}( r_{\rm in} )$ by marginalizing over $f_{\rm col}$. We omit the ``CF'' superscript for probability densities $f_{K}( K_{\rm flux} )$ and $f_{M, D, i}( M, D, i_{\rm disc} )$, which are the same between the continuum fitting literature and our treatment.

Let us take a brief detour here to explain some notation. Standard convention for a probability density $f_{X}(x)$ is to adopt an uppercase letter (e.g., $X$) for the random variable associated with the density (denoted as a subscript) and its corresponding lowercase form (e.g., $x$) to specify the individual point to be evaluated (denoted as an argument). In our convention, probability densities have subscripts and arguments with these same meanings, but we forego the upper/lower case convention because it is not compatible with the symbols we use for variables (i.e., $r_{\rm in}$, $K_{\rm flux}$, $f_{\rm col}$, $M$, $D$, $i_{\rm disc}$). Here are examples of several probability densities the reader might encounter below, using the random variables $x$ and $y$: The marginal density of $x$ evaluated at $x = x$ is $f_{x}(x)$; the marginal density of $x$ evaluated at $x = y$ is $f_{x}(y)$; the conditional density of $x$ given $y$ evaluated at $x = x$ and $y = y$ is $f_{x | y}{( x | y )}$; and the joint density of $x$ and $y$ evaluated at $x = x$ and $y = y$ is $f_{x, y}( x, y )$.

\subsection{Change of Variables Method: $f_{a}\left( a \right) \leftrightarrow f_{\circ}\left( r_{\rm ISCO} \right)$}
\label{app:a2risco}
The derived measurement from disc continuum fitting is the inner disc radius marginal density $f_{r}( r_{\rm in} )$. This then gets converted to the black hole spin parameter marginal density $f_{a}( a )$ under the assumption that $f_{r}( r_{\rm in} ) = f_{\circ}( r_{\rm ISCO} )$, which is the marginal density for the innermost stable circular orbit (ISCO). The transformation from $f_{\circ}( r_{\rm ISCO})$ to $f_{a}( a )$ follows from the change of variables
\begin{equation}
f_{a}\left( a \right) = f_{\circ}\left( h_{\circ}^{-1}\left( a \right) \right) \left| J\left( a \right) \right|, \label{eqn:f_a}
\end{equation}
where $r_{\rm ISCO} = h_{\circ}^{-1}( a )$ is the inverse transformation function, given by equation \eqref{eqn:risco}, and $J( a )$ is the Jacobian of the transformation from $r_{\rm ISCO}$ to $a$, given by
\begin{equation}
J\left( a \right) = \frac{d r_{\rm ISCO}}{d a}
= \left\{
    \begin{array}{@{\hspace{0mm}}l@{\hspace{1mm}}l@{\hspace{0mm}}}
        \dfrac{d Z_{2}}{d a} - \dfrac{\left( 3 - Z_{1} \right) \left( d Z_{2} / d a \right) - \left( Z_{1} + Z_{2} \right) \left( d Z_{1} / d a \right)}{\sqrt{\left( 3 - Z_{1} \right) \left( 3 + Z_{1} + 2 Z_{2} \right)}} &,~\mathrm{for}~0 \le a \le 1 \\ \\
        \dfrac{d Z_{2}}{d a} + \dfrac{\left( 3 - Z_{1} \right) \left( d Z_{2} / d a \right) - \left( Z_{1} + Z_{2} \right) \left( d Z_{1} / d a \right)}{\sqrt{\left( 3 - Z_{1} \right) \left( 3 + Z_{1} + 2 Z_{2} \right)}} &,~\mathrm{for}~-1 \le a < 0,
    \end{array}
    \right. \label{eqn:Jrisco}
\end{equation}
where from differentiating equation \eqref{eqn:Z1Z2},
\begin{align}
\frac{d Z_{1}}{d a} &= \frac{1}{3} \left[ \left( 1 + a \right)^{-2/3} - \left( 1 - a \right)^{-2/3} \right] \left( 1- a^{2} \right)^{1/3} - \frac{2}{3} a \left[ \left( 1 + a \right)^{1/3} + \left( 1 - a \right)^{1/3} \right] \left( 1 - a^{2} \right)^{-2/3} \\
\frac{d Z_{2}}{d a} &= \frac{3 a + Z_{1} \left( d Z_{1} / d a \right)}{\sqrt{3 a^{2} + Z_{1}^{2}}}.
\end{align}
By using equation \eqref{eqn:f_a} with the Jacobian $J( a )$ from equation \eqref{eqn:Jrisco}, we can transform from $f_{\circ}( r_{\rm ISCO} )$ to $f_{a}( a )$.

Our initial task is to obtain the $f_{r}^{\rm CF}( r_{\rm in} )$ measurement from the disc continuum fitting literature. In some cases, this is provided upfront, while in other cases the published distribution is $f_{a}^{\rm CF}( a )$ and we must perform the reverse process described above to convert $f_{a}^{\rm CF}( a )$ to $f_{\circ}^{\rm CF}( r_{\rm ISCO} )$ under the transformation
\begin{equation}
f_{\circ}\left( r_{\rm ISCO} \right) = f_{a}\left( h_{a}^{-1}\left( r_{\rm ISCO} \right) \right) \left| J\left( r_{\rm ISCO} \right) \right|.\label{eqn:f_risco}
\end{equation}
By numerically solving equation \eqref{eqn:risco} for $a$, we obtain the inverse transformation function $a = h_{a}^{-1}( r_{\rm ISCO} )$ and numerically differentiating yields the Jacobian $J( r_{\rm ISCO} ) = d a / d r_{\rm ISCO}$ of the transformation from $a$ to $r_{\rm ISCO}$. Making the usual, but crucial, assumption that $f_{\circ}^{\rm CF}( r_{\rm ISCO} ) = f_{r}^{\rm CF}( r_{\rm in} )$, we can proceed to \S \ref{app:rin2Kflux}.

\subsection{Multi-Variate Change of Variables Method: $f_{r}^{\rm CF}( r_{\rm in} ) \rightarrow f_{K}( K_{\rm flux} )$}
\label{app:rin2Kflux}
The ultimate goal of Appendix \ref{app:meth_PDFs} is to obtain the marginal density $f_{r}( r_{\rm in} )$ for the \textit{dependent} variable $r_{\rm in}$, which is the measurement derived from our re-parametrized version of the disc continuum fitting technique (see \S\ref{sec:meth}), given the marginal density $f_{K}( K_{\rm flux} )$ for the \textit{independent} variable $K_{\rm flux}$, which is the actual observable. In this section \ref{app:rin2Kflux}, we accomplish the intermediate goal of reverse-engineering the published measurement $f_{r}^{\rm CF}( r_{\rm in} )$ to obtain $f_{K}( K_{\rm flux} )$. We are also given the independent variable\footnote{There is speculation that $f_{\rm col}$ could have an $i_{\rm disc}$ dependence \citep[e.g.,][]{MillerMiller2015}, so we look to existing calculations for guidance \citep{Davis2005}. From fitting \texttt{kerrbb} to \texttt{bhspec} models with a fixed choice for the Eddington-scaled disc luminosity spanning $l =0.03$--0.3, we find $f_{\rm col}$ only increases by as much as $\sim0.1$ when the disc goes from being face-on to edge-on. There is a similarly weak trend of $f_{\rm col}$ to increase with $a$ by as much as $\sim0.1$ when the black hole goes from non-spinning to near-maximal. These weak dependencies justify treating $f_{\rm col}$ is an independent parameter over the narrow ranges of $i_{\rm disc}$ and $a$ reported for any given source in Table \ref{tab:obs}. \label{foot:fcol}} $f_{\rm col}$ with marginal density $f_{f}( f_{\rm col} )$ and the set of independent variables $\{ M, D, i_{\rm disc} \}$ with joint density $f_{M, D, i}( M, D, i_{\rm disc} )$. In cases where the observables $M$, $D$, $i_{\rm disc}$ are obtained independently, their joint density can be replaced with the product of their marginals.

Given the set of variables $\{ K_{\rm flux}, f_{\rm col}, M, D, i_{\rm disc} \}$ with joint density $f_{K, f, M, D, i}( K_{\rm flux}, f_{\rm col}, M, D, i_{\rm disc} )$, we can construct the joint density for the transformed set of variables $\{ r_{\rm in}, f_{\rm col}, M, D, i_{\rm disc} \}$ as
\begin{equation}
f_{r, f, M, D, i}\left( r_{\rm in}, f_{\rm col}, M, D, i_{\rm disc} \right) = f_{K, f, M, D, i}\left( h^{-1}\left( r_{\rm in}, f_{\rm col}, M, D, i_{\rm disc} \right), f_{\rm col}, M, D, i_{\rm disc} \right) \left| J\left( r_{\rm in}, f_{\rm col}, M, D, i_{\rm disc} \right) \right|, \label{eqn:mvcov}
\end{equation}
where $J( r_{\rm in}, f_{\rm col}, M, D, i_{\rm disc} )$ is the Jacobian of the $K_{\rm flux} \rightarrow r_{\rm in}$ transformation and the inverse transformation function is (see equation \ref{eqn:Kflux})
\begin{equation}
K_{\rm flux} = h^{-1}\left( r_{\rm in}, f_{\rm col}, M, D, i_{\rm disc} \right) = \frac{r_{\rm in}^{2}}{f_{\rm col}^{4}} \left( \frac{G M / c^{2}}{D} \right)^{2} \cos\left( i_{\rm disc} \right) \Upsilon\left( i_{\rm disc} \right) g_{\rm GR}\left( r_{\rm in}, i_{\rm disc} \right) g_{\rm NT}\left( r_{\rm in} \right). \label{eqn:Kflux_inv}
\end{equation}
Because the variables $\{ K_{\rm flux}, f_{\rm col}, ( M, D, i_{\rm disc} ) \}$ are independent, their joint density is equivalent to the product of their marginals,\footnote{Because $r_{\rm in}$ is a derived variable, \textit{not} an independent variable, its joint density with other variables cannot be decomposed into the product of their marginals; that is, $f_{r, f, M, D, i}( r_{\rm in}, f_{\rm col}, M, D,  i_{\rm disc} ) \ne f_{r}( r_{\rm in} ) f_{f}( f_{\rm col} ) f_{M, D, i}( M, D, i_{\rm disc} )$. Therefore, we cannot directly calculate the marginal density $f_{K}( K_{\rm flux} )$ using a multi-variate change of variables approach analogous to that used here to isolate $f_{r}( r_{\rm in} )$.}
\begin{equation}
f_{K, f, M, D, i}\left( h^{-1}\left( r_{\rm in}, f_{\rm col}, M, D, i_{\rm disc} \right), f_{\rm col}, M, D, i_{\rm disc} \right) = f_{K}\left( h^{-1}\left( r_{\rm in}, f_{\rm col}, M, D, i_{\rm disc} \right) \right) f_{f}\left( f_{\rm col} \right) f_{M, D, i}\left( M, D, i_{\rm disc} \right). \label{eqn:joint2marg}
\end{equation}
Substituting equation \eqref{eqn:joint2marg} into \eqref{eqn:mvcov} and marginalizing over $f_{\rm col}$, $M$, $D$, $i_{\rm disc}$, we arrive at the integral transform
\begin{equation}
f_{r}\left( r_{\rm in} \right) = \iiint \limits_{R_{\mathbf{x}}} \int_{f_{\rm col} = f_{\rm col}^{\rm min}}^{f_{\rm col}^{\rm max}} f_{K}\left( h^{-1}\left( r_{\rm in}, f_{\rm col}, \mathbf{x} \right) \right) f_{f}\left( f_{\rm col} \right) f_{\mathbf{x}}\left( \mathbf{x} \right) \left| J\left( r_{\rm in}, f_{\rm col}, \mathbf{x} \right) \right| d f_{\rm col}~d \mathbf{x}, \label{eqn:f_rin}
\end{equation}
where we adopt the shorthand notation
\begin{align}
\iiint \limits_{R_{\mathbf{x}}} &\equiv \int^{M^{\rm max}}_{M = M^{\rm min}} \int^{D^{\rm max}}_{D = D^{\rm min}} \int^{i_{\rm disc}^{\rm max}}_{i_{\rm disc} = i_{\rm disc}^{\rm min}} \\
f_{\mathbf{x}}\left( \mathbf{x} \right) &\equiv f_{M, D, i}\left( M, D, i_{\rm disc} \right) \\
d \mathbf{x} &\equiv d M~d D~d i_{\rm disc}.
\end{align}
The integration bounds span the variable domains $f_{\rm col} \in [ f_{\rm col}^{\rm min}, f_{\rm col}^{\rm max} ]$, $M \in [ M^{\rm min}, M^{\rm max} ]$, $D \in [ D^{\rm min}, D^{\rm max} ]$, and $i_{\rm disc} \in [ i_{\rm disc}^{\rm min}, i_{\rm disc}^{\rm max} ]$.\footnote{In practice, we truncate the domain of a variable to enclose the inter-99.994\% (i.e., $\pm 4\sigma$) region of its marginal density. This truncation is inconsequential because our focus is to characterize the bulk of a distribution, not its extremities of negligibly small probability density. If necessary, we enforce $f_{\rm col}^{\rm min} = 1$ and truncate $f_{f}( f_{\rm col} )$ appropriately due to values of $f_{\rm col} < 1$ being unphysical.} The Jacobian of the $K_{\rm flux} \rightarrow r_{\rm in}$ transformation follows from the determinant of the matrix of partial derivatives,
\begin{align}
J &= {\rm det} \left[ \frac{\partial \left( K_{\rm flux}, f_{\rm col}, M, D, i_{\rm disc} \right)}{\partial \left( r_{\rm in}, f_{\rm col}, M, D, i_{\rm disc} \right)} \right] \nonumber \\
&= {\rm det} \begin{bmatrix}
\left. \frac{\partial K_{\rm flux}}{\partial r_{\rm in}} \right|_{f_{\rm col}, M, D, i_{\rm disc}} & \left. \frac{\partial K_{\rm flux}}{\partial f_{\rm col}} \right|_{r_{\rm in}, M, D, i_{\rm disc}} & \left. \frac{\partial K_{\rm flux}}{\partial M} \right|_{r_{\rm in}, f_{\rm col}, D, i_{\rm disc}} & \left. \frac{\partial K_{\rm flux}}{\partial D} \right|_{r_{\rm in}, f_{\rm col}, M, i_{\rm disc}} & \left. \frac{\partial K_{\rm flux}}{\partial i_{\rm disc}} \right|_{r_{\rm in}, f_{\rm col}, M, D} \smallskip \\
\left. \frac{\partial f_{\rm col}}{\partial r_{\rm in}} \right|_{f_{\rm col}, M, D, i_{\rm disc}} & \left. \frac{\partial f_{\rm col}}{\partial f_{\rm col}} \right|_{r_{\rm in}, M, D, i_{\rm disc}} & \left. \frac{\partial f_{\rm col}}{\partial M} \right|_{r_{\rm in}, f_{\rm col}, D, i_{\rm disc}} & \left. \frac{\partial f_{\rm col}}{\partial D} \right|_{r_{\rm in}, f_{\rm col}, M, i_{\rm disc}} & \left. \frac{\partial f_{\rm col}}{\partial i_{\rm disc}} \right|_{r_{\rm in}, f_{\rm col}, M, D} \smallskip \\
\left. \frac{\partial M}{\partial r_{\rm in}} \right|_{f_{\rm col}, M, D, i_{\rm disc}} & \left. \frac{\partial M}{\partial f_{\rm col}} \right|_{r_{\rm in}, M, D, i_{\rm disc}} & \left. \frac{\partial M}{\partial M} \right|_{r_{\rm in}, f_{\rm col}, D, i_{\rm disc}} & \left. \frac{\partial M}{\partial D} \right|_{r_{\rm in}, f_{\rm col}, M, i_{\rm disc}} & \left. \frac{\partial M}{\partial i_{\rm disc}} \right|_{r_{\rm in}, f_{\rm col}, M, D} \smallskip \\
\left. \frac{\partial D}{\partial r_{\rm in}} \right|_{f_{\rm col}, M, D, i_{\rm disc}} & \left. \frac{\partial D}{\partial f_{\rm col}} \right|_{r_{\rm in}, M, D, i_{\rm disc}} & \left. \frac{\partial D}{\partial M} \right|_{r_{\rm in}, f_{\rm col}, D, i_{\rm disc}} & \left. \frac{\partial D}{\partial D} \right|_{r_{\rm in}, f_{\rm col}, M, i_{\rm disc}} & \left. \frac{\partial D}{\partial i_{\rm disc}} \right|_{r_{\rm in}, f_{\rm col}, M, D} \smallskip \\
\left. \frac{\partial i_{\rm disc}}{\partial r_{\rm in}} \right|_{f_{\rm col}, M, D, i_{\rm disc}} & \left. \frac{\partial i_{\rm disc}}{\partial f_{\rm col}} \right|_{r_{\rm in}, M, D, i_{\rm disc}} & \left. \frac{\partial i_{\rm disc}}{\partial M} \right|_{r_{\rm in}, f_{\rm col}, D, i_{\rm disc}} & \left. \frac{\partial i_{\rm disc}}{\partial D} \right|_{r_{\rm in}, f_{\rm col}, M, i_{\rm disc}} & \left. \frac{\partial i_{\rm disc}}{\partial i_{\rm disc}} \right|_{r_{\rm in}, f_{\rm col}, M, D}
\end{bmatrix} \nonumber \\
&= {\rm det} \begin{bmatrix}
\left. \frac{\partial K_{\rm flux}}{\partial r_{\rm in}} \right|_{f_{\rm col}, M, D, i_{\rm disc}} & 0 & 0 & 0 & 0 \\
0 & 1 & 0 & 0 & 0 \\
0 & 0 & 1 & 0 & 0 \\
0 & 0 & 0 & 1 & 0 \\
0 & 0 & 0 & 0 & 1
\end{bmatrix} \nonumber \\
&= \left. \frac{\partial K_{\rm flux}}{\partial r_{\rm in}} \right|_{f_{\rm col}, M, D, i_{\rm disc}}, \label{eqn:J_Kflux2rin_part}
\end{align}
which is the partial derivative of $K_{\rm flux}$ with respect to $r_{\rm in}$, holding constant $f_{\rm col}, M, D, i_{\rm disc}$. Using the inverse transformation function (equation \ref{eqn:Kflux_inv}) to relate $K_{\rm flux}$ and $r_{\rm in}$, we calculate this partial derivative to find
\begin{equation}
J\left( r_{\rm in}, f_{\rm col}, M, D, i_{\rm disc} \right) = h^{-1}\left( r_{\rm in}, f_{\rm col}, M, D, i_{\rm disc} \right) \left[ \frac{2}{r_{\rm in}} + \frac{\partial}{\partial r_{\rm in}} \ln\left( g_{\rm GR}\left( r_{\rm in}, i_{\rm disc} \right) \right) + \frac{d}{d r_{\rm in}} \ln\left( g_{\rm NT}\left( r_{\rm in} \right) \right) \right]. \label{eqn:J_Kflux2rin}
\end{equation}

To recap, taking advantage of the independence of the variables $\{ K_{\rm flux}, f_{\rm col}, ( M, D, i_{\rm disc} ) \}$, we used a multi-variate change of variables approach to derive the integral transform equation \eqref{eqn:f_rin} that relates the marginal densities $f_{r}( r_{\rm in} )$ and $f_{K}( K_{\rm flux} )$. At this stage, and as discussed in \S\ref{sec:fcolErr}, we replace $f_{r}( r_{\rm in} )$ with the inner disc radius marginal density $f_{r}^{\rm CF}( r_{\rm in} )$ measured by the disc continuum fitting practitioners. We also replace the colour correction factor marginal density $f_{f}( f_{\rm col} )$ with a Dirac delta function centered on the assumed value $f_{\rm col}^{\rm CF}$ corresponding to the $f_{r}^{\rm CF}( r_{\rm in} )$ measurement. With these replacements, equation \eqref{eqn:f_rin} becomes
\begin{align}
f_{r}^{\rm CF}\left( r_{\rm in} \right) &= \iiint \limits_{R_{\mathbf{x}}} \int_{f_{\rm col} = f_{\rm col}^{\rm CF} - \epsilon}^{f_{\rm col}^{\rm CF} + \epsilon} f_{K}\left( h^{-1}\left( r_{\rm in}, f_{\rm col}, \mathbf{x} \right) \right) \delta\left( f_{\rm col} - f_{\rm col}^{\rm CF} \right) f_{\mathbf{x}}\left( \mathbf{x} \right) \left| J\left( r_{\rm in}, f_{\rm col}, \mathbf{x} \right) \right| d f_{\rm col}~d \mathbf{x} \nonumber \\
&= \iiint \limits_{R_{\mathbf{x}}} f_{K}\left( h^{-1}\left( r_{\rm in}, f_{\rm col}^{\rm CF}, \mathbf{x} \right) \right) f_{\mathbf{x}}\left( \mathbf{x} \right) \left| J\left( r_{\rm in}, f_{\rm col}^{\rm CF}, \mathbf{x} \right) \right| d \mathbf{x}. \label{eqn:f_rinCF}
\end{align}
The objective now is to perform an inverse integral transform; that
is, to solve for the disc flux normalization marginal density
$f_{K}( K_{\rm flux} )$ in the integrand of equation
\eqref{eqn:f_rinCF}. In other words, given
$f_{r}^{\rm CF}( r_{\rm in} )$, we wish to reverse-engineer this to
$f_{K}( K_{\rm flux} )$, which is the observable in our
re-parametrized version of the disc continuum fitting
technique. Inserting this $f_{K}( K_{\rm flux} )$ into equation
\eqref{eqn:f_rin}, along with a marginal density
$f_{f}( f_{\rm col} )$ that reflects realistic $f_{\rm col}$
uncertainties, we can make a revised measurement of
$f_{r}( r_{\rm in} )$ and consequently the black hole spin parameter.

\subsubsection{Inverse Integral Transform Method Using Orthonormal Basis Functions}
\label{app:basisfunc}
Picking up from where \S\ref{app:rin2Kflux} left off, we seek the marginal density $f_{K}( K_{\rm flux} )$ of the disc flux normalization that is consistent with the reported measurement for the marginal density $f_{r}^{\rm CF}( r_{\rm in} )$ of the inner disc radius. From independent measurements, we also know the joint density $f_{M, D, i}( M, D, i_{\rm disc} )$ of the black hole mass $M$, distance $D$, and inner disc inclination $ i_{\rm disc}$. Our starting point is equation \eqref{eqn:f_rinCF}, the integral transform for $f_{r}^{\rm CF}( r_{\rm in} )$, and our problem amounts to inverting this expression for $f_{K}( K_{\rm flux} ) = f_{K}( h^{-1}( r_{\rm in}, f_{\rm col}^{\rm CF}, \mathbf{x} ) )$, which we will accomplish below by approximating $f_{r}^{\rm CF}( r_{\rm in} )$ and $f_{K}( K_{\rm flux} )$ as linear combinations of orthonormal basis functions. Broadly, these techniques are called \textit{spectral methods}. See \citet{BoydSpectral} for a review.\footnote{We acknowledge a StackExchange post by user David Zaslavsky, who outlined the method adopted here to solve a similar problem: \href{https://scicomp.stackexchange.com/questions/10161/numerical-methods-for-inverting-integral-transforms}{https://scicomp.stackexchange.com/questions/10161/numerical-methods-for-inverting-integral-transforms}}

The marginalization over a random variable removes information. Therefore the inversion of equation \eqref{eqn:f_rinCF} is under-determined. Given an $f_{r}^{\rm CF}( r_{\rm in} )$, there may be more than one $f_{K}( K_{\rm flux} )$ that satisfies equation \eqref{eqn:f_rinCF}. Since, in the end, we are interested in building a new $f_{r}( r_{\rm in} )$ based on $f_{K}( K_{\rm flux} )$, this ambiguity is not present in our main results. We will therefore develop and apply a spectral method for regression, where we find one of many optimal fits of a model for our data, rather than seeking a single exact solution.

Given a set of orthonormal basis functions $T_{\ell}( z^{\prime} )$, any function $f( z^{\prime} )$ can be expressed (approximated) as an infinite (truncated) linear combination of those basis functions and the coefficients $a_{\ell}$,
\begin{equation}
f\left( z^{\prime} \right) = \sum_{\ell=0}^{\infty} a_{\ell} T_{\ell}\left( z^{\prime} \right) \simeq \sum_{\ell=0}^{L} a_{\ell} T_{\ell}\left( z^{\prime} \right). \label{eqn:lincomb}
\end{equation}
Orthonormal basis functions also have the useful property that their inner product with respect to their weighting function $W( z^{\prime} )$ on the interval $z^{\prime} \in [ a, b ]$ evaluates to the Kronecker delta,
\begin{equation}
\left\langle T_{\ell}\left( z^{\prime} \right), T_{k}\left( z^{\prime} \right) \right\rangle \equiv \int_{a}^{b} W\left( z^{\prime} \right) T_{\ell}\left( z^{\prime} \right) T_{k}\left( z^{\prime} \right) dz^{\prime} = \delta_{\ell k}, \label{eqn:innerprod}
\end{equation}
where both $W( z^{\prime} )$ and $[ a, b ]$ are determined by the choice of $T_{\ell}( z^{\prime} )$, which we leave unspecified for now. We will also exploit the ``coefficient formula'' for orthonormal functions,
\begin{equation}
a_{\ell} = \left\langle f\left( z^{\prime} \right), T_{\ell}\left( z^{\prime} \right) \right\rangle. \label{eqn:coefform}
\end{equation}
Importantly, our notation of placing a prime on a variable (e.g., $z^{\prime}$) now signifies that its domain conforms to $[ a, b ]$.

To take advantage of the inner product property of orthonormal basis functions (equation \ref{eqn:innerprod}), we linearly re-scale $K_{\rm flux} \in [ K_{\rm flux}^{\rm min}, K_{\rm flux}^{\rm max} ]$ and $r_{\rm in} \in [ r_{\rm in}^{\rm min}, r_{\rm in}^{\rm max} ]$ to $K_{\rm flux}^{\prime} \in [ a, b ]$ and $r_{\rm in}^{\prime} \in [ a, b ]$ using the relation
\begin{equation}
z^{\prime} = (b - a) \frac{z - z^{\rm min}}{z^{\rm max} - z^{\rm min}} + a, \label{eqn:zprime}
\end{equation}
where $z^{\prime} \in [ a, b ]$ and $z \in [ z^{\rm min}, z^{\rm max} ]$. Next, using the invariance of the probability contained in a differential element,
\begin{equation}
\left| f_{z}\left( z \right) dz \right| = \left| f_{z^{\prime}}\left( z^{\prime} \right) dz^{\prime} \right|, \label{eqn:eqprob}
\end{equation}
we replace the marginal densities $f_{K}( K_{\rm flux} )$ and $f_{r}^{\rm CF}( r_{\rm in} )$ with $f_{K^{\prime}}( K_{\rm flux}^{\prime} )$ and $f_{r^{\prime}}^{\rm CF}( r_{\rm in}^{\prime} )$, and equation \eqref{eqn:f_rinCF} becomes
\begin{equation}
f_{r^{\prime}}^{\rm CF}\left( r_{\rm in}^{\prime} \right) = \frac{r_{\rm in}^{\rm max} - r_{\rm in}^{\rm min}}{K_{\rm flux}^{\rm max} - K_{\rm flux}^{\rm min}} \iiint \limits_{R_{\mathbf{x}}} f_{K^{\prime}}\left( h^{-1}\left( r_{\rm in}^{\prime}, f_{\rm col}^{\rm CF}, \mathbf{x} \right) \right) f_{\mathbf{x}}\left( \mathbf{x} \right) \left| J\left( r_{\rm in}^{\prime}, f_{\rm col}^{\rm CF}, \mathbf{x} \right) \right| d \mathbf{x}. \label{eqn:f_rprime}
\end{equation}
At this stage, equation \eqref{eqn:f_rprime} expresses the integral transform in terms of the marginal densities $f_{K^{\prime}}( K_{\rm flux}^{\prime} )$ and $f_{r^{\prime}}^{\rm CF}( r_{\rm in}^{\prime} )$, both to be approximated with equation \eqref{eqn:lincomb}, and whose variables $K_{\rm flux}^{\prime} \in [ a, b ]$ and $r_{\rm in}^{\prime} \in [ a, b ]$ span the appropriate range to invoke equation \eqref{eqn:innerprod}. We discuss our choice of basis functions in \S\ref{app:chebyshev}.

Approximating $f_{K^{\prime}}(K_{\rm flux}^{\prime})$ as a linear combination of orthonormal basis functions,
\begin{equation}
f_{K^{\prime}}\left( K_{\rm flux}^{\prime} \right) \simeq \sum_{n=0}^{N} c_{n} T_{n}\left( K_{\rm flux}^{\prime} \right). \label{eqn:fKflux_lincomb}
\end{equation}
The ultimate goal of everything to come is to find the coefficients $c_{n}$. Replacing the argument $K_{\rm flux}^{\prime}$ of equation \eqref{eqn:fKflux_lincomb} with the inverse transformation function $K_{\rm flux}^{\prime} = h^{-1}( r_{\rm in}^{\prime}, f_{\rm col}^{\rm CF}, \mathbf{x} )$ of equation \eqref{eqn:Kflux_inv}, we can approximate the marginal density of the disc flux normalization as it appears in the integrand of equation \eqref{eqn:f_rprime} with the linear combination
\begin{equation}
f_{K^{\prime}}\left( h^{-1}\left( r_{\rm in}^{\prime}, f_{\rm col}^{\rm CF}, \mathbf{x} \right) \right) \simeq \sum_{n=0}^{N} c_{n} T_{n}\left( h^{-1}\left( r_{\rm in}^{\prime}, f_{\rm col}^{\rm CF}, \mathbf{x} \right) \right). \label{eqn:fKflux_inv_lincomb}
\end{equation}
Equation \eqref{eqn:f_rprime} then becomes
\begin{align}
  f_{r^{\prime}}^{\rm CF}\left( r^{\prime}_{\rm in} \right)
  &\simeq \frac{r_{\rm in}^{\rm max} - r_{\rm in}^{\rm min}}{K_{\rm flux}^{\rm max} - K_{\rm flux}^{\rm min}} \iiint \limits_{R_{\mathbf{x}}} \left[ \sum_{n=0}^{N} c_{n} T_{n}\left( h^{-1}\left( r_{\rm in}^{\prime}, f_{\rm col}^{\rm CF}, \mathbf{x} \right) \right) \right] f_{\mathbf{x}}\left( \mathbf{x} \right) \left| J\left( r_{\rm in}^{\prime}, f_{\rm col}^{\rm CF}, \mathbf{x} \right) \right| d \mathbf{x} \nonumber \\
  \label{eqn:fr:fk:inv:lincomb}
  &\simeq \sum_{n=0}^{N} c_{n} \frac{r_{\rm in}^{\rm max} - r_{\rm in}^{\rm min}}{K_{\rm flux}^{\rm max} - K_{\rm flux}^{\rm min}} \iiint \limits_{R_{\mathbf{x}}} T_{n}\left( h^{-1}\left( r_{\rm in}^{\prime}, f_{\rm col}^{\rm CF}, \mathbf{x} \right) \right) f_{\mathbf{x}}\left( \mathbf{x} \right) \left| J\left( r_{\rm in}^{\prime}, f_{\rm col}^{\rm CF}, \mathbf{x} \right) \right| d \mathbf{x}.
\end{align}
Importantly, we can evaluate this integral expression,\footnote{When doing the numerical integration in practice, we set $T_{n}( h^{-1}( r_{\rm in}^{\prime}, f_{\rm col}^{\rm CF}, \mathbf{x} ) ) = 0$ if the argument of the basis function lies outside of its domain; that is, if $h^{-1}( r_{\rm in}^{\prime}, f_{\rm col}^{\rm CF}, \mathbf{x} ) = K_{\rm flux}^{\prime} \not\in [ a, b ]$.} which we denote as
\begin{equation}
\widetilde{T}_{n}\left( r_{\rm in}^{\prime} \right) \equiv \frac{r_{\rm in}^{\rm max} - r_{\rm in}^{\rm min}}{K_{\rm flux}^{\rm max} - K_{\rm flux}^{\rm min}} \iiint \limits_{R_{\mathbf{x}}} T_{n}\left( h^{-1}\left( r_{\rm in}^{\prime}, f_{\rm col}^{\rm CF}, \mathbf{x} \right) \right) f_{\mathbf{x}}\left( \mathbf{x} \right) \left| J\left( r_{\rm in}^{\prime}, f_{\rm col}^{\rm CF}, \mathbf{x} \right) \right| d \mathbf{x}. \label{eqn:Tn_trans}
\end{equation}
Substituting equation \eqref{eqn:Tn_trans} into \eqref{eqn:fr:fk:inv:lincomb}, we find that
\begin{equation}
  \label{eqn:for:cn}
    f_{r^{\prime}}^{\rm CF}(r^{\prime}_{\rm in}) \simeq \sum_{n=0}^N c_n \widetilde{T}_{n}\left( r_{\rm in}^{\prime} \right),
\end{equation}
where the unknowns $c_n$ now translate between the known functions $\widetilde{T}_{n}( r_{\rm in}^{\prime} )$ and $f_{r^{\prime}}^{\rm CF}( r_{\rm in}^{\prime} )$.

Next, we consider two options for converting equation \eqref{eqn:for:cn} into a set of $M+1 > N$ equations for the $N+1$ unknowns $c_{n}$:
\begin{enumerate}
\item \label{itm:modes} Perform a mode decomposition for $f_{r^{\prime}}^{\rm CF}(r^{\prime}_{\rm in})$ using equation \eqref{eqn:lincomb}, as we did for $f_{K^{\prime}}( K_{\rm flux}^{\prime} )$.
\item \label{itm:nodes} Discretize $r^{\prime}_{\rm in}$ directly as a grid of points $r^{\prime}_{i}$, where $i = 0, 1, \ldots, M$.
\end{enumerate}
We call case \ref{itm:modes} the \textit{modal} approach. The modal approach can also be generalized to use cumulative distribution functions, rather than probability distribution functions. We call this the \textit{CDF} approach. We describe the modal and CDF approaches in \S\ref{app:modes}. We call case \ref{itm:nodes} the \textit{nodal} approach and describe it in \S\ref{app:nodes}. The modal and CDF approaches generally perform better when $f_{r^{\prime}}^{\rm CF}( r_{\rm in}^{\prime} )$ is smooth and slowly varying. On the other hand, the nodal approach performs best when $f_{r^{\prime}}^{\rm CF}( r_{\rm in}^{\prime} )$ has features that the modal and CDF approaches fail to capture. In practice, we use the CDF and nodal approaches as appropriate on a case-by-case basis.

\subsubsection{The Modal Approach}
\label{app:modes}
In the modal approach, we approximate $f_{r^{\prime}}^{\rm CF}( r_{\rm in}^{\prime} )$ using a linear combination of basis functions with coefficients $C_{m}$,
\begin{equation}
f_{r^{\prime}}^{\rm CF}\left( r_{\rm in}^{\prime} \right) \simeq \sum_{m=0}^{M} C_{m} T_{m}\left( r_{\rm in}^{\prime} \right) \label{eqn:frin_lincomb}
\end{equation}
and then equation \eqref{eqn:for:cn} becomes
\begin{align}
\sum_{m=0}^{M} C_{m} T_{m}\left( r_{\rm in}^{\prime} \right)  \simeq \sum_{n=0}^N c_n \widetilde{T}_{n}\left( r_{\rm in}^{\prime} \right). \label{eqn:linsys}
\end{align}
Now, the trick is to realize that $\widetilde{T}_{n}( r_{\rm in}^{\prime} )$ itself is just a function, meaning that it, too, can be expressed as a linear combination of orthonormal basis functions,
\begin{equation}
\widetilde{T}_{n}\left( r_{\rm in}^{\prime} \right) = \sum_{k=0}^{\infty} A_{n k} T_{k}\left( r_{\rm in}^{\prime} \right), \label{eqn:Tn_trans_lincomb}
\end{equation}
where $A_{n k}$ is a coefficient matrix. Equation \eqref{eqn:linsys}
thus becomes,
\begin{equation}
\sum_{m=0}^{M} C_{m} T_{m}\left( r_{\rm in}^{\prime} \right) \simeq \sum_{n=0}^{N} c_{n} \sum_{k=0}^{\infty} A_{n k} T_{k}\left( r_{\rm in}^{\prime} \right).
\end{equation}
Taking the inner product of both sides with $T_{\ell}( r_{\rm in}^{\prime} )$,
\begin{align}
\left\langle \sum_{m=0}^{M} C_{m} T_{m}\left( r_{\rm in}^{\prime} \right), T_{\ell}\left( r_{\rm in}^{\prime} \right) \right\rangle &\simeq \left\langle \sum_{n=0}^{N} c_{n} \sum_{k=0}^{\infty} A_{n k} T_{k}\left( r_{\rm in}^{\prime} \right), T_{\ell}\left( r_{\rm in}^{\prime} \right) \right\rangle \nonumber \\
\int_{a}^{b} W\left( r_{\rm in}^{\prime} \right) \sum_{m=0}^{M} C_{m} T_{m}\left( r_{\rm in}^{\prime} \right) T_{\ell}\left( r_{\rm in}^{\prime} \right) d r_{\rm in}^{\prime} &\simeq \int_{a}^{b} W\left( r_{\rm in}^{\prime} \right) \sum_{n=0}^{N} c_{n} \sum_{k=0}^{\infty} A_{n k} T_{k}\left( r_{\rm in}^{\prime} \right) T_{\ell}\left( r_{\rm in}^{\prime} \right) d r_{\rm in}^{\prime} \nonumber \\
\sum_{m=0}^{M} C_{m} \int_{a}^{b} W\left( r_{\rm in}^{\prime} \right) T_{m}\left( r_{\rm in}^{\prime} \right) T_{\ell}\left( r_{\rm in}^{\prime} \right) d r_{\rm in}^{\prime} &\simeq \sum_{n=0}^{N} c_{n} \sum_{k=0}^{\infty} A_{n k} \int_{a}^{b} W\left( r_{\rm in}^{\prime} \right) T_{k}\left( r_{\rm in}^{\prime} \right) T_{\ell}\left( r_{\rm in}^{\prime} \right) d r_{\rm in}^{\prime} \nonumber \\
\sum_{m=0}^{M} C_{m} \delta_{m \ell} &\simeq \sum_{n=0}^{N} c_{n} \sum_{k=0}^{\infty} A_{n k} \delta_{k \ell} \nonumber \\
C_{\ell} &\simeq \sum_{n=0}^{N} c_{n} A_{n \ell}, \label{eqn:linsys_CcA}
\end{align}
which is the desired linear system of $M+1$ equations and $N+1$ unknowns in terms of the coefficients $C_{\ell}$, $c_{n}$, and $A_{n \ell}$. Note that because $\ell$ and $m$ are dummy indices, they can be swapped out for each other. We can solve this system of equations for the $c_{n}$ coefficients --- and therefore obtain $f_{K^{\prime}}( K_{\rm flux}^{\prime} )$ from equation \eqref{eqn:fKflux_lincomb} --- if we know $C_{\ell}$ and $A_{n \ell}$, both of which are accessible from applications of the inner product (see equation \ref{eqn:coefform}), as follows.

The coefficient formula for $C_{\ell}$ follows from taking the inner product $\langle f_{r^{\prime}}^{\rm CF}( r_{\rm in}^{\prime} ), T_{\ell}( r_{\rm in}^{\prime} ) \rangle$ and substituting in equation \eqref{eqn:frin_lincomb} for $f_{r^{\prime}}^{\rm CF}( r_{\rm in}^{\prime} )$,
\begin{align}
\int_{a}^{b} W\left( r_{\rm in}^{\prime} \right) f_{r^{\prime}}^{\rm CF}\left( r_{\rm in}^{\prime} \right) T_{\ell}\left( r_{\rm in}^{\prime} \right) d r_{\rm in}^{\prime} &\simeq \int_{a}^{b} W\left( r_{\rm in}^{\prime} \right) \left[ \sum_{m=0}^{M} C_{m} T_{m}\left( r_{\rm in}^{\prime} \right) \right] T_{\ell}\left( r_{\rm in}^{\prime} \right) d r_{\rm in}^{\prime} \nonumber \\
&\simeq \sum_{m=0}^{M} C_{m}  \int_{a}^{b} W\left( r_{\rm in}^{\prime} \right) T_{m}\left( r_{\rm in}^{\prime} \right) T_{\ell}\left( r_{\rm in}^{\prime} \right) d r_{\rm in}^{\prime} \nonumber \\
&\simeq \sum_{m=0}^{M} C_{m} \delta_{m \ell} \nonumber \\
&\simeq C_{\ell}, \label{eqn:coefC}
\end{align}
where because $f_{r^{\prime}}^{\rm CF}( r_{\rm in}^{\prime} )$ is known, we can evaluate the left-hand side of equation \eqref{eqn:coefC} to obtain the coefficients $C_{\ell}$.

\begin{figure*}
    \begin{center}
        \includegraphics[width=0.495\textwidth]{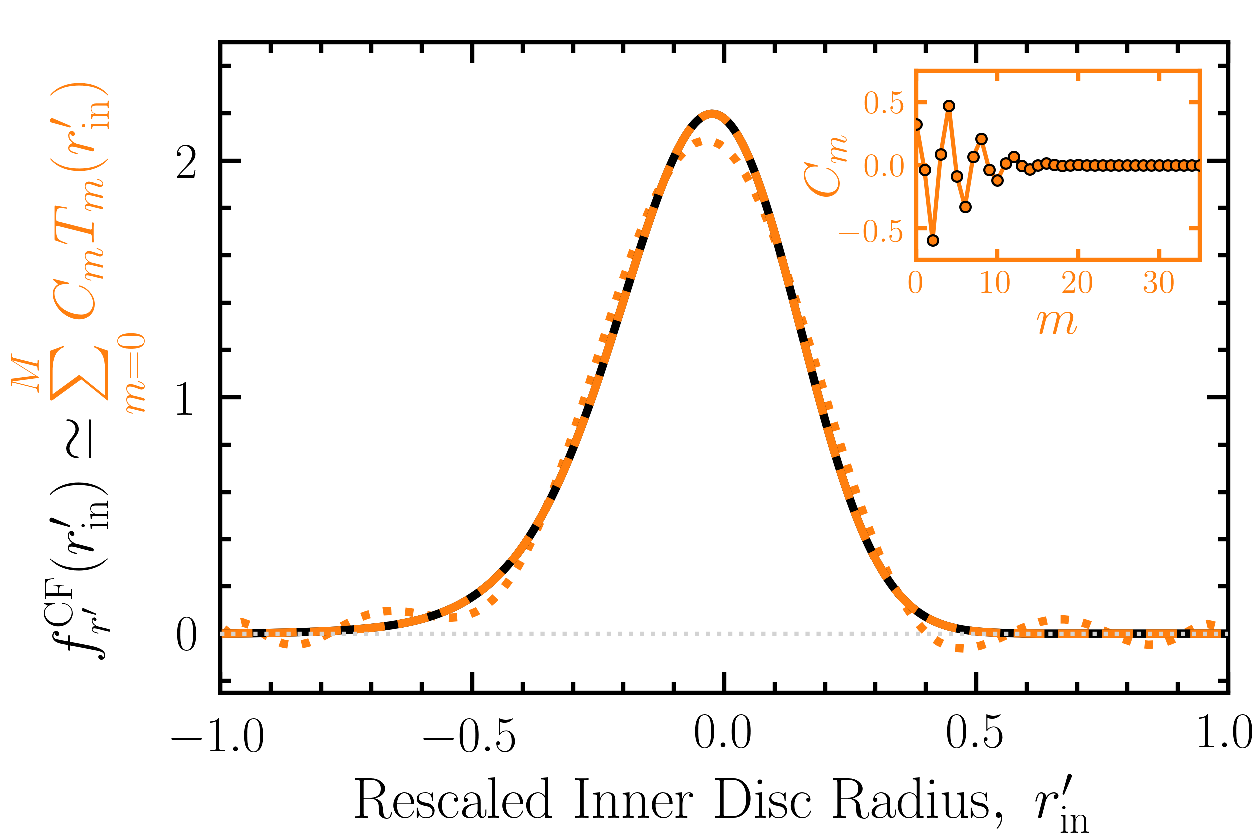}
        \hfill
        \includegraphics[width=0.495\textwidth]{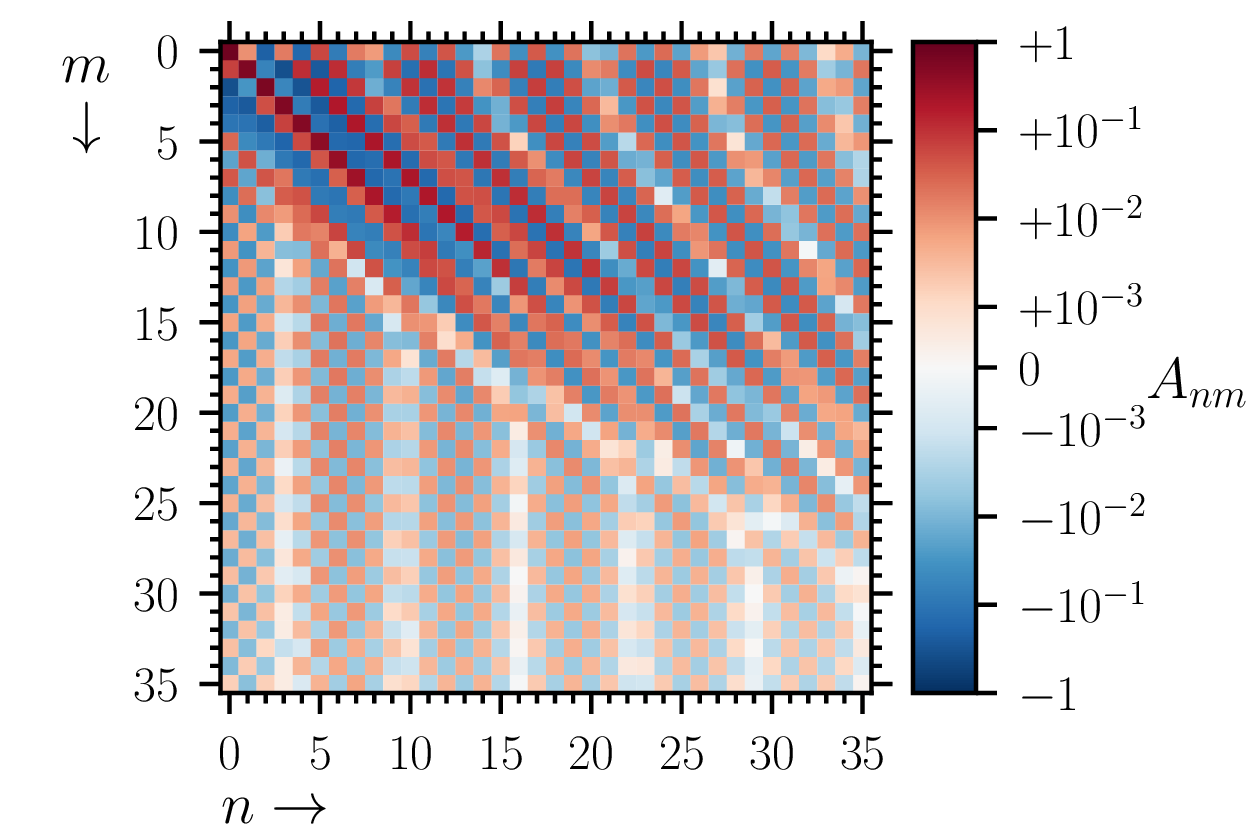}        
        \hfill
        \includegraphics[width=1.0\textwidth]{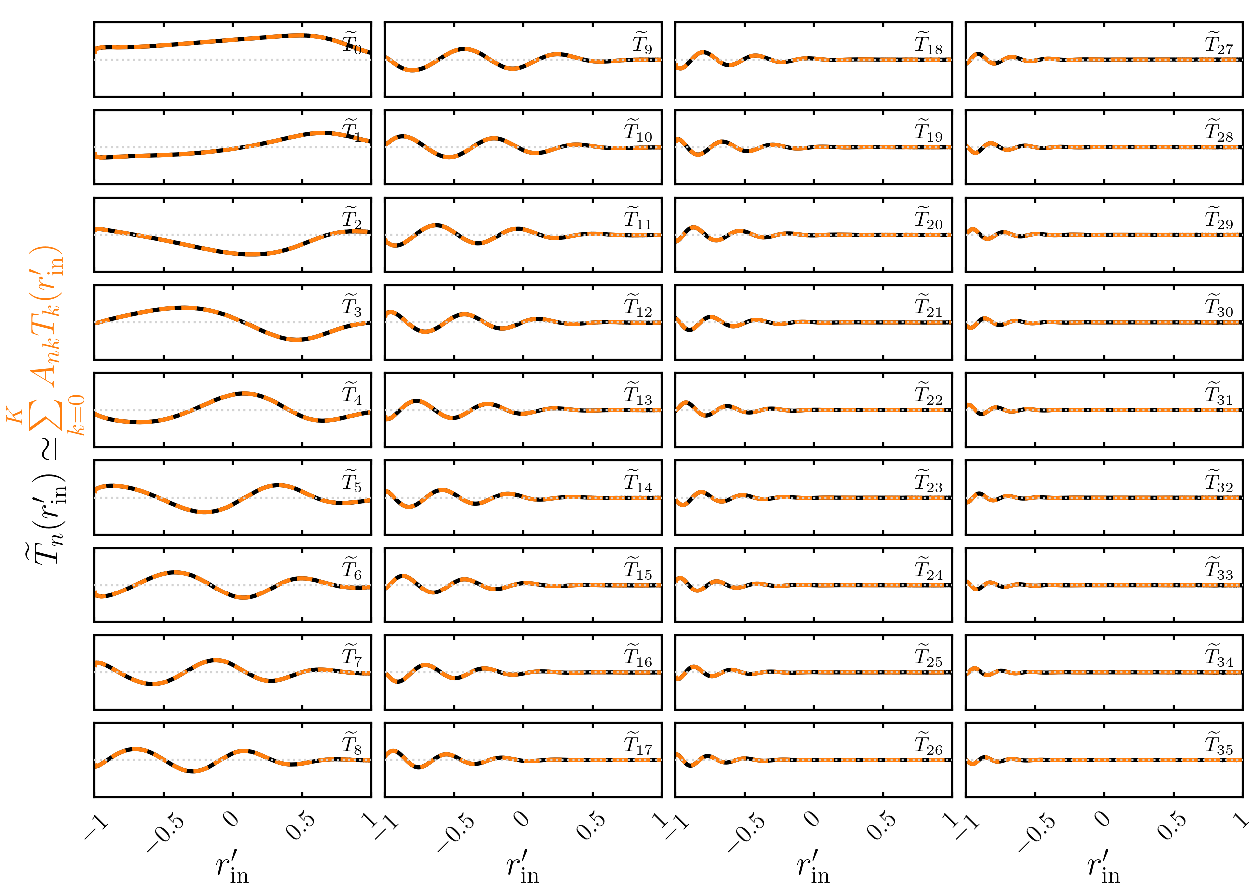}
    \end{center}
    \vspace{-4mm}
    \caption{Workflow to obtain the coefficients $C_{m}$ and $A_{nm}$, as needed to solve the linear system \eqref{eqn:linsys_CcA} for the coefficients $c_{n}$, using GRO J1655--40 as an example. \textit{Top left}: We re-scale the marginal density $f_{r}^{\rm CF}( r_{\rm in} )$ with $r_{\rm in} \in [ r_{\rm in}^{\rm min}, r_{\rm in}^{\rm max} ]$ to the marginal density $f_{r^{\prime}}^{\rm CF}( r_{\rm in}^{\prime} )$ with $r_{\rm in}^{\prime} \in [ -1, 1 ]$ shown by the \textit{solid black line}. We approximate the known $f_{r^{\prime}}^{\rm CF}( r_{\rm in}^{\prime} )$ as the linear combination $\sum\limits_{m=0}^{M} C_{m} T_{m}( r_{\rm in}^{\prime} )$ with $M = 35$ following equation \eqref{eqn:frin_lincomb} and shown by the \textit{dashed orange line}. The approximation is excellent and we use equation \eqref{eqn:coefC} to calculate the coefficients $C_{m}$ shown in the \textit{inset}. The \textit{dotted orange line} demonstrates a poor approximation using $M = 11$ for the linear combination. \textit{Bottom}: Using equation \eqref{eqn:Tn_trans}, we calculate $\widetilde{T}_{n}( r_{\rm in}^{\prime} )$ shown by the \textit{solid black lines} (the $y$-axes linearly span $[-2,2]$), which we approximate as the linear combination $\sum\limits_{k=0}^{K} A_{n k} T_{k}( r_{\rm in}^{\prime} )$ with $K = 35$ following equation \eqref{eqn:Tn_trans_lincomb} and shown by the \textit{dashed orange lines}. \textit{Top right}: In doing so, we use equation \eqref{eqn:coefA} to calculate the coefficient matrix $A_{n m}$. When calculating $C_{m}$ and $A_{n m}$, we incorporate the appropriate normalization factors (see equation \ref{eqn:an}).}
    \label{fig:frP_Anm_Tnr}
\end{figure*}

Similarly, the coefficient formula for the matrix $A_{n \ell}$ follows from taking the inner product $\langle \widetilde{T}_{n}( r_{\rm in}^{\prime} ), T_{\ell}( r_{\rm in}^{\prime} ) \rangle$ and substituting in equation \eqref{eqn:Tn_trans_lincomb} for $\widetilde{T}_{n}( r_{\rm in}^{\prime} )$,
\begin{align}
\int_{a}^{b} W\left( r_{\rm in}^{\prime} \right) \widetilde{T}_{n}\left( r_{\rm in}^{\prime} \right) T_{\ell}\left( r_{\rm in}^{\prime} \right) d r_{\rm in}^{\prime} &= \int_{a}^{b} W\left( r_{\rm in}^{\prime} \right) \left[ \sum_{k=0}^{\infty} A_{n k} T_{k}\left( r_{\rm in}^{\prime} \right) \right] T_{\ell}\left( r_{\rm in}^{\prime} \right) d r_{\rm in}^{\prime} \nonumber \\
&= \sum_{k=0}^{\infty} A_{n k} \int_{a}^{b} W\left( r_{\rm in}^{\prime} \right) T_{k}\left( r_{\rm in}^{\prime} \right) T_{\ell}\left( r_{\rm in}^{\prime} \right) d r_{\rm in}^{\prime} \nonumber \\
&= \sum_{k=0}^{\infty} A_{n k} \delta_{k \ell} \nonumber \\
&= A_{n \ell}, \label{eqn:coefA}
\end{align}
where because $\widetilde{T}_{n}( r_{\rm in}^{\prime} )$ is known (equation \ref{eqn:Tn_trans}), we can evaluate the left-hand side of equation \eqref{eqn:coefA} to obtain the coefficient matrix $A_{n \ell}$. Each element of the $(N+1) \times (M+1)$ matrix $A_{n \ell}$ requires numerically calculating a quadruple integral. Figure \ref{fig:frP_Anm_Tnr} shows an example of the workflow for calculating these $C_{\ell}$ and $A_{n \ell}$ coefficients.

Having calculated each of the $C_{\ell}$ coefficients (equation \ref{eqn:coefC}) and each element of the $A_{n \ell}$ coefficient matrix (equation \ref{eqn:coefA}), we solve the linear system \eqref{eqn:linsys_CcA} for the $c_{n}$ coefficients following \S\ref{app:constopt}, where we describe solution methods. Typically, we set $N = M = K$ so there are an equal number of $C_{m}$ and $c_{n}$ coefficients and such that $A_{nk}$ is square.

Both sides of equation \eqref{eqn:for:cn} can be integrated to achieve an expression in terms of the \textit{cumulative distribution function} for $r^{\prime}_{\rm in}$, $\text{CDF}^{\rm CF}_{r^{\prime}}(r^{\prime}_{\rm in})$:
\begin{equation}
  \label{eq:for:cn:CDF}
  \text{CDF}_{r^{\prime}}^{\rm CF}(r^{\prime}_{\rm in}) \simeq \sum_{n=0}^{N} c_{n} \int_{-1}^{r^{\prime}_{\rm in}} \widetilde{T}_{n}\left( x \right) dx.
\end{equation}
The discussion above and in \S\ref{app:constopt} proceeds identically, except that the left-hand side is the cumulative rather than the probability distribution function and the right-hand side integrates over $\widetilde{T}_{n}$. In practice, we only combined the CDF with the modal approach. We call this the \textit{CDF} approach.

\subsubsection{The Nodal Approach}
\label{app:nodes}
In the nodal approach, we investigate equation \eqref{eqn:for:cn} restricted to $M \geq N$ specific values of $r^{\prime}_{\rm in}$: $r^{\prime}_{i}$, for $i = 0,1,\ldots,M$. In keeping with best practices for spectral methods, we choose $r^{\prime}_{i}$ to be the Gauss quadrature points, which are the $M+1$ nodes of $T_{M}( r^{\prime}_{\rm in})$.\footnote{Note that the Gauss points do not include the boundary points, $r^{\prime}_{\rm in} = -1,1$. An alternative choice is the Gauss-Lobatto points, which \textit{do} include the boundaries. See \citet{BoydSpectral} for a more detailed discussion.} So restricted, equation \eqref{eqn:for:cn} becomes a set of $M+1$ linear equations
\begin{equation}
  \label{eqn:for:cn:nodes}
  f_{r^{\prime}}^{\rm CF}\left( r^{\prime}_{i} \right) \simeq \sum_{n=0}^{N} c_{n} \widetilde{T}_{n}\left( r_{i}^{{\prime}} \right)\ \forall\ i = 0, 1, \ldots, M
\end{equation}
for the $N+1$ unknown $c_{n}$ coefficients. Equation \eqref{eqn:for:cn:nodes} can be cast in matrix form as
\begin{equation}
  \label{eq:for:cn:nodes:matrix:form}
  f_{r^{\prime}}^{\rm CF}\left( r^{\prime}_{m} \right) \simeq \sum_{n=0}^{N} B_{mn} c_{n},
\end{equation}
where $B_{mn} = \widetilde{T}_{n}( r^{\prime}_{m} )$ and we swapped the dummy index $i$ for $m$. Inverting $B_{mn}$ in equation \eqref{eq:for:cn:nodes:matrix:form} yields the vector of unknowns $c_{n}$. We typically set $M = N$.

\subsubsection{Imposing Constraints on the Linear System}
\label{app:constopt}
At this stage, we must solve either equation \eqref{eqn:linsys_CcA} or equation \eqref{eq:for:cn:nodes:matrix:form} for the coefficients $c_{n}$. For the purposes of this discussion, we write both equations in the more generic form of equation \eqref{eqn:for:cn}
\begin{displaymath}
  f_{r^{\prime}}^{\rm CF}\left( r^{\prime}_{\rm in} \right) \simeq \sum_{n=0}^{N} c_{n} \widetilde{T}_{n}\left( r_{\rm in}^{\prime} \right).
\end{displaymath}
Since the inversion of equation \eqref{eqn:f_rinCF} is under-determined, the linear systems \eqref{eqn:linsys_CcA} and \eqref{eq:for:cn:nodes:matrix:form} as written are ill-conditioned,\footnote{That is, the matrices $A_{n \ell}$ and $B_{mn}$ have large condition numbers.} meaning the system is effectively under-constrained and matrix inversion is not a numerically well-defined procedure. We therefore shift our perspective from finding a \textit{unique} solution to finding one of potentially many \textit{physically correct} solutions. We seek to find the set of $c_{n}$ that \textit{best} satisfies \textit{both} our linear equation (either \ref{eqn:linsys_CcA} or \ref{eq:for:cn:nodes:matrix:form}) and a set of physically meaningful constraints.

In particular, $f_{K^{\prime}}( K_{\rm flux}^{\prime} )$ defined in equation \eqref{eqn:fKflux_lincomb} is a probability distribution, so must be positive definite, integrate to unity, and vanish at $K_{\rm flux}^{\prime} = 0, 1$. We encode these constraints on $c_{n}$ as
\begin{equation}
  \label{eq:cn:positivity:condition}
  \sum_{n=0}^{N} c_{n} T_{n}\left( K_{\rm flux}^{\prime} \right) - \left| \sum_{n=0}^{N} c_{n} T_{n}\left( K_{\rm flux}^{\prime} \right) \right| = 0\ \forall\ K_{\rm flux}^{\prime} \in \left[ 0, 1 \right]
\end{equation}
for positivity, as
\begin{equation}
  \label{eq:cn:unitarity}
  1 = \sum_{n=0}^{N} c_{n} \int_{-1}^1 \widetilde{T}_{n}\left( r^{\prime}_{\rm in} \right) dr^{\prime}_{\rm in}
\end{equation}
for unitarity, and as
\begin{equation}
  \label{eq:cn:boundary:condition}
  \sum_{n=0}^{N} c_{n} T_{n}\left( 0 \right) = \sum_{n=0}^{N} c_{n} T_{n}\left( 1 \right) = 0
\end{equation}
for the vanishing boundaries condition. Given a monotonically increasing grid of $M+1$ points $0 \leq K_{i}^{\prime} \leq 1$, such that $K_{0}^{\prime} = 0$ and $K_{M}^{\prime} = 1$, we can approximate condition \eqref{eq:cn:positivity:condition} as
\begin{equation}
  \label{eq:positicvity:condition:discrete}
    \sum_{n=0}^{N} c_{n} T_{n}\left( K_{i}^{\prime} \right) - \left| \sum_{n=0}^{N} c_{n} T_{n}\left( K_{i}^{\prime} \right) \right| = 0\ \forall\ i = 0, 1, ..., M,
\end{equation}
which is easier to impose numerically. We choose this grid of $K_{i}^{\prime}$ points to be the Gauss points described in \S\ref{app:nodes}. 

To impose these constraints in our numerical method, we transform the linear system into a \textit{constrained optimization problem}. We define the \textit{residual}, which takes the vector of coefficients $c_{n}$ and returns a scalar measure of how close $c_{n}$ is to satisfying equations \eqref{eqn:linsys_CcA}, \eqref{eq:cn:unitarity}, \eqref{eq:cn:boundary:condition}, and \eqref{eq:positicvity:condition:discrete}:
\begin{eqnarray}
  \label{eq:residual:main}
  \mathcal{R}[c_{n}] &=& \sum_{m=0}^{M} \left( f^{\rm CF}_{r^{\prime}}(r^{\prime}_{\rm in}) - \sum_{n=0}^{N} c_{n} \widetilde{T}_{n}(r^{\prime}_{\rm in}) \right)^{2} \\
  \label{eq:residual:left}
                   &&\qquad + \lambda_{l} \left( \sum_{n=0}^{N} c_{n} T_{n}\left( 0 \right) \right)^{2} \\
  \label{eq:residual:right}
                     &&\qquad + \lambda_{r} \left( \sum_{n=0}^{N} c_{n} T_{n}\left( 1 \right) \right)^{2} \\
  \label{eq:residual:unit}
                     &&\qquad + \lambda_{1} \left( 1 - \sum_{n=0}^{N} c_{n} \int_{-1}^1 \widetilde{T}_{n}(r^{\prime}_{\rm in}) dr^{\prime}_{\rm in} \right)^{2} \\
  \label{eq:residual:pos}
                     &&\qquad + \lambda_{p} \sum_{i=0}^{M} \left( \sum_{n=0}^{N} c_{n} T_{n}\left( K_{i}^{\prime} \right) - \left| \sum_{n=0}^{N} c_{n} T_{n}\left( K_{i}^{\prime} \right) \right| \right)^{2}\\
  \label{eq:residual:filter}
                       &&\qquad + \lambda_{e} \sum_{n=0}^{N} c_{n}^{2},
\end{eqnarray}
where contribution \eqref{eq:residual:main} is the main linear system, depending on whether the modal or nodal method is used; contribution \eqref{eq:residual:left} is the left boundary; contribution \eqref{eq:residual:right} is the right boundary; contribution \eqref{eq:residual:unit} is the unitarity condition; contribution \eqref{eq:residual:pos} is the positivity condition; and contribution \eqref{eq:residual:filter} is an exponential \textit{filter} condition found to reduce spurious oscillations in the solution \citep{DeepLearning}. $\lambda_{l}$, $\lambda_{r}$, $\lambda_{1}$, $\lambda_{p}$, and $\lambda_{e}$ are Lagrange multipliers. We then search for the global minimum of this residual
\begin{eqnarray}
  \label{eq:c_n:numerical}
  c_{n} := \argmin_{c_{n}^{\prime}} \mathcal{R}\left[ c_{n}^{\prime} \right]
\end{eqnarray}
via a standard optimization algorithm such as Newton's method \citep{Press1986}.

In practice, the Lagrange multiplier approach outlined here has the problem that the constraints may only be \textit{approximately} satisfied, not \textit{exactly} satisfied. The Newton-Raphson iteration in a standard solver may be modified to enforce that the constraints be \textit{exactly} satisfied by performing a linear search only in constraint-satisfying directions of the parameter-space of possible values of $c_{n}$. This has the disadvantage that the solver can less easily explore the space, and it may be stuck in a local minimum that it could otherwise escape if it were allowed to temporarily violate the constraints. We experimented with both Lagrange-multiplier and exact constraints and chose the best combination on a system-by-system basis. If only Lagrange multipliers are used, we use the limited-memory variant of Newton's method first presented in \citet{BFGS}. In the case of exact constraints, we use the sequential least squares programming algorithm originally implemented in \citet{KraftSLSQP}. In both cases, we use the \href{https://docs.scipy.org/doc/scipy/reference/generated/scipy.optimize.minimize.html}{minimize function} in \texttt{SciPy} \citep{scipy}, which wraps both solvers. \texttt{SciPy} calls the former \texttt{BFGS} and the latter \texttt{SLSQP}.

\setlength{\tabcolsep}{10pt}
\begin{table*}
\centering
\begin{tabular}{c c c c c c c}
\hline
\hline
Source & $N+1$ & $[K_{\rm flux}^{\rm min}, K_{\rm flux}^{\rm max}]$ & $[r_{\rm in}^{\rm min}, r_{\rm in}^{\rm max}]$ & Method & Lagrange & Exact Constraints \\
\smallskip
 & & $[{\rm km^{2} / (kpc / 10)^{2}}]$ & $[R_{\rm g}]$ & & $\lambda_{l}$, $\lambda_{r}$, $\lambda_{p}$, $\lambda_{1}$, $\lambda_{e}$ & \texttt{bnds}, \texttt{unit}, \texttt{pos} \\
\hline
\smallskip
LMC X--1 & 18 & $[0, 25]$ & $[0, 5]$ & CDF & 0, 0, 0, 0, 0 & 1, 1, 1 \\
\smallskip
4U 1543--47 & 32 & $[0, 600]$ & $[0.5, 5]$ & CDF & 0, 0, 0, 1, 1 & 1, 1, 1 \\
\smallskip
GRO J1655--40 & 36 & $[0, 800]$ & $[1, 6]$ & CDF & 5, 5, 1, 1, 1 & 0, 0, 0 \\
\smallskip
XTE J1550--564 & 164 & $[0, 1250]$ & $[1, 9]$ & Nodal & 0, 0, 1, 0, 0 & 1, 1, 0 \\
\smallskip
M33 X--7 & 24 & $[0.005, 0.0325]$ & $[1.5, 4.5]$ & CDF & 0, 0, 2, 0, 1 & 1, 1, 0 \\
\smallskip
  LMC X--3 & 28 & $[3, 6]$ & $[3, 8]$ & Nodal & 0, 0, 10, 0, 10 & 1, 1, 0 \\
\smallskip
H1743--322 & 36 & $[0, 250]$ & $[0, 10]$ & Nodal & 0, 0, 10, 0, 0 & 1, 1, 0 \\
\smallskip
A0620--00 & 24 & $[1.5, 2] \times 10^{4}$ & $[3, 8]$ & Nodal & 0, 0, 0, 0, 0.1 & 1, 1, 1 \\
\hline
\end{tabular}
\caption{Inputs and constraints used to calculate the disc flux normalization marginal density $f_{K}( K_{\rm flux} )$ for each X-ray binary (see Figure \ref{fig:fK}). From \textit{left} to \textit{right}, the columns are: source name, $N+1$ number of $c_{n}$ coefficients used to approximate $f_{K^{\prime}}( K_{\rm flux}^{\prime} )$, $K_{\rm flux}$ range, $r_{\rm in}$ range, adopted method, values of Lagrange multipliers used, and imposed exact constraints (see \S\ref{app:constopt}). Common to all calculations, we used 18 grid zones each for $\{ M, D, i_{\rm disc} \}$, 180 grid zones each for $\{ r_{\rm in}, K_{\rm flux} \}$, and a Gaussian smoothing kernel of $25~R_{\rm g}$ for the $\partial g_{\rm GR} / \partial r_{\rm in}$ grid (see Figure \ref{fig:gGR}). LCM X--3 required smoothing $f^{\rm CF}_{r^{\prime}}( r^{\prime}_{\rm in} )$ by a boxcar kernel of width 11 grid points to achieve convergence.}
\label{tab:fK}
\end{table*}

\subsubsection{Our Basis Functions}
\label{app:chebyshev}
Up until now, we left the orthonormal basis functions $T_{\ell}( z^{\prime} )$ unspecified. After experimenting with many different orthogonal basis functions, we ultimately settled on the Chebyshev polynomials of the first kind. The first two polynomials are $T_{0}( z^{\prime} ) = 1$ and $T_{1}( z^{\prime} ) = z^{\prime}$, and a recurrence relation defines all the others,
\begin{equation}
T_{n}\left( z^{\prime} \right) = 2 z^{\prime} T_{n-1}\left( z^{\prime} \right) - T_{n-2}\left( z^{\prime} \right),~\mathrm{for}~n \ge 2,
\end{equation}
where $n$ is an integer. The Chebyshev polynomials are orthogonal, but \textit{not} normalized, with respect to the weighting function $W( z^{\prime} ) = 1 / \sqrt{1 - ( z^{\prime} )^{2}}$ on the interval $z^{\prime} \in[ a, b ] = [ -1, 1 ]$,
\begin{equation}
\left\langle T_{n}\left( z^{\prime} \right), T_{m}\left( z^{\prime} \right) \right\rangle = \int_{-1}^{1} \frac{T_{n}\left( z^{\prime} \right) T_{m}\left( z^{\prime} \right)}{\sqrt{1 - \left( z^{\prime} \right)^{2}}} dz^{\prime}
= \left\{
    \begin{array}{@{\hspace{0mm}}l@{\hspace{1mm}}l@{\hspace{0mm}}}
        \pi &,~\mathrm{for}~n = 0, m = 0 \\ \\
        \dfrac{\pi}{2} \delta_{n m} &,~\mathrm{for}~n \ne 0, m \ne 0.
    \end{array}
    \right. \label{eqn:cheby}
\end{equation}
Consequently, we must modify the ``coefficient formula'' (equation \ref{eqn:coefform}) accordingly,
\begin{align}
\left\langle f\left( z^{\prime} \right), T_{n}\left( z^{\prime} \right) \right\rangle &= \int_{-1}^{1} \frac{f\left( z^{\prime} \right) T_{n}\left( z^{\prime} \right)}{\sqrt{1 - \left( z^{\prime} \right)^{2}}} dz^{\prime} \nonumber \\
&= \int_{-1}^{1} \left[ \sum_{m=0}^{\infty} a_{m} T_{m}\left( z^{\prime} \right) \right] \frac{T_{n}\left( z^{\prime} \right)}{\sqrt{1 - \left( z^{\prime} \right)^{2}}} dz^{\prime} \nonumber \\
&= \sum_{m=0}^{\infty} a_{m} \int_{-1}^{1} \frac{T_{n}\left( z^{\prime} \right) T_{m}\left( z^{\prime} \right)}{\sqrt{1 - \left( z^{\prime} \right)^{2}}} dz^{\prime} \nonumber \\
&= \left\{
    \begin{array}{@{\hspace{0mm}}l@{\hspace{1mm}}l@{\hspace{0mm}}}
        \pi \sum_{m=0}^{\infty} a_{m} \delta_{n m} &,~\mathrm{for}~n = 0, m = 0 \\ \\
        \dfrac{\pi}{2} \sum_{m=0}^{\infty} a_{m} \delta_{n m} &,~\mathrm{for}~n \ne 0, m \ne 0.
    \end{array}
    \right. \nonumber \\
&= \left\{
    \begin{array}{@{\hspace{0mm}}l@{\hspace{1mm}}l@{\hspace{0mm}}}
        \pi a_{n} &,~\mathrm{for}~n = 0 \\ \\
        \dfrac{\pi}{2} a_{n} &,~\mathrm{for}~n \ne 0.
    \end{array}
    \right. \label{eqn:an}
\end{align}
For our specific problem, this amounts to multiplying the left-hand sides of equations \eqref{eqn:coefC} and \eqref{eqn:coefA} by the appropriate factor of $1 / \pi$ or $2 / \pi$ when calculating the coefficients $C_{\ell}$ and $A_{n \ell}$. We also invoked the coefficient formula on both sides when deriving the linear system of equations \eqref{eqn:linsys_CcA}, so the normalization factors cancel out in this case.

\subsubsection{$f_{K}( K_{\rm flux} )$ At Last}
\label{app:fk:at:last}
We experimented with the modal, nodal, and CDF approaches and with both the Lagrange-multiplier and exact constraints to achieve $c_{n}$ on a system-by-system basis. We tabulate these choices in Table \ref{tab:fK}.

\begin{figure*}
    \begin{center}
        \includegraphics[width=0.495\textwidth]{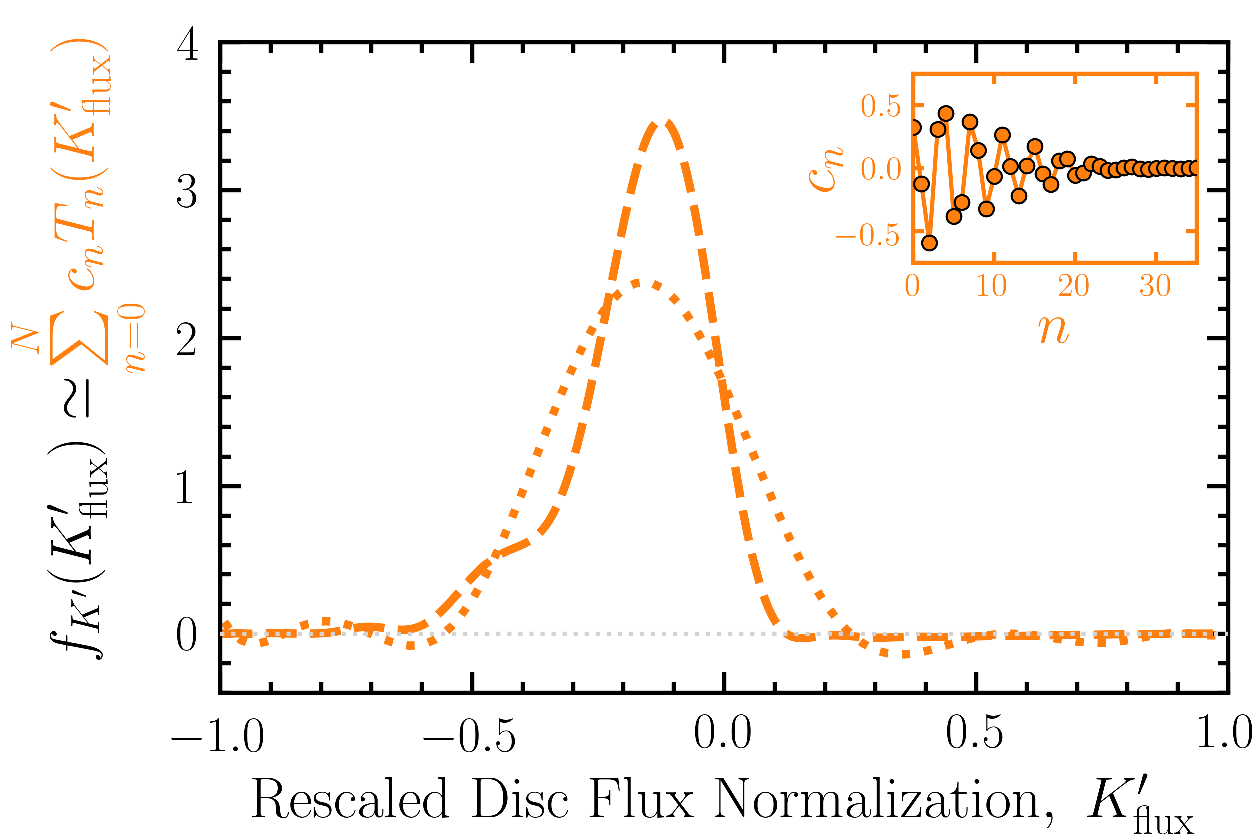}
        \hfill
        \includegraphics[width=0.495\textwidth]{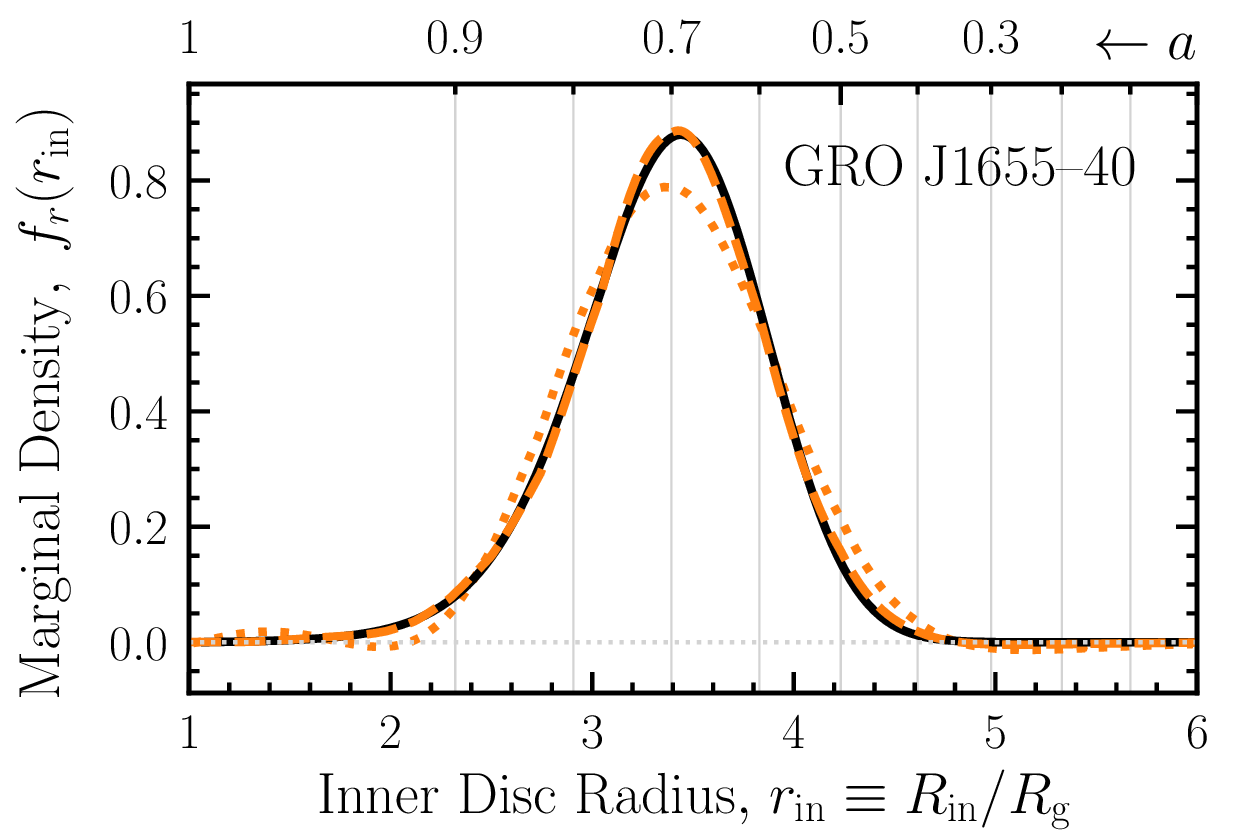}
    \end{center}
    \vspace{-2mm}
    \caption{Workflow to approximate the marginal density $f_{K^{\prime}}( K_{\rm flux}^{\prime} )$ and a demonstration of our method's validity by recovering the input marginal density $f_{r}^{\rm CF}( r_{\rm in} )$, using GRO J1655--40 as an example. \textit{Left}: With the coefficients $C_{m}$ and $A_{n m}$ from Figure \ref{fig:frP_Anm_Tnr}, we solve the linear system \eqref{eqn:linsys_CcA} for the coefficients $c_{n}$ (\textit{inset}), which we use to approximate $f_{K^{\prime}}( K_{\rm flux}^{\prime} )$ as the linear combination $\sum\limits_{n=0}^{N} c_{n} T_{n}( K_{\rm flux}^{\prime} )$ with $N = 35$ shown by the \textit{dashed orange line} following equation \eqref{eqn:fKflux_lincomb}. Next, we re-scale the marginal density $f_{K^{\prime}}( K_{\rm flux}^{\prime} )$ with $K_{\rm flux}^{\prime} \in [ -1, 1 ]$ to the marginal density $f_{K}( K_{\rm flux} )$ with $K_{\rm flux} \in [ r_{\rm in}^{\rm min}, r_{\rm in}^{\rm max} ]$ shown in Figure \ref{fig:fK}. \textit{Right}: To test the backwards compatibility of our method, we attempt to recover the marginal density $f_{r}^{\rm CF}( r_{\rm in} )$ measured by the disc continuum fitting practitioners, shown by the \textit{black solid line}. With $f_{K}( K_{\rm flux} )$ in hand, we follow \S\ref{app:Kflux2rin} and incorporate no $f_{\rm col}$ uncertainties (i.e., $f_{f}( f_{\rm col} ) = \delta( f_{\rm col} - f_{\rm col}^{\rm CF} )$) to arrive at the recovered $f_{r}( r_{\rm in} )$ shown by the \textit{dashed orange line}. While not perfect, the agreement is very good in this case and all others (compare the \textit{black} and \textit{grey lines} in Figure \ref{fig:spinPDFs}), which demonstrates the validity of our methods. We note that both $f_{K^{\prime}}( K_{\rm flux}^{\prime} )$ and $f_{r}( r_{\rm in} )$ are converged, while the \textit{dotted orange lines} show unconverged results using $N = 11$ for the linear combination.}
    \label{fig:fKP_fr}
\end{figure*}

With $c_{n}$ in hand, we plug the coefficients into equation \eqref{eqn:fKflux_lincomb} and finally arrive at the marginal density $f_{K^{\prime}}( K_{\rm flux}^{\prime} )$, with an example shown in the \textit{left panel} of Figure \ref{fig:fKP_fr}. With the help of equations \eqref{eqn:zprime} and \eqref{eqn:eqprob}, the final step is to re-scale the disc flux normalization and its marginal density back to $K_{\rm flux} \in [ K_{\rm flux}^{\rm min}, K_{\rm flux}^{\rm max} ]$,
\begin{align}
K_{\rm flux} &= \left( K_{\rm flux}^{\prime} - a \right) \frac{K_{\rm flux}^{\rm max} - K_{\rm flux}^{\rm min}}{b - a} + K_{\rm flux}^{\rm min} \\
f_{K}\left( K_{\rm flux} \right) &= f_{K^{\prime}}\left( K_{\rm flux}^{\prime} \right) \frac{b - a}{K_{\rm flux}^{\rm max} - K_{\rm flux}^{\rm min}}.
\end{align}

\begin{figure*}
    \begin{center}
        \includegraphics[width=0.495\textwidth]{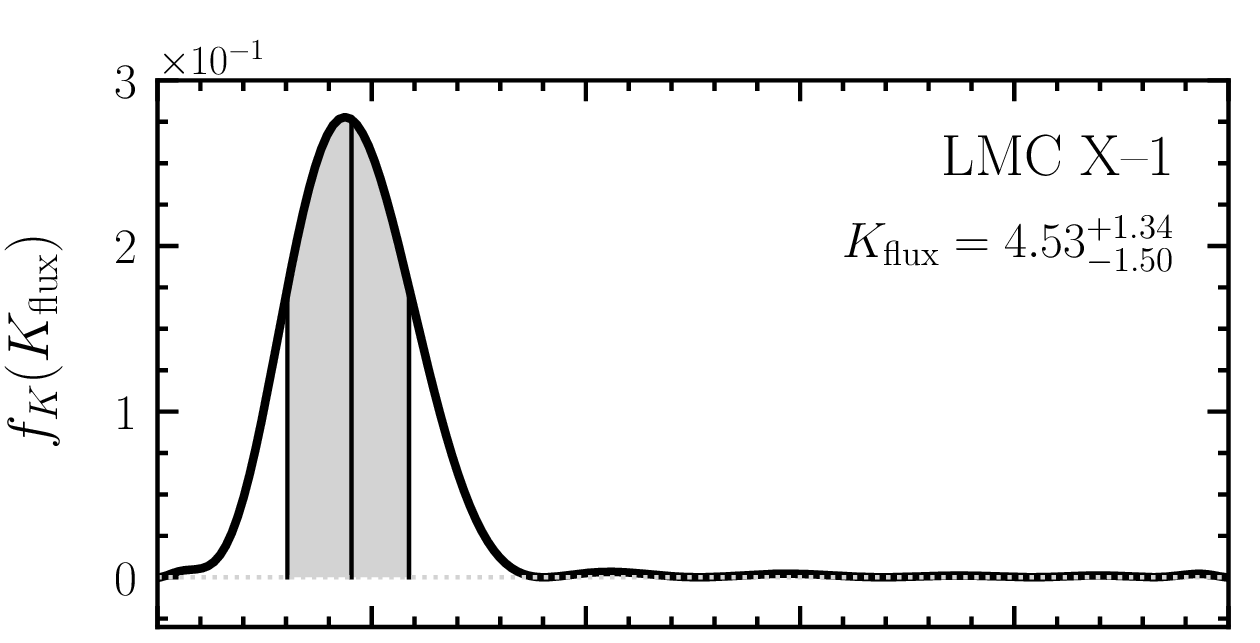}
        \includegraphics[width=0.495\textwidth]{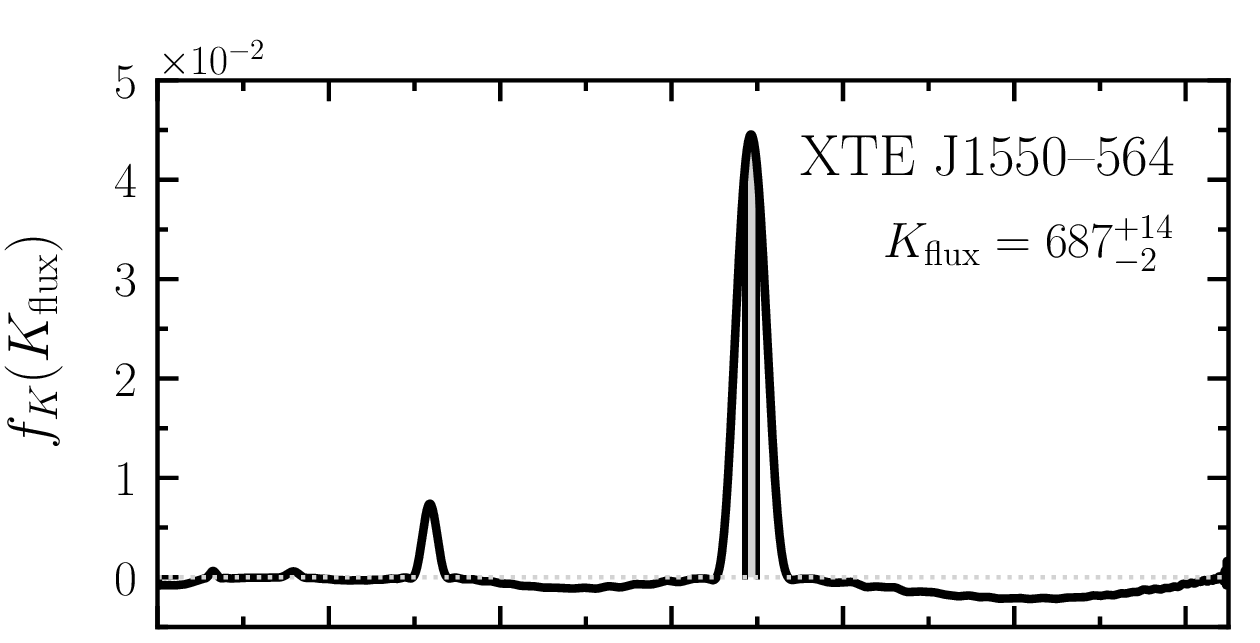}
        \includegraphics[width=0.495\textwidth]{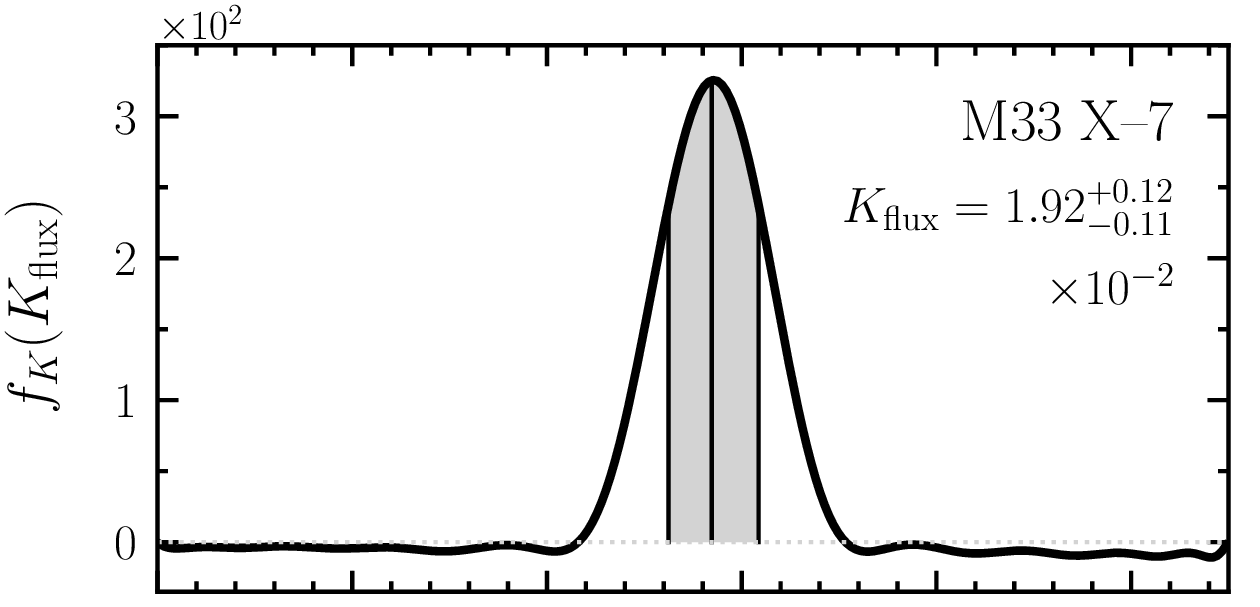}
        \includegraphics[width=0.495\textwidth]{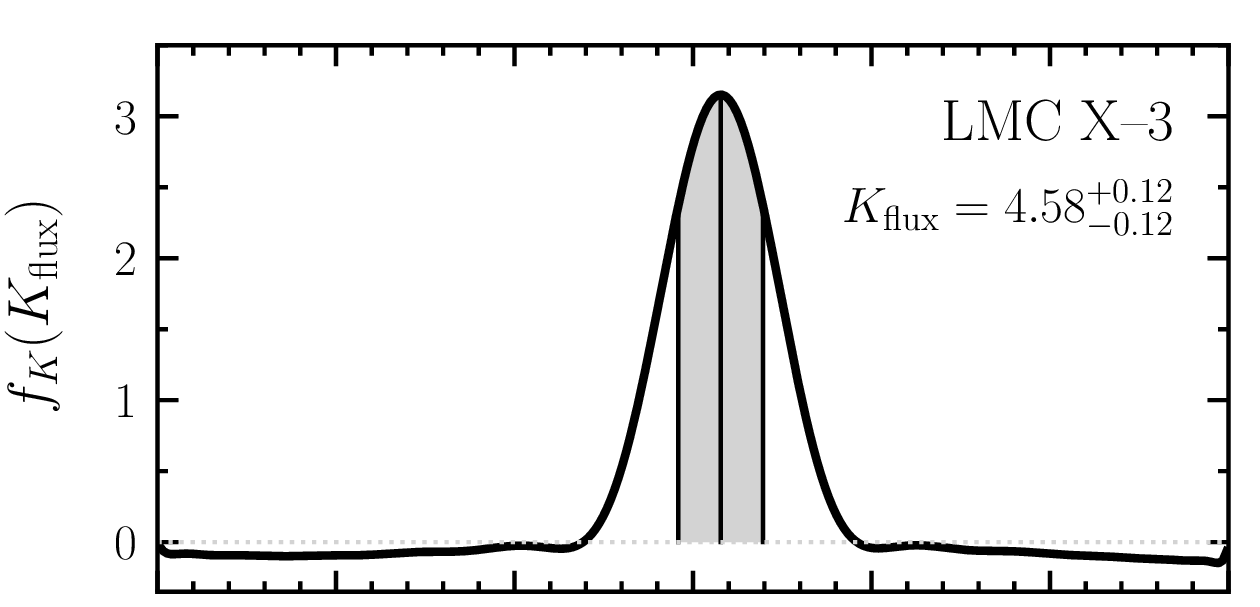}
        \includegraphics[width=0.495\textwidth]{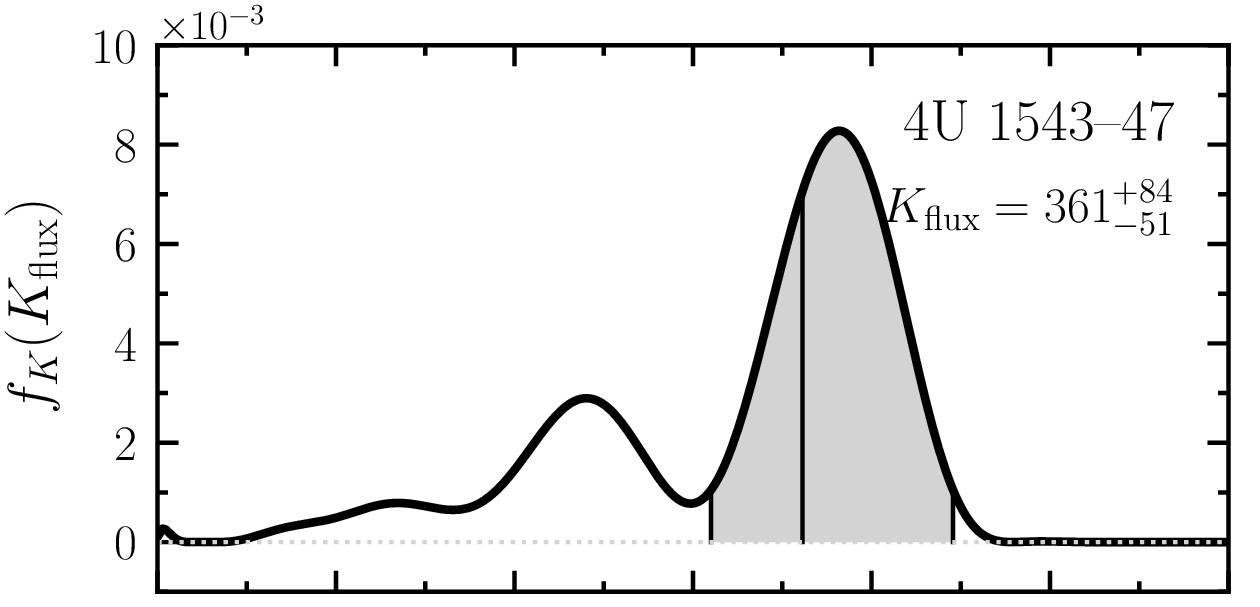}
        \includegraphics[width=0.495\textwidth]{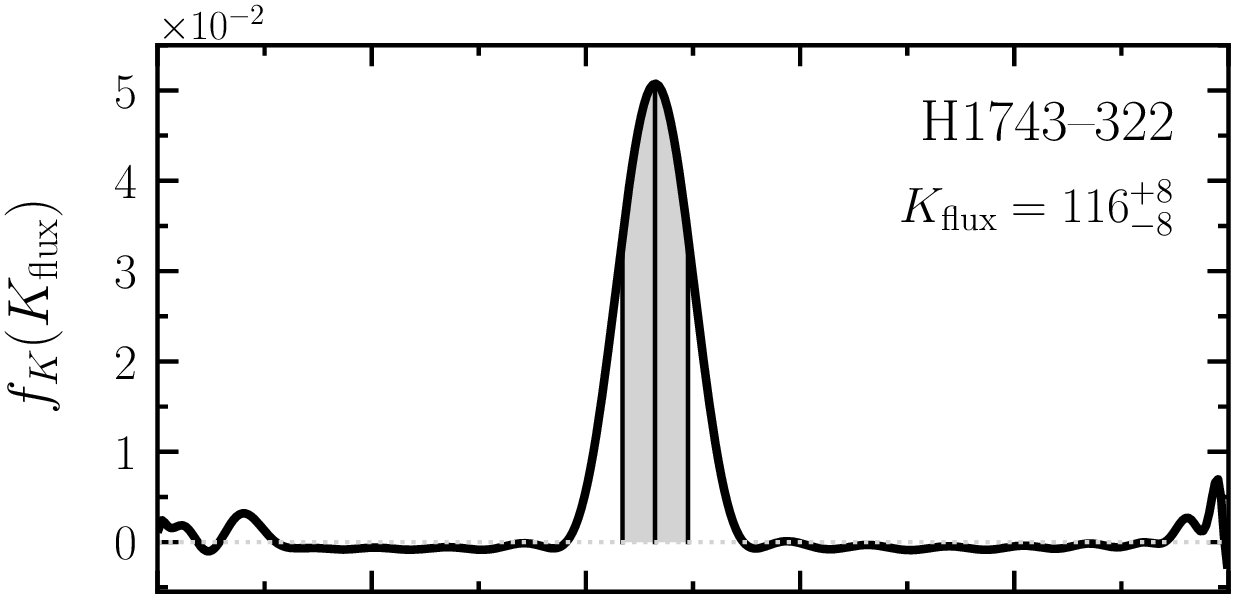}
        \includegraphics[width=0.495\textwidth]{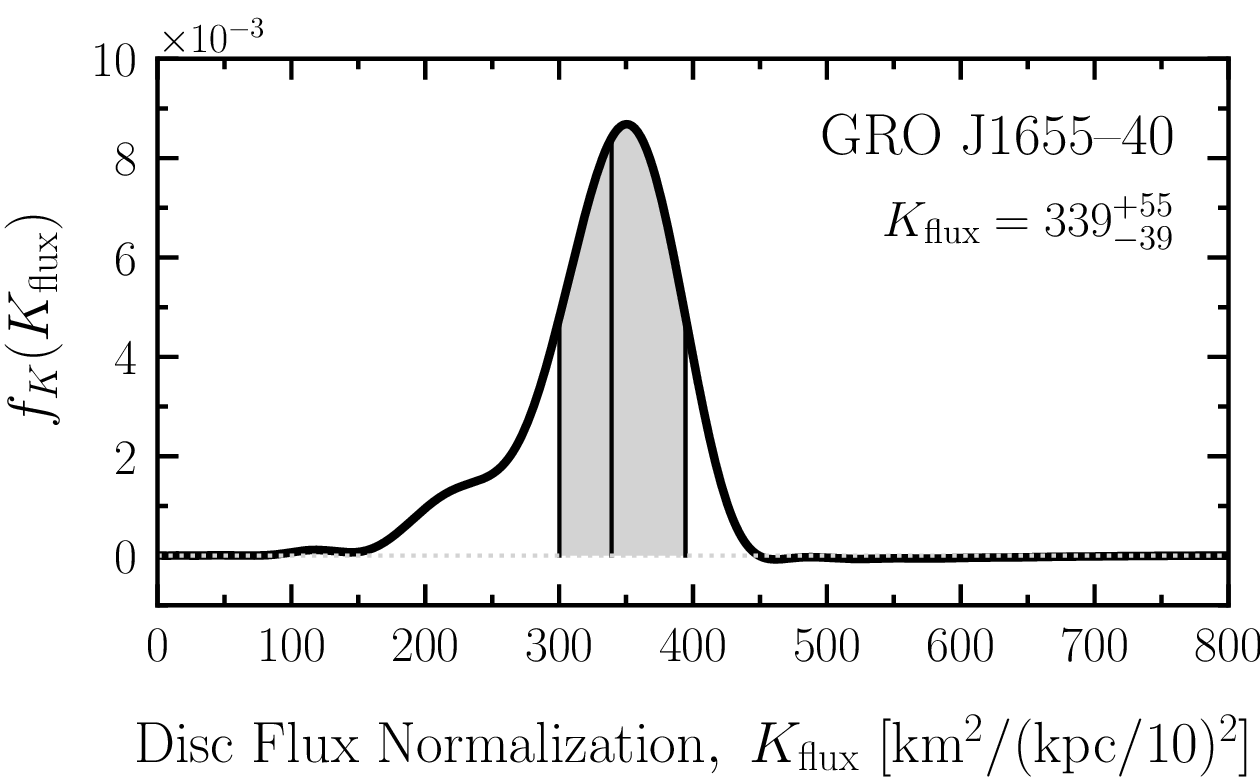}
        \includegraphics[width=0.495\textwidth]{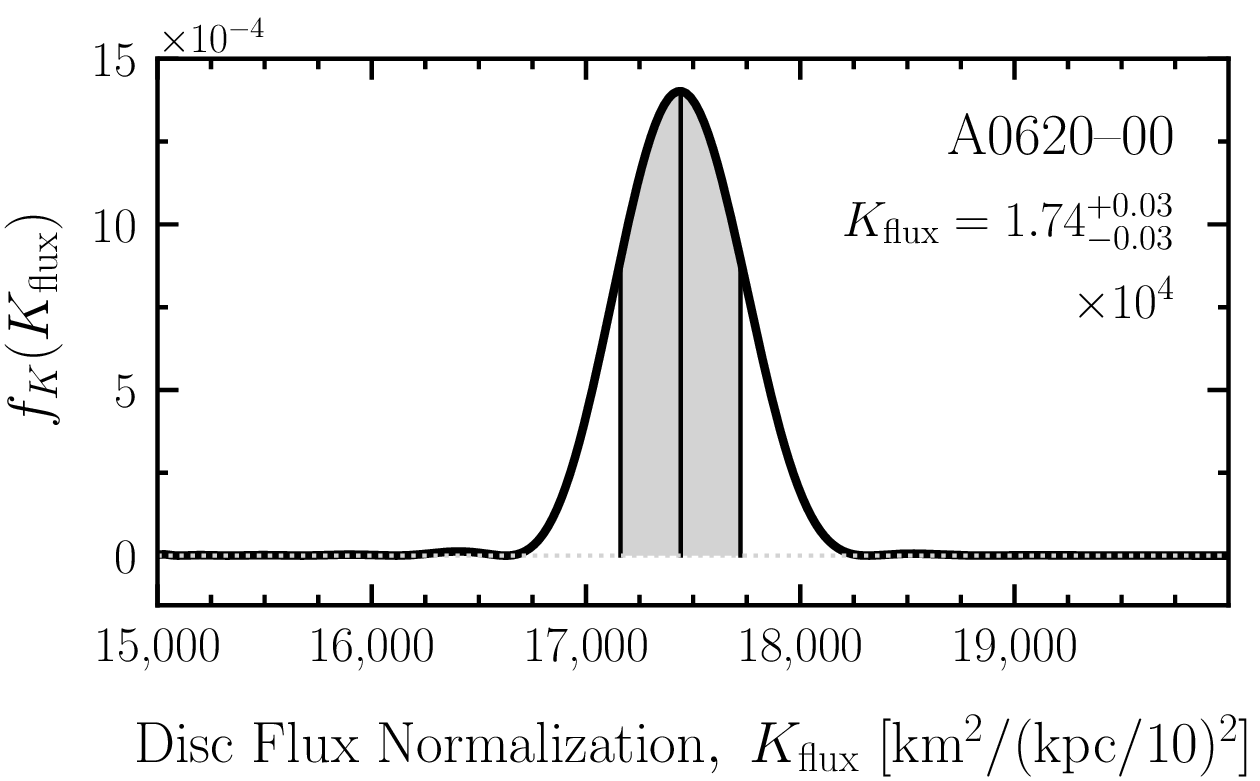}
    \end{center}
    \vspace{-2mm}
    \caption{Disc flux normalization marginal density $f_{K}( K_{\rm flux} )$, as calculated from inverting equation \eqref{eqn:f_rinCF} using the methods from \S\ref{app:basisfunc} and the constraints from \S\ref{app:constopt}, for each X-ray binary in Table \ref{tab:fK}, which also lists the inputs and imposed constraints used in each calculation. Each \textit{legend} gives the $K_{\rm flux}$ value where the cumulative density $F_{K}( K_{\rm flux} ) = \sfrac{1}{2}$ (i.e., the median), along with the inter-68\% $K_{\rm flux}$ range of the marginal density $f_{K}( K_{\rm flux} )$. The \textit{horizontal dotted line} shows that, in some cases, we are unable to enforce the positivity constraint in the tails of the distribution, but the offense is relatively minor.}
    \label{fig:fK}
\end{figure*}

Figure \ref{fig:fK} shows each disc flux normalization marginal density $f_{K}( K_{\rm flux} )$ calculated as described above using the inputs and constraints (see \S\ref{app:constopt}) listed in Table \ref{tab:fK}. With $f_{K}( K_{\rm flux} )$ finally in hand, we can proceed to \S\ref{app:Kflux2rin}.

\subsection{Updating the Black Hole Spin Measurement to Include $f_{\rm col}$ Uncertainties: $f_{K}( K_{\rm flux} ) \rightarrow f_{r}( r_{\rm in} ) \rightarrow  f_{a}( a )$}
\label{app:Kflux2rin}
The final step is to determine the marginal density $f_{a}( a )$ of the black hole spin parameter that would have been measured had the continuum fitting practitioners used a generic marginal density $f_{f}( f_{\rm col} )$ of the colour correction factor. We calculate the revised $f_{r}( r_{\rm in} )$ that incorporates $f_{\rm col}$ uncertainties by inserting into equation \eqref{eqn:f_rin} both a non-trivial $f_{f}( f_{\rm col} )$ and the marginal density $f_{K}( K_{\rm flux} )$ of the disc flux normalization from \S\ref{app:rin2Kflux}, \S\ref{app:basisfunc}, and \S\ref{app:constopt}. Finally, we assume $f_{r}( r_{\rm in} ) = f_{\circ}( r_{\rm ISCO} )$ and use equation \eqref{eqn:f_a} to transform $f_{\circ}( r_{\rm ISCO} )$ to $f_{a}( a )$.

The \textit{right panel} of Figure \ref{fig:fKP_fr} validates our methods by demonstrating that we recover the input marginal density $f_{r}^{\rm CF}( r_{\rm in} )$ if we ignore $f_{\rm col}$ uncertainties, while Figure \ref{fig:spinPDFs} showcases our final product: black hole spin measurements that include $f_{\rm col}$ uncertainties.

Equivalently, we could obtain $f_{r}( r_{\rm in} )$ by sampling many sets of the \textit{independent} variables $\{ K_{\rm flux}, f_{\rm col}, ( M, D, i_{\rm disc}) \}$ from their respective probability densities, calculating each corresponding $r_{\rm in}$ from the transformation function $r_{\rm in} = h( K_{\rm flux}, f_{\rm col}, M, D, i_{\rm disc} )$, histogramming the results, and normalizing this distribution to get the marginal density $f_{r}( r_{\rm in} )$. Notably, a numerical root finder is required to calculate the transformation function because $r_{\rm in}$ cannot be isolated in equation \eqref{eqn:Kflux}. Although $g_{\rm GR}( r_{\rm in}, i_{\rm disc} )$ can be double-valued in $r_{\rm in}$ (see Figure \ref{fig:gGR}), in practice we always find one unique root for $r_{\rm in}$ because $K_{\rm flux} \propto r_{\rm in}^{2} g_{\rm GR}( r_{\rm in}, i_{\rm disc} )$ is not double-valued in $r_{\rm in}$.


\label{lastpage}
\end{document}